\documentclass[structabstract]{aa}  
\usepackage[pdftex]{color,xcolor}
\usepackage[pdftex]{graphicx}
\pdfoutput=1
\usepackage{graphicx}
\usepackage{amstext} 
\usepackage{amsmath} 
\usepackage{amssymb} 
\usepackage[figuresright]{rotating}
\usepackage{natbib} 
\usepackage{afterpage}
\usepackage{txfonts}
\usepackage[urlcolor=blue,colorlinks=true,linkcolor=blue,citecolor=blue]{hyperref}
\usepackage{hyperref} 
\hypersetup{pdfauthor=Johannes Sahlmann}

\begin{document} 
   \title{Search for brown-dwarf companions of stars\thanks{Based on observations collected with the CORALIE Echelle spectrograph on the Swiss telescope at the European Southern Observatory in La Silla, Chile, on observations made with the ESA Hipparcos astrometry satellite, and on observations made with the HARPS instrument on the ESO 3.6-m telescope (GTO programme 072.C-0488). \newline The CORALIE and HARPS radial-velocity measurements discussed in this paper are only available in electronic format at CDS via \url{http://cdsweb.u-strasbg.fr/cgi-bin/qcat?J/A+A/}.}}
\author{J. Sahlmann \inst{1,2}
		\and D. S\'egransan \inst{1}
		\and D.~Queloz\inst{1} 
		\and S.~Udry\inst{1}
	       \and N.C.~Santos\inst{3,4} 
              \and M.~Marmier\inst{1} 
	       \and M.~Mayor\inst{1} 
              \and D.~Naef\inst{1} 
              \and F.~Pepe\inst{1} 
	       \and S.~Zucker\inst{5}			}		
\institute{Observatoire de Gen\`eve, Universit\'e de Gen\`eve, 51 Chemin Des Maillettes, 1290 Sauverny, Switzerland\\
		\email{johannes.sahlmann@unige.ch}	
	\and	
		European Southern Observatory, Karl-Schwarzschild-Str. 2, 85748 Garching bei M\"unchen, Germany
	 \and 
	 Centro de Astrof\'{\i}sica, Universidade do Porto, Rua das Estrelas, 4150-762 Porto, Portugal
      \and
      Departamento de F\'{\i}sica e Astronomia, Faculdade de Ci\^encias, Universidade do Porto, Portugal
       \and       	
		Department of Geophysics and Planetary Sciences, Tel Aviv University, Tel Aviv 69978, Israel 
	      }

\date{Received 19 July 2010 / Accepted 23 September 2010} 			

\abstract
{The frequency of brown-dwarf companions in close orbit around Sun-like stars is low compared to the frequency of planetary and stellar companions. Currently, there is no comprehensive explanation for this lack of brown-dwarf companions.}
{Through the combination of the orbital solution obtained from a stellar radial-velocity curve and Hipparcos astrometric measurements, we attempt to determine the orbit inclination and therefore the mass of the orbiting companion. By determining the masses of potential brown-dwarf companions, we can refine the properties of the companion mass-function.}
{The radial-velocity solutions revealing potential brown-dwarf companions are obtained for stars from the CORALIE and HARPS planet-search surveys or from the literature. The best Keplerian fit to our radial-velocity measurements is found by adjustment with the Levenberg-Marquardt algorithm. The spectroscopic elements of the radial-velocity solution constrain the fit to the Intermediate Astrometric Data of the new Hipparcos reduction. The astrometric solution and the orbit inclination are found using non-linear $\chi^2$-minimisation on a two-parameter search grid. The statistical confidence of the adopted orbital solution is evaluated based on the distribution-free permutation test.}
{The discovery of 9 new brown-dwarf candidates orbiting stars in the CORALIE and HARPS radial-velocity surveys is reported. New CORALIE radial velocities yielding accurate orbits of 6 previously-known hosts of potential brown-dwarf companions are presented. Including the literature targets, 33 hosts of potential brown-dwarf companions are examined. Employing innovative methods, we use the new reduction of the Hipparcos data to fully characterise the astrometric orbits of 6 objects, revealing M-dwarf companions with masses between $90\,M_J$ and $0.52\,M_\odot$. Additionally, the masses of two companions can be restricted to the stellar domain. The companion to HD~137510 is found to be a brown dwarf. At 95~\% confidence, the companion of HD~190228 is also a brown dwarf. The remaining 23 companions persist as brown-dwarf candidates. Based on the {\footnotesize CORALIE} planet-search sample, we obtain an upper limit of 0.6~\% for the frequency of brown-dwarf companions around Sun-like stars. We find that the companion-mass distribution function is rising at the lower end of the brown-dwarf mass range, suggesting that in fact we are detecting the high-mass tail of the planetary distribution.}
{Our findings agree with the results of previous similar studies and confirm the pronounced paucity of brown-dwarf companions around Sun-like stars. They are affected by the Hipparcos astrometric precision and mission duration, which limits the minimum detectable companion mass, and some of the remaining candidates probably are brown-dwarf companions.} 

\maketitle
\section{Introduction}
The mass distribution of close companions to Sun-like stars exhibits a clear deficit of brown dwarfs compared to planets and stellar binaries \citep{Grether:2006kx}. With a frequency of less than 1~\% \citep{Marcy:2000pb}, brown-dwarf companions to solar-type stars at separations below 10 AU are much less common than planetary companions with a frequency of $\sim$7~\% (e.g. \citealt{Udry:2007sf}) and stellar binaries with a frequency of $\sim$13~\% \citep{Duquennoy:1991kx, Halbwachs:2003kx} at similar separations. Brown dwarfs are substellar objects, massive enough to burn deuterium but too light to permit hydrogen burning, which places them in the mass range of approximately $13-80$ Jupiter masses \citep{Burrows:1997qe, Chabrier:2000kl}. They therefore constitute objects at the transition between planets and stars, though the lower end of the brown-dwarf mass range overlaps with the one of massive planets (e.g. \citealt{Luhman:2009oq}), and the distinction between planets and brown dwarfs may require to trace the individual formation process. A comprehensive explanation for the formation of low-mass objects, covering the range of planets to very low-mass stars, remains to be found, although the progress in this direction is formidable as reported in the reviews of \cite{Whitworth:2007pi}, \cite{Luhman:2007zm} and \cite{Burgasser:2007ix}. 

Recently, a few companions with masses of $18-60$ Jupiter masses ($M_J$) were determined through astrometry (e.g. \citealt{Martioli:2010kx, Benedict:2010ph}) or discovered in transiting systems  (e.g. \citealt{Deleuil:2008it}, Anderson et al., \emph{in preparation}, Bouchy et al., \emph{in preparation}). Yet, most potential brown dwarf companions are discovered in radial velocity surveys (e.g. \citealt{Nidever:2002vn}). In these high-precision programs, aiming at planet detection, they are easily spotted and characterised because of their large signatures, by far exceeding the radial velocity measurement precision (e.g. \citealt{Patel:2007ys, Bouchy:2009rt}). However, radial-velocity measurements alone do not constrain the orbit inclination. Therefore, they can not reveal the companion mass, but yield a lower limit to it. Observations of complementary effects like the transit light-curve \citep{Deleuil:2008it} or the astrometric motion of the host star are required to solve this ambiguity and to determine the companion mass. 

High-precision astrometry with the Hubble space telescope ({\footnotesize HST}) has been used to find the masses of planetary companions discovered by radial velocity (\citealt{Benedict:2010ph}, and references therein) and has demonstrated the power of astrometry to distinguish between planetary, brown-dwarf, or stellar companions. Ground-based optical interferometers achieve the necessary precision to detect orbital motions of binary and multiple systems and to solve for the orbital parameters together with radial-velocity measurements \citep{Hummel:1993rz, Muterspaugh:2006eu, Lane:2007sf}. The astrometric detection of planetary orbits using IR-interferometry is the aim of the extrasolar-planet search with {\footnotesize PRIMA} project ({\footnotesize ESPRI}, \citealt{Launhardt2008}). On a larger scale, the {\footnotesize GAIA} satellite (e.g. \citealt{Lindegren:2010kx}) will advance the field of astrometric detection and characterisation of planetary systems, due to its outstanding measurement precision \citep{Casertano:2008th}.

Hipparcos astrometry has been extensively used to constrain the masses of sub-stellar and stellar companions in multiple systems \citep{Perryman:1996vn, Pourbaix:2000sf, Zucker:2001ve, Torres:2007kx}. \cite{Han:2001kx} used Hipparcos astrometry to analyse 30 extrasolar-planet candidates and concluded that half of these companions were rather brown of M dwarfs. With the consequent rectification by \cite{Pourbaix:2001rt}, showing that the results of \cite{Han:2001kx} had been biased by the fitting procedure, it became clear that the fitting of astrometric data has to be done very carefully, especially when the size of the orbital signature approaches the measurement precision of the instrument \citep{Pourbaix:2001qe}. All of these past works used the original Hipparcos data \citep{Perryman:1997kx}.

In 2007, the new reduction of the raw Hipparcos data \citep{:2007kx} has been released, which represents a significant improvement in quality compared to the original data \citep{van-Leeuwen:2007yq}. It also includes a revised version of the intermediate astrometric data in a new format that facilitates the search for the signature of orbital motion. We use the new Hipparcos reduction for our study, which aims at determining the masses of companions to stars, that were detected by high-precision radial velocity measurements and found to have minimum masses in the brown-dwarf domain. Our target sample is composed of stars from the {\footnotesize CORALIE} and  {\footnotesize HARPS} planet surveys. In addition, we included a list of targets selected from the literature. To validate our method of astrometric orbit determination, we also analysed a few comparison targets with published astrometric orbits. This study further explores the \emph{brown dwarf desert} and is in line with the foregoing works by \cite{Halbwachs:2000rt} and \cite{Zucker:2001ve}. 

The paper is organised as follows. The brown-dwarf candidates from the {\footnotesize CORALIE} and  {\footnotesize HARPS} surveys are presented in Sect. \ref{sec:cortargets} together with 18 candidates selected from the literature. The method of combining radial-velocity orbits and Hipparcos astrometry is described in Sect. \ref{sec:method}. The results are summarised in Sect. \ref{sec:results} and discussed in Sect. \ref{sec:discussion}. We conclude in Sect. \ref{sec:conclusions}.

\section{Brown dwarf candidates}\label{sec:cortargets}
Our target sample consists of 33 stars, that exhibit radial velocity variations caused by a companion with minimum mass $M_2 \sin i$ in the brown-dwarf mass range of $13 - 80\, M_J$. It contains 14 stars from the {\footnotesize CORALIE} survey \citep{Udry:2000kx} and one star from the {\footnotesize HARPS} planet search program \citep{Mayor:2003cs}. To date, the {\footnotesize CORALIE} planet survey has contributed to the discovery of more than 50 extrasolar planets (e.g. \citealt{Naef:2001nx, Mayor:2004oq, Tamuz:2008qe}). {\footnotesize CORALIE} is an optical echelle spectrograph mounted on the 1.2~m Swiss Telescope located at the European Southern Observatory in La Silla, Chile. A description of the instrument can be found in \cite{Queloz:2000ad} and the references therein. With its unmatched precision, the {\footnotesize HARPS} instrument permitted the discovery of more than 70 planetary companions to date, including earth-mass planets (e.g. \citealt{Pepe:2007pi, Santos:2007hb, Mayor:2009ph, Forveille:2009jt}). The characteristics, radial velocities, and orbital solutions of the 15 {\footnotesize CORALIE} and {\footnotesize HARPS} stars are presented and discussed in Sects. \ref{sec:stellarChar}-\ref{sec:objnotes}. Additionally, 18 potential brown-dwarf host stars with radial-velocity orbits were selected from the literature and are listed in Sect.~\ref{sec:littargets}. These include HD~190228, initially a planet-host star \citep{Sivan:2004rm}, whose importance we recognised during the analysis.

 \begin{table}\caption{Hipparcos parameters of the surveyed stars}
\label{tab:stellarCharHIP} 
\centering  
\begin{tabular}{r r r c r r r } 	
\hline\hline %
Nr. & \multicolumn{2}{c}{Object} & Sp. T. & $V$ & \hbox{$B\!-\!V$} & $\varpi$   \\  
    &HD        &  HIP     &    &     &         & (mas)                                 \\
\hline 
1&{3277} & 2790 & G8V & 7.59 & 0.73 & $34.7 \pm 0.7$  \\ 
2&{4747} & 3850 & G8 & 7.30 & 0.77  & $ 53.5 \pm 0.5$  \\ 
3&{17289} & 12726 & G0V & 7.56 & 0.59 & $20.8 \pm 1.2$  \\ 
4&{30501} & 22122 & K0V & 7.73 & 0.88 & $47.9 \pm 0.5$  \\ 
5&{43848} & 29804 & K1IV & 8.80 & 0.93 & $26.4 \pm 0.8$  \\
6&{52756} & 34052 & K1V & 8.80 & 1.18 & $57.4 \pm 1.2$  \\ 
7&{53680} & 33736 & K5V & 8.61 & 0.90 & $30.1 \pm 0.6$  \\ 
8&{74014} & 42634 & K0 & 7.73 & 0.76 & $29.6 \pm 0.7$  \\ 
9&{89707} & 50671 & G1V & 7.29 & 0.55 & $30.6 \pm 0.6$  \\ 
10&{154697} & 83770 & G5 & 7.97 & 0.73 & $30.6 \pm 1.0$  \\ 
11&{164427A} & 88531 & G4IV & 7.01 & 0.62 & $26.3 \pm 0.7$  \\ 
12&{167665} & 89620 & F8V & 6.48 & 0.54 & $32.8 \pm 0.5$  \\ 
13&{189310} & 99634 & K2V & 8.61 & 0.90 & $31.0 \pm 0.6$  \\ 
14&{211847} & 110340 & G5V & 8.78 & 0.66 & $19.8 \pm 1.3$  \\ 
15&{$\cdots$}& 103019& K5  &  10.45 	 &	1.33& $ 28.3 \pm	2.3  $  \\
\hline
\end{tabular} 
\end{table}

\begin{table*}[ht]
\caption{Derived stellar parameters of the surveyed stars}
\label{tab:stellarChar} 
\centering  
\begin{tabular}{r r r r r r r r r} 	
\hline\hline %
Object &  Instr.\tablefootmark{b} &$M_\mathrm{V}$ & $T_\mathrm{eff}$ & $\log g$ & [Fe/H] & $\nu \sin i$ & $M_{1}$  & Age \\  
            &           &                   &  (K)          &   (cgs)    &  (dex) & (kms$^{-1}$)& ($M_{\odot}$) & (Gyr) \\
\hline 
\object{HD~3277} & C&5.30 & $5539 \pm 50$ & $4.36 \pm 0.10$ & $-0.06 \pm 0.08$ & $<2.0$ & $0.91 \pm 0.03$ & $0.4 - 8.0$  \\ 
\object{HD~4747} & C&5.94 & $5316 \pm 50$\tablefootmark{a} & $4.48 \pm 0.10$\tablefootmark{a} & $-0.21 \pm 0.05$\tablefootmark{a} & $0.79 \pm 0.06$\tablefootmark{a} & $0.81 \pm 0.02$ & $0.1 - 7.3$  \\ 
\object{HD~17289} &C& 4.15 & $5924 \pm 50$ & $4.37 \pm 0.10$ & $-0.11 \pm 0.06$ & $<2.0$ & $1.01 \pm 0.03$ & $6.0 - 9.3$  \\ 
\object{HD~30501} & C&6.13 & $5223 \pm 50$ & $4.56 \pm 0.10$ & $-0.06 \pm 0.06$ & $<2.0$ & $0.81 \pm 0.02$ & $0.8 - 7.0$  \\ 
\object{HD~43848} & C&5.91 & $5334 \pm 92$ & $4.56 \pm 0.15$ & $0.22 \pm 0.06$ & $<2.0$ & $0.89 \pm 0.02$ & $0.0 - 5.1$  \\ 
\object{HD~52756} & C&7.60 & $5216 \pm 65$ & $4.47 \pm 0.11$ & $0.13 \pm 0.05$ & $<2.0$ & $0.83 \pm 0.01$ & $0.2 - 2.7$  \\ 
\object{HD~53680} & C&6.00 & $5167 \pm 94$ & $5.37 \pm 0.29$ & $-0.29 \pm 0.08$ & $2.08 \pm 0.31$ & $0.79 \pm 0.02$ & $0.7 - 9.4$  \\ 
\object{HD~74014} & C&5.08 & $5662 \pm 55$ & $4.39 \pm 0.10$ & $0.26 \pm 0.08$ & $<2.0$ & $1.00 \pm 0.03$ & $0.1 - 3.8$  \\ 
\object{HD~89707} & C&4.71 & $6047 \pm 50$ & $4.52 \pm 0.10$ & $-0.33 \pm 0.06$ & $<2.0$ & $0.96 \pm 0.04$ & $0.5 - 6.0$  \\ 
\object{HD~154697} & C&5.40 & $5648 \pm 50$ & $4.42 \pm 0.10$ & $0.13 \pm 0.06$ & $<2.0$ & $0.96 \pm 0.02$ & $0.0 - 3.3$  \\ 
\object{HD~164427A} &C& 4.11 & $6003 \pm 50$ & $4.35 \pm 0.10$ & $0.19 \pm 0.06$ & $<2.0$ & $1.15 \pm 0.03$ & $1.2 - 4.1$  \\ 
\object{HD~167665} & C&4.06 & $6224 \pm 50$ & $4.44 \pm 0.10$ & $-0.05 \pm 0.06$ & $<2.0$ & $1.14 \pm 0.03$ & $0.7 - 3.6$  \\ 
\object{HD~189310} & C&6.07 & $5188 \pm 50$ & $4.49 \pm 0.10$ & $-0.01 \pm 0.06$ & $<2.0$ & $0.83 \pm 0.02$ & $0.0 - 7.5$  \\ 
\object{HD~211847} & C&5.26 & $5715 \pm 50$ & $4.49 \pm 0.10$ & $-0.08 \pm 0.06$ & $<2.0$ & $0.94 \pm 0.04$ & $0.1 - 6.0$  \\ 
\object{HIP~103019}&  H&7.61& $4913\pm 115$ &$4.45\pm0.28$ & $-0.30\pm0.06$& \multicolumn{1}{c}{$\cdots$} & $0.70 \pm 0.01$ & $0.0-7.9$ \\ 
\hline
\end{tabular} 
\tablefoot{\tablefoottext{a}  {Data from \citet{Santos:2005tg}.} \tablefoottext{b} {C and H stand for CORALIE and HARPS, respectively.}}
\end{table*}

\subsection{Stellar characteristics}\label{sec:stellarChar}
The identifiers and basic stellar characteristics of the {\footnotesize CORALIE} stars and of HIP~103019 from the {\footnotesize HARPS} survey are listed in Table \ref{tab:stellarCharHIP}. The apparent visual magnitude $V$, the colour \hbox{$B\!-\!V$}, and the parallax $\varpi$ are from the new Hipparcos reduction \citep{:2007kx}, whereas the spectral type is from the original Hipparcos catalogue \citep{Perryman:1997kx}. Table \ref{tab:stellarChar} displays the derived stellar parameters. The uncertainty of $V$ is below 0.002 mag. The absolute magnitude $M_V$ is derived from the Hipparcos magnitude and parallax. The effective temperature $T_{\mathrm{eff}}$, the surface gravity $\log g$, and the metallicity [Fe/H] are derived from the spectroscopic analysis of high-signal-to-noise spectra with the method presented in \cite{Santos:2004zl} and using the \ion{Fe}{I} and \ion{Fe}{II} lines listed in \citet{Sousa:2008nx}. The stellar rotation parameter $\nu \sin i$ is derived from the calibration of the {\footnotesize CORALIE} or {\footnotesize HARPS} cross correlation function by \cite{Santos:2002ad}. Finally, the stellar mass of the primary $M_1$ and the age are estimated from the theoretical isochrones of \cite{Girardi:2000bf} and a Bayesian estimation method described in \cite{da-Silva:2006eu}\footnote{The web interface is \url{http://stev.oapd.inaf.it/cgi-bin/param}}. Stellar ages of main-sequence dwarfs are usually not well constrained and we quote the 1-$\sigma$ confidence interval. The spectroscopic analysis was problematic for HIP~103019, which may be due to its faintness and late spectral type. Errors on the parameters of this star are possibly underestimated. The stellar characteristics of the targets from the literature can be found in the respective references given in Sect. \ref{sec:littargets}.

\subsection{Radial-velocity measurements and orbital solutions}\label{sec:rvsol}
Optical high-resolution spectra of the stars in Table \ref{tab:stellarChar} were collected with the {\footnotesize CORALIE} and {\footnotesize HARPS} spectrographs over time spans extending to 11 years. Radial velocities are estimated from the cross-correlations of the extracted stellar spectra with numerical templates, which depend on the target's spectral type. During {\footnotesize CORALIE} observations, a reference Thorium-Argon spectrum is recorded simultaneously with the stellar spectrum and is used to measure and correct for residual zero-point drifts \citep{Baranne:1996rc}. Photon noise limits the obtained precision to typically 2-4~ms$^{-1}$ per epoch. To account for systematic drifts, an external error of 5~ms$^{-1}$ is quadratically added to {\footnotesize CORALIE} radial-velocity uncertainties before performing the period search and the model adjustment. The abbreviations C98 and C07 refer to the {\footnotesize CORALIE} instrument before and after its upgrade in 2007, respectively \citep{Segransan:2010xr}. The two parameters $\gamma_{\mathrm{C98}}$ and $\gamma_{\mathrm{C07}}$ are introduced to account for the possibly differing velocity offsets of these instrument configurations. 

Because of the stability of the {\footnotesize HARPS} instrument and the targeted precision of 2~ms$^{-1}$ per epoch for the volume-limited programme, HIP~103019 is observed without Thorium-Argon reference (cf. Naef et al. 2010). To account for possible drifts, an additional error of 0.5~ms$^{-1}$ is quadratically added to the  {\footnotesize HARPS} radial-velocity uncertainties. 

The Keplerian-orbit solution is found by imposing the model function Eq. \ref{eq:vrad} with 6 parameters ($K_1$, $e$, $\omega$, $T$, $P$, and $\gamma$) to the radial velocities of a given star. The fit determines the values of the period $P$, the eccentricity $e$, the time of periastron passage $T_0$, the longitude of periastron $\omega$, the radial-velocity semi-amplitude $K_1$, and the systemic velocity $\gamma$. In general, measurements with uncertainties larger than $15$ ms$^{-1}$ are not used for the model adjustments.

A single-companion model fits well the data of 11 stars with reduced-chi values of $\chi_\mathrm{r} = 0.7-1.3$. For 4 stars, HD~17289, HD~89707, HD~211847, and HIP~103019, the fit is not as good with $\chi_\mathrm{r} = 1.7-2.0$ and produces residuals of unusually large amplitude, without contesting the presence of a large orbital motion. Introducing additional model parameters, corresponding to a linear or quadratic drift or a second companion does not significantly improve the fit. Therefore, the moderate fit quality has to be attributed to either underestimated radial velocity errors or to additional noise sources, such as stellar activity. This is discussed for the individual targets in the next Section. 

The orbital periods of HD~4747 and HD~211847 are very long and not covered by the measurement time-span. Hence, the best-fit period is poorly constrained and realistic confidence intervals have to be derived through a different method. We chose a criterion based on the fit-quality metric $\chi_r$ and its error $\sigma_{\chi}$ obtained for the best fit. During a second fit, we fix the period during the adjustment, such that the resulting $\chi$-value increases to $\chi_\mathrm{r} + 3 \sigma_{\chi}$. In this way, we obtain upper and lower limits for the orbital parameters. They are listed at the bottom of Tables \ref{tab:KeplerOrbits} and \ref{tab:KeplerParams}. The method to search for the astrometric signature in the Hipparcos data presented in Sect~\ref{sec:method} relies on the precise knowledge of the radial-velocity parameters. Because the periods of HD~4747 and HD~211847 are poorly constrained and additionally the orbital coverage of Hipparcos is very low, we do not perform the combined astrometric analysis for these targets. They are presented as part of the brown-dwarf candidates characterised with {\footnotesize CORALIE}. 

\begin{figure}\begin{center} 
\includegraphics[width= \linewidth, trim = 0cm 2cm 0cm 1.8cm, clip=true]{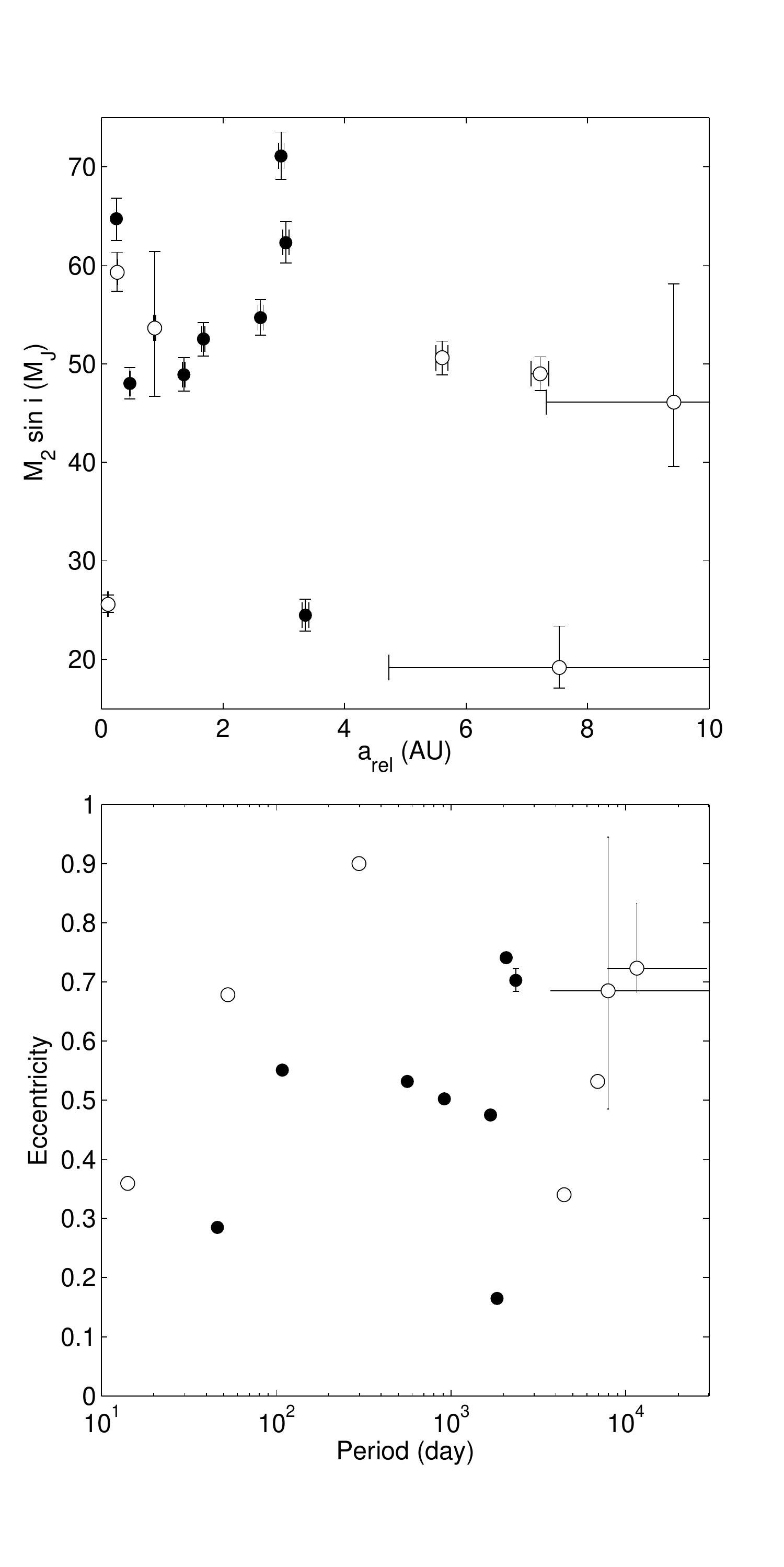} 
\caption{Orbital characteristics of the 15 brown-dwarf candidates observed with {\footnotesize CORALIE} and {\footnotesize HARPS}. Filled symbols indicate the companions with stellar mass identified through this study. The error bars are often smaller than the symbol size, though the orbits of HD~4747 and HD~211847 are not covered and therefore less constrained, see the text for details. \emph{Top}: Minimum companion mass and separation. \emph{Bottom}: Eccentricity and orbital period.}
\label{fig:orbchar} 
  \end{center}\end{figure}

\begin{figure*}\begin{center} 
\includegraphics[width= 0.49\linewidth]{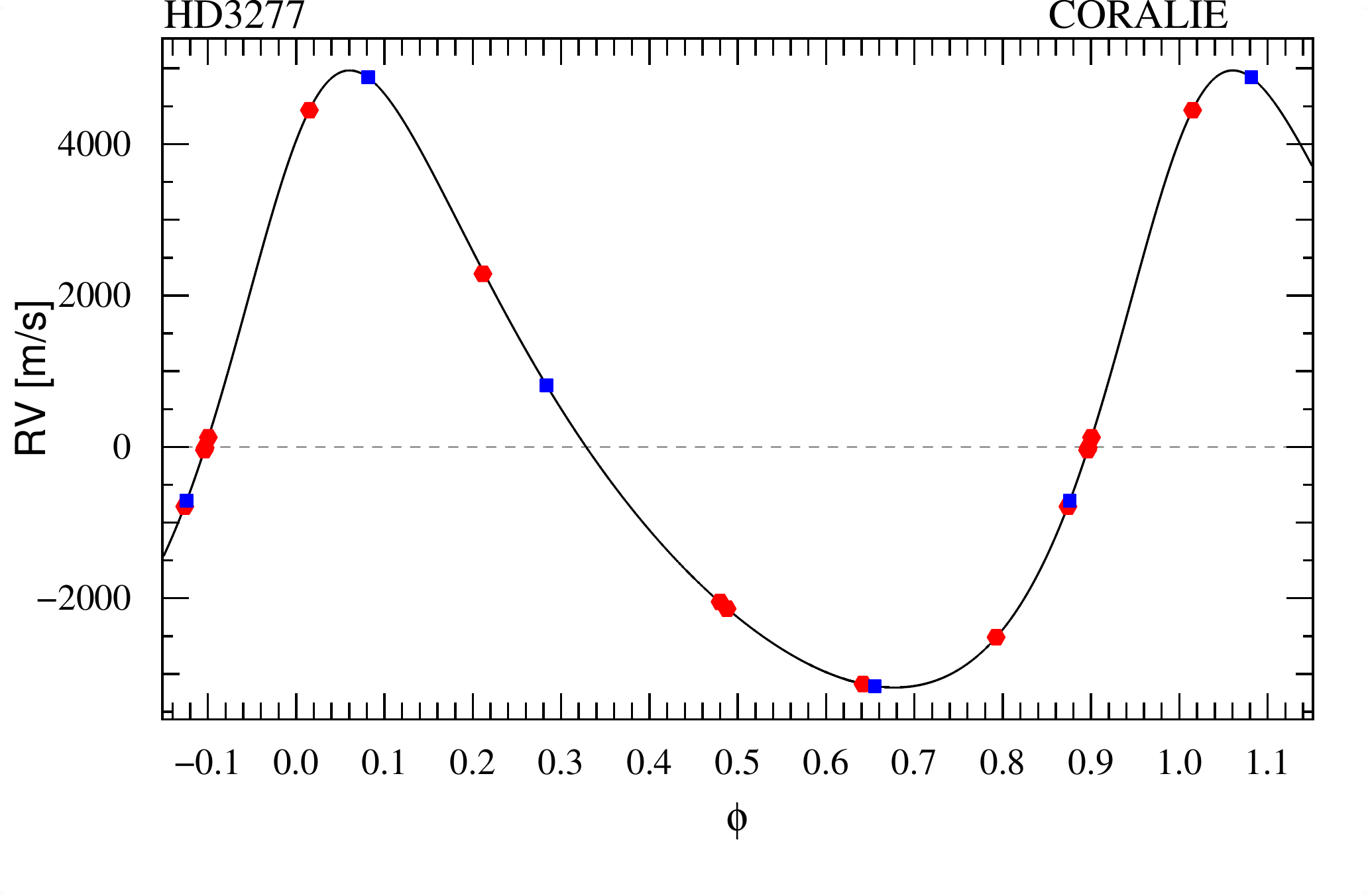} 
\includegraphics[width= 0.49\linewidth]{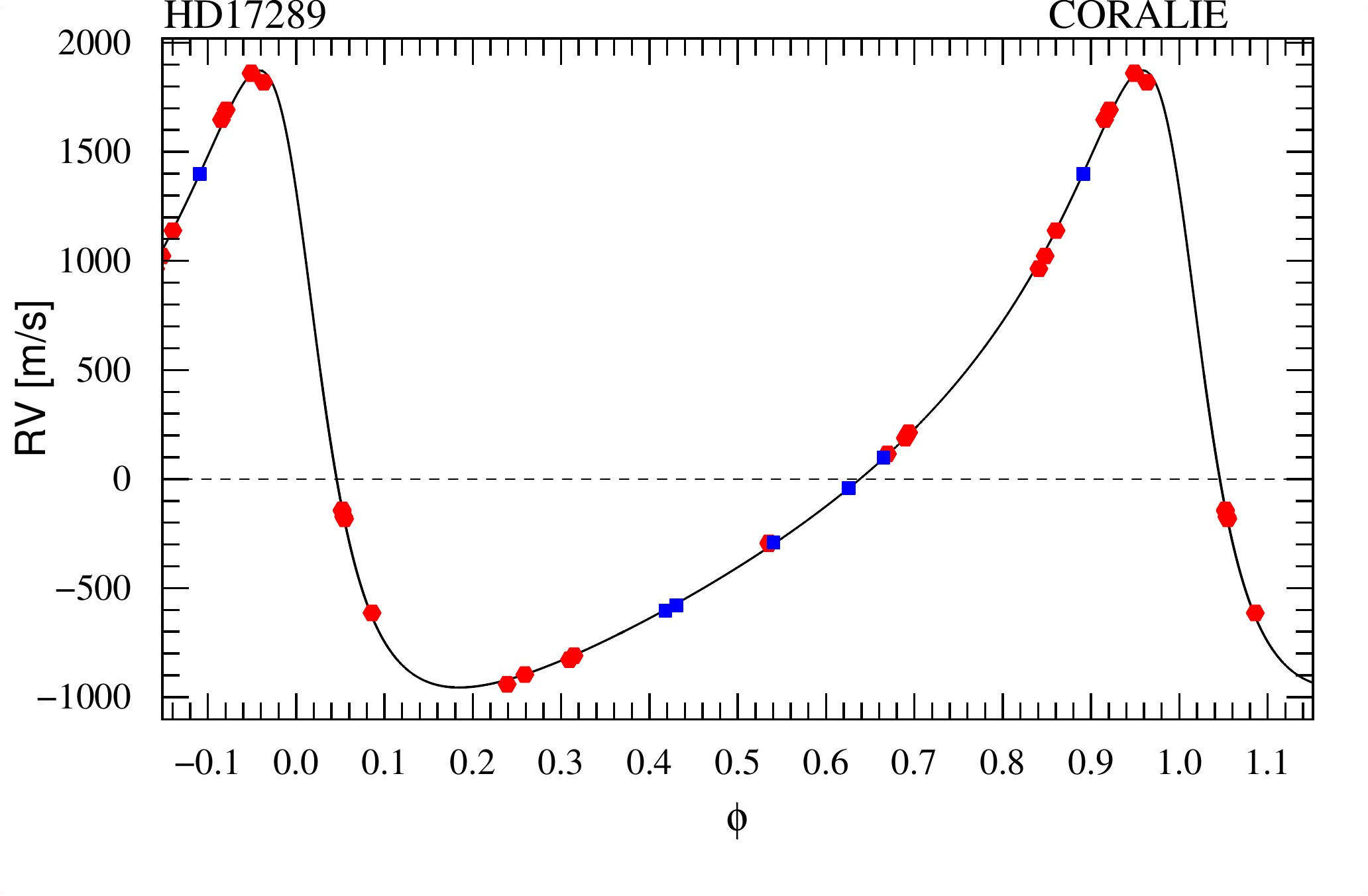} 
\includegraphics[width= 0.49\linewidth]{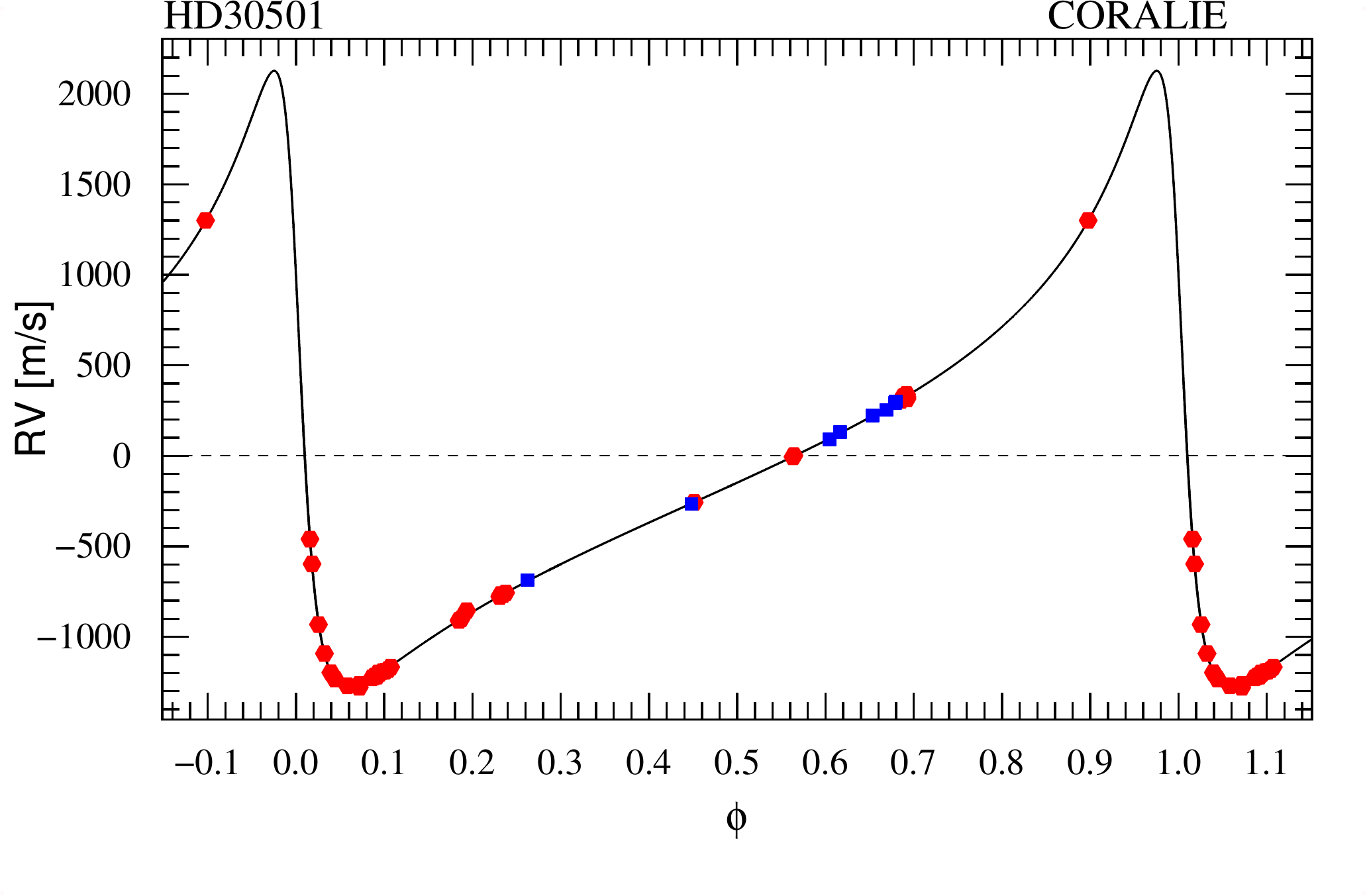} 
\includegraphics[width= 0.49\linewidth]{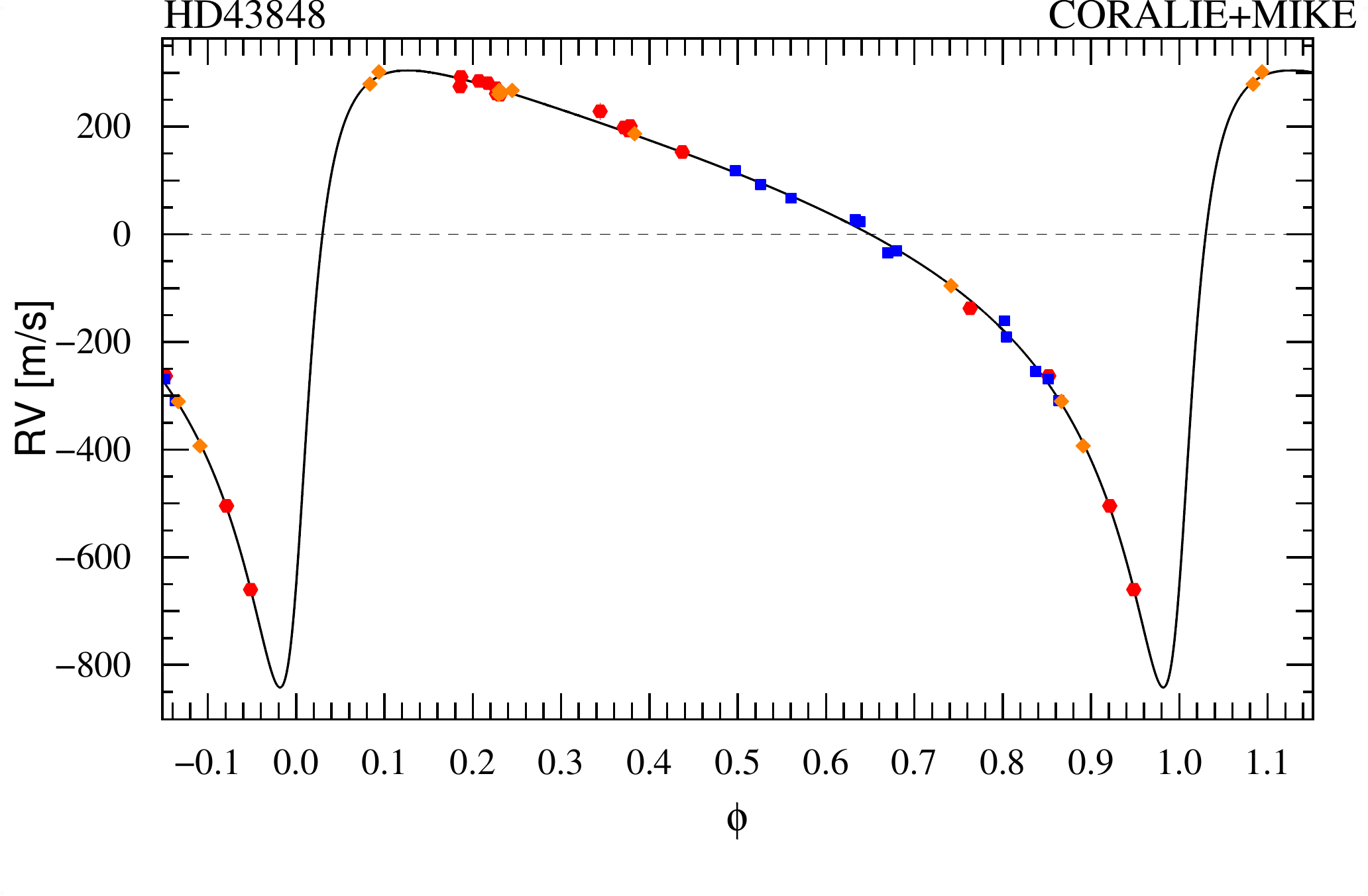}
\includegraphics[width= 0.49\linewidth]{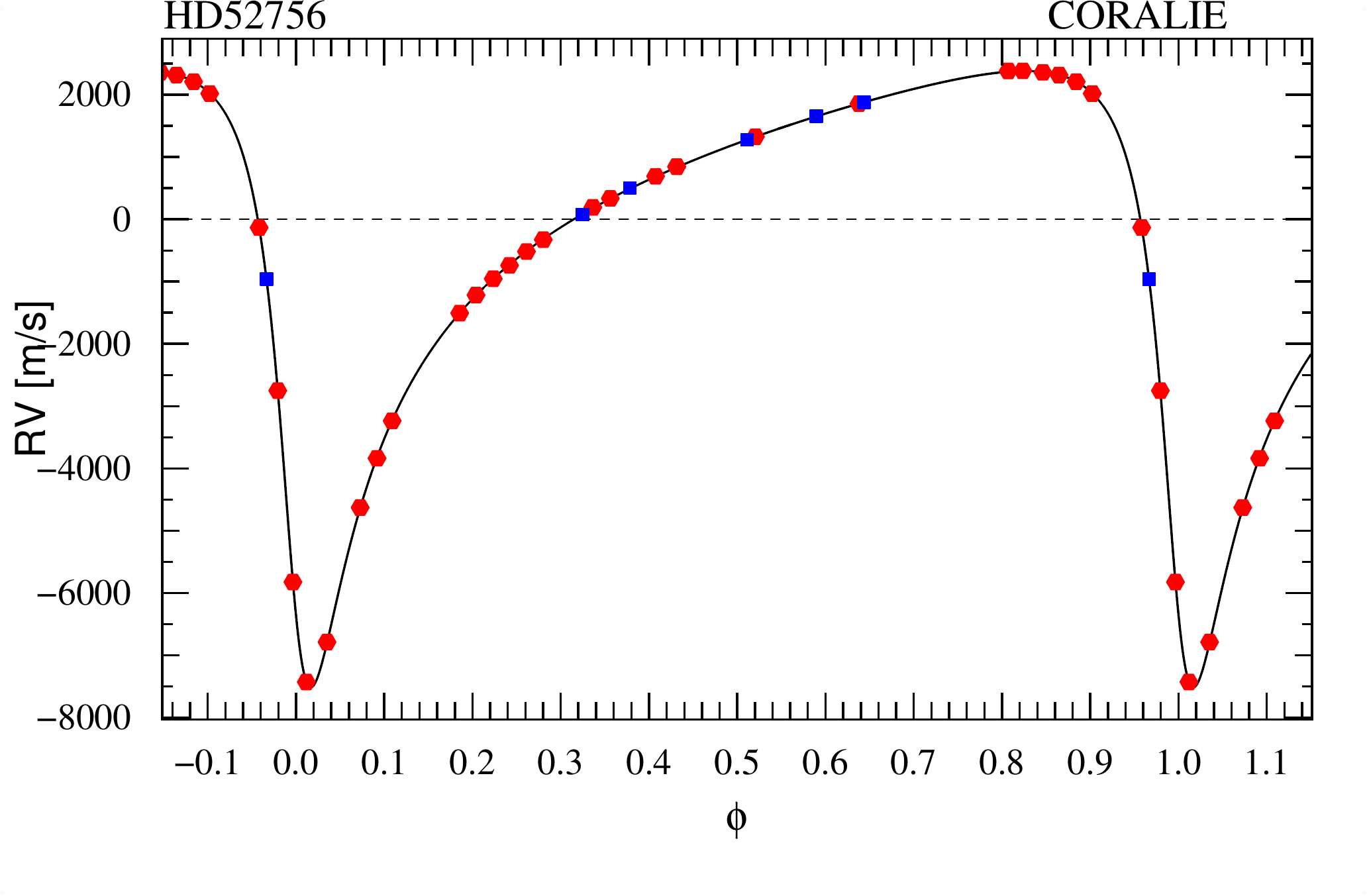} 
\includegraphics[width= 0.49\linewidth]{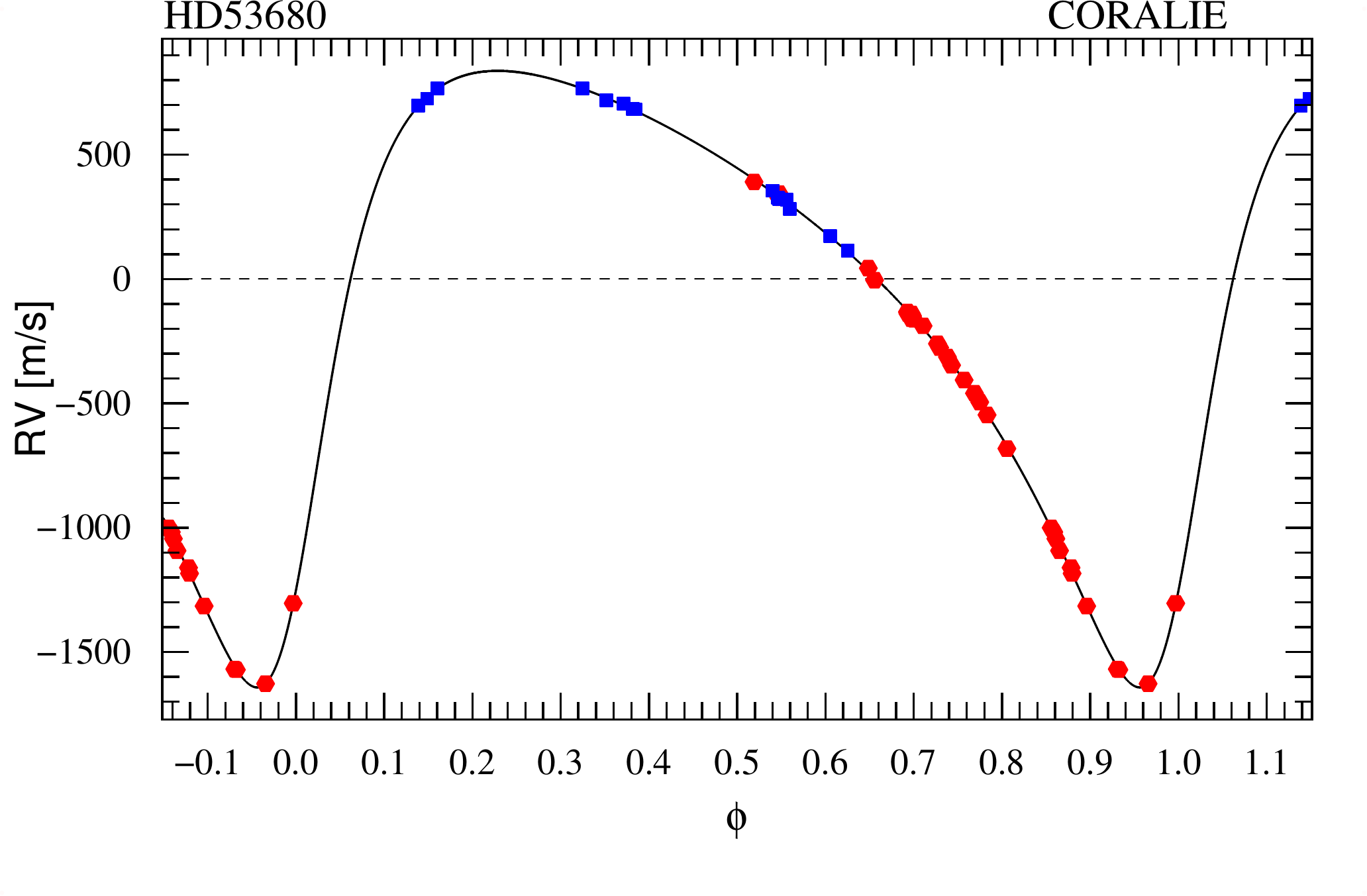} 
\includegraphics[width= 0.49\linewidth]{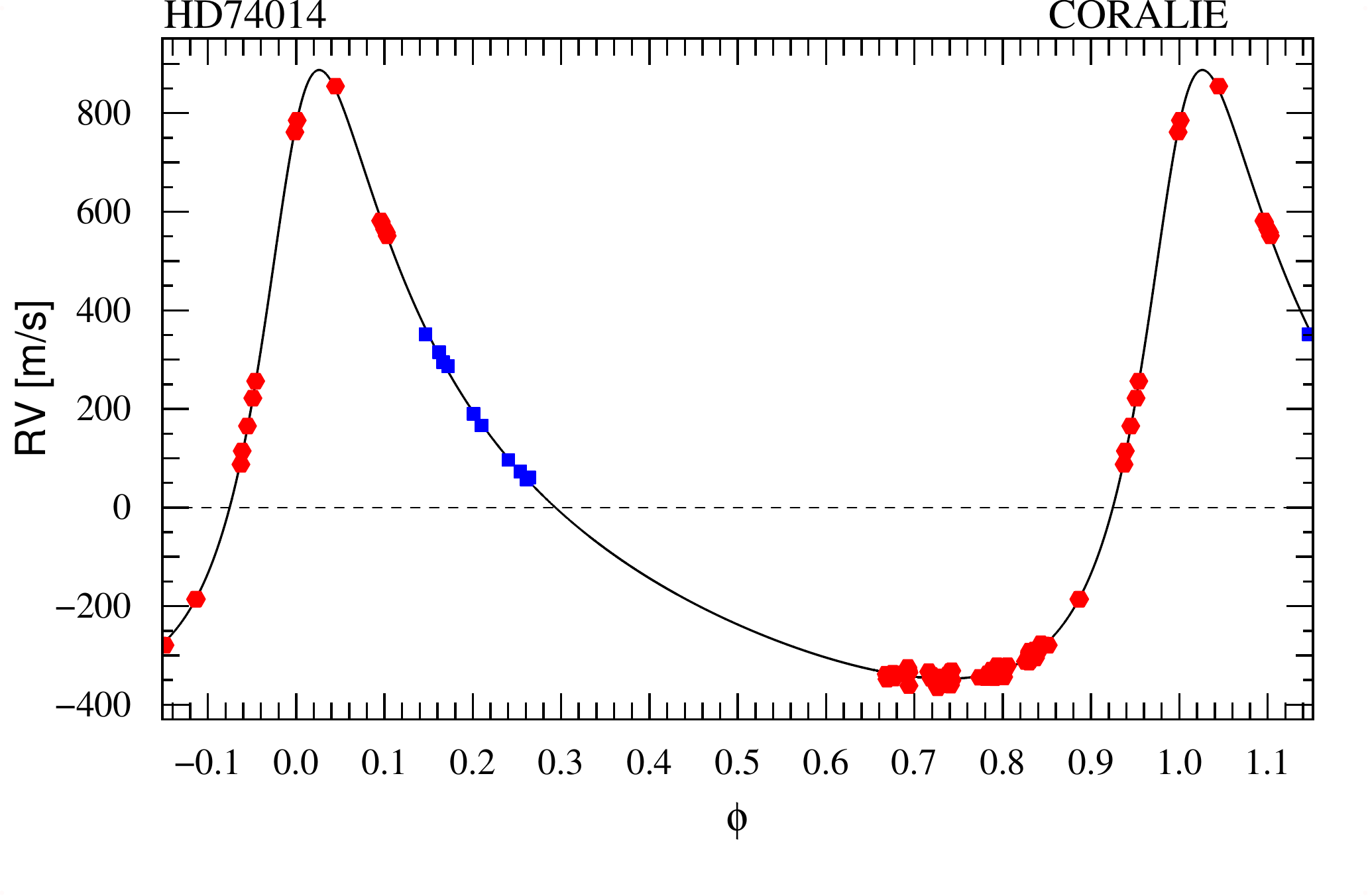} 
\includegraphics[width= 0.49\linewidth]{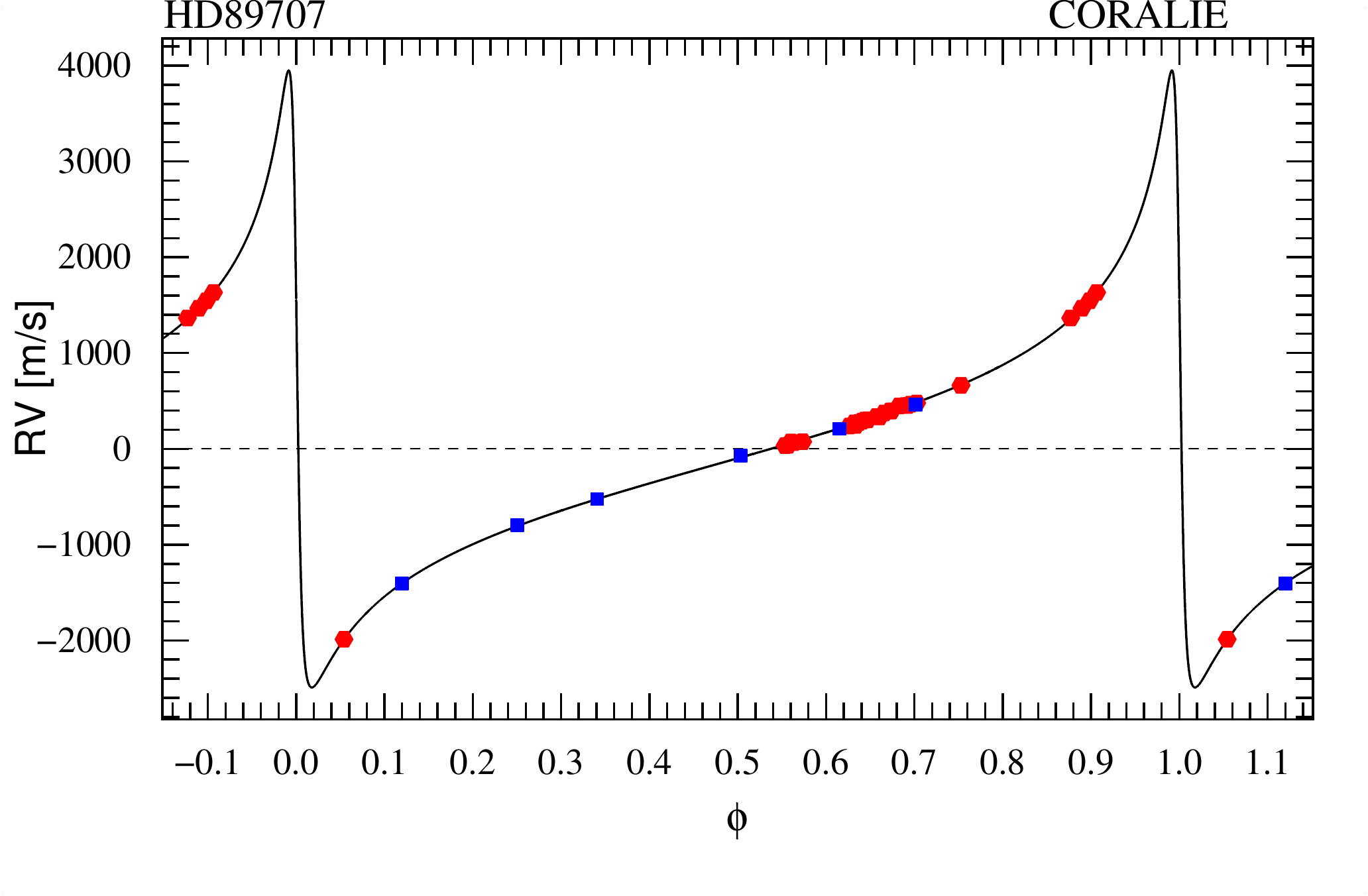} 
\caption{Phase-folded radial velocities of 8 stars with potential brown-dwarf companions. Red circles and blue squares indicate measurements with {\footnotesize CORALIE} C98 and C07, respectively. For HD~43848, the orange diamonds show the {\footnotesize MIKE} measurements. The solid lines correspond to the best-fit solutions. The error bars are smaller than the symbol size.}  \label{fig:rvfirst}\end{center} \end{figure*}

\begin{figure*}\begin{center} 
\includegraphics[width= 0.49\linewidth]{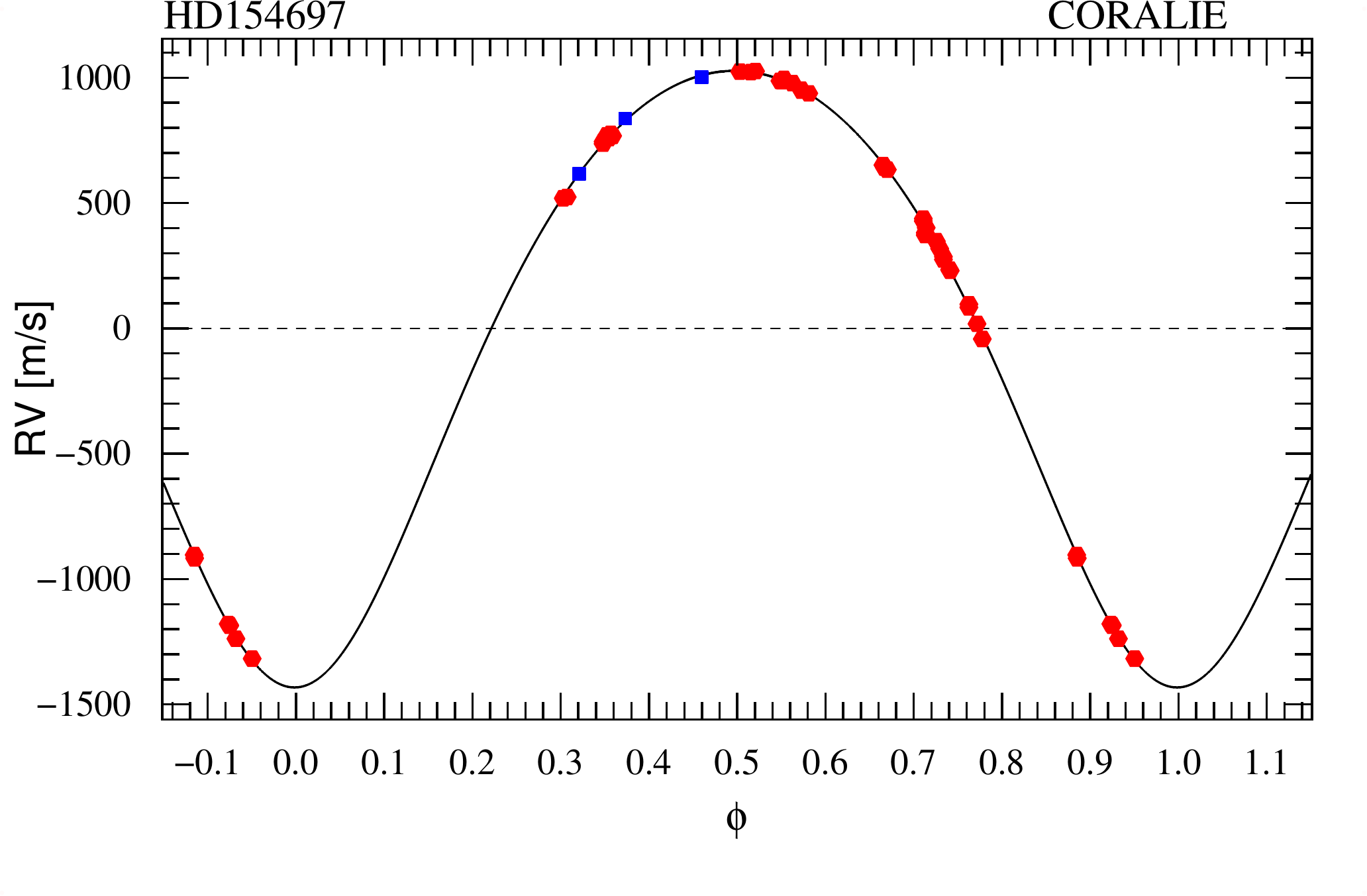} 
\includegraphics[width= 0.49\linewidth]{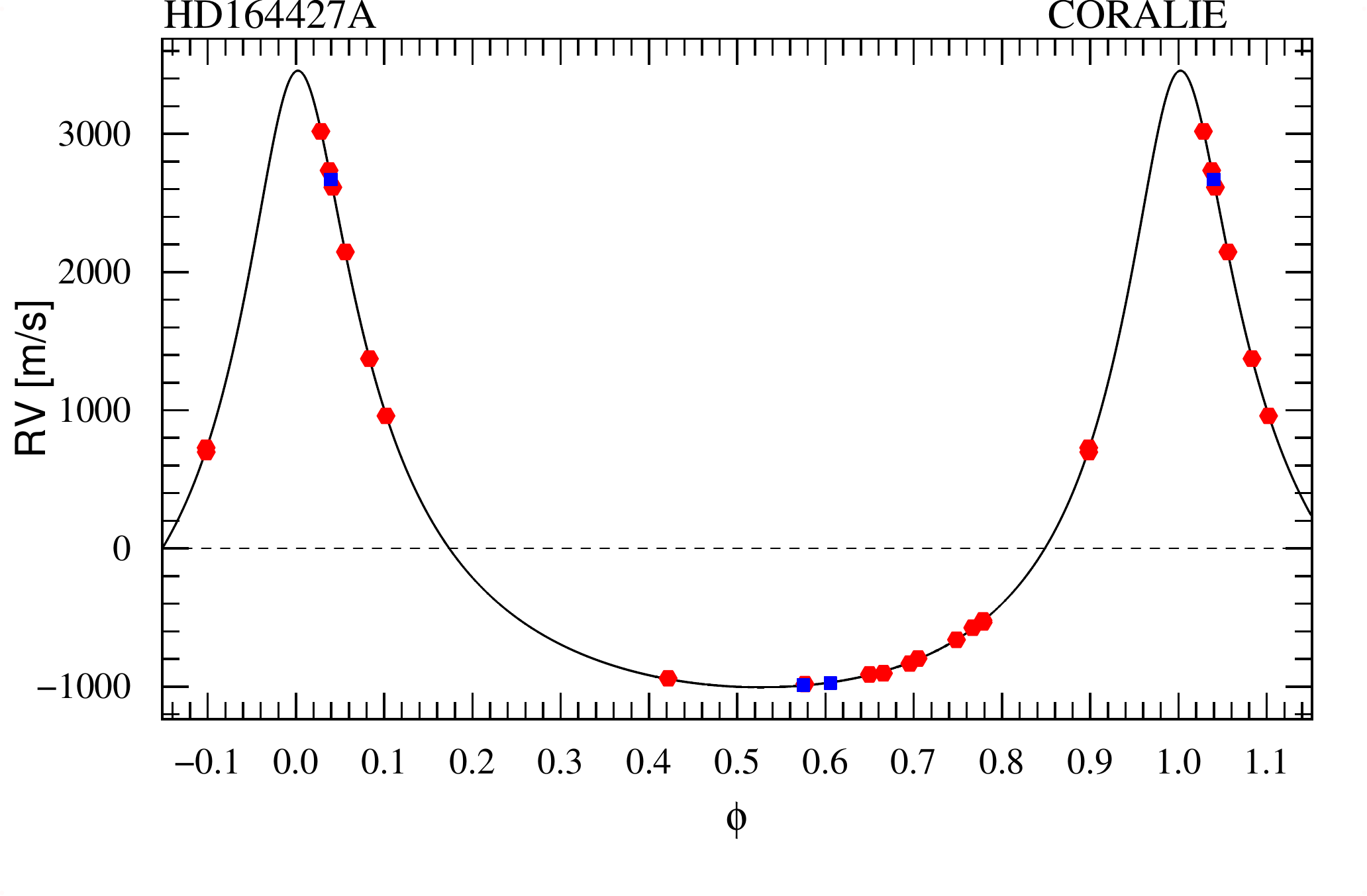} 
\includegraphics[width= 0.49\linewidth]{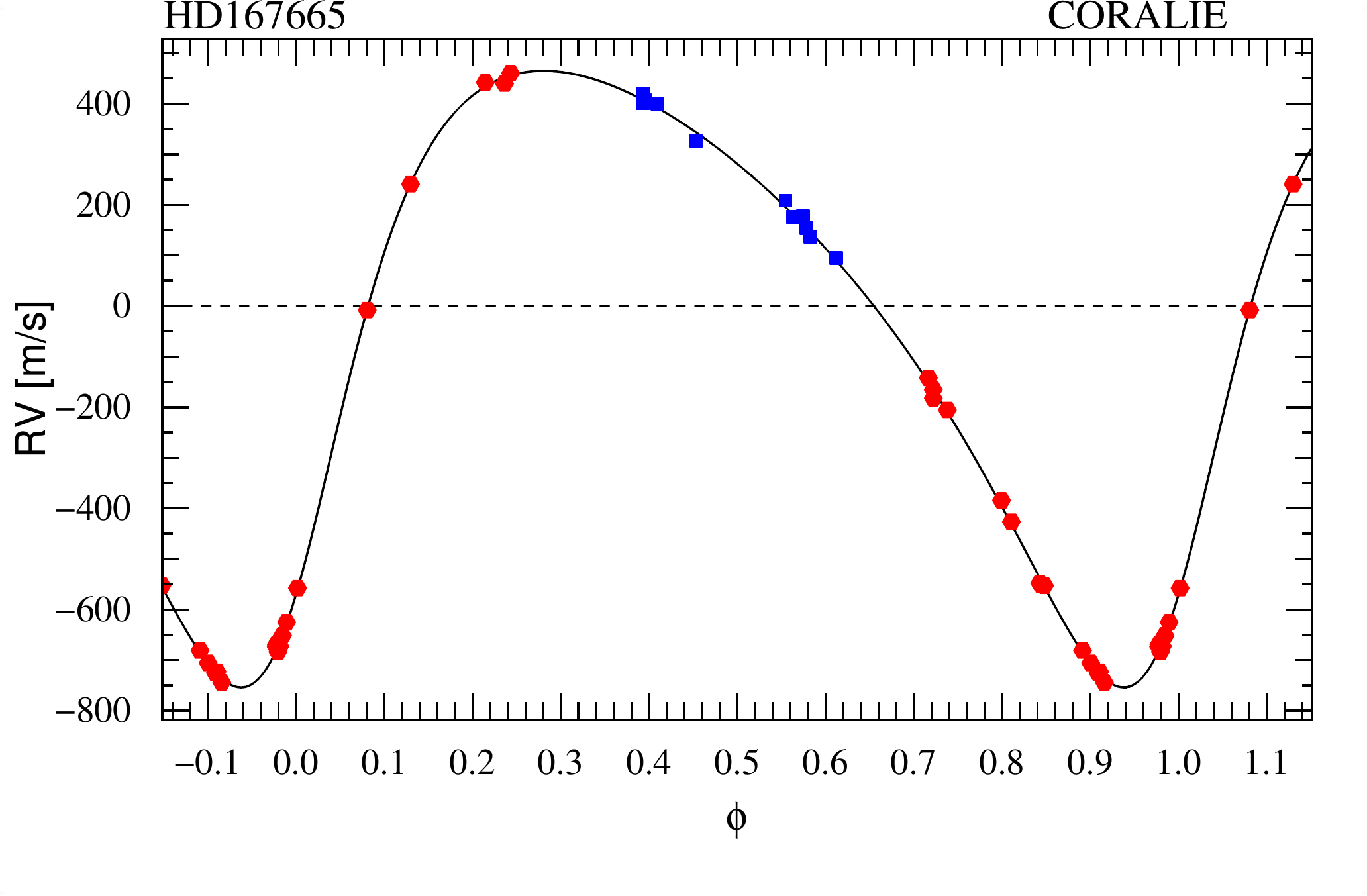} 
\includegraphics[width= 0.49\linewidth]{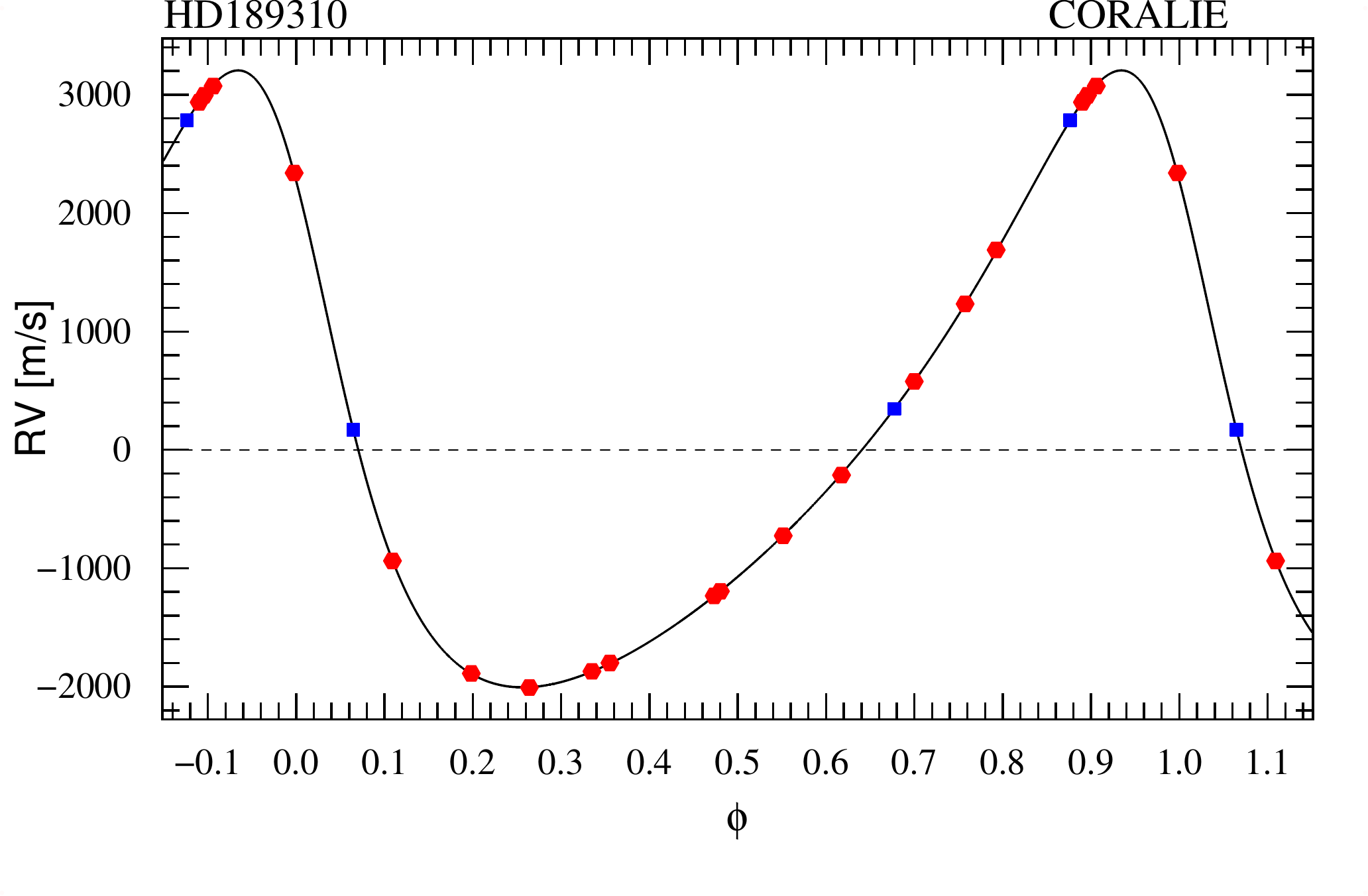} 
\includegraphics[width= 0.49\linewidth]{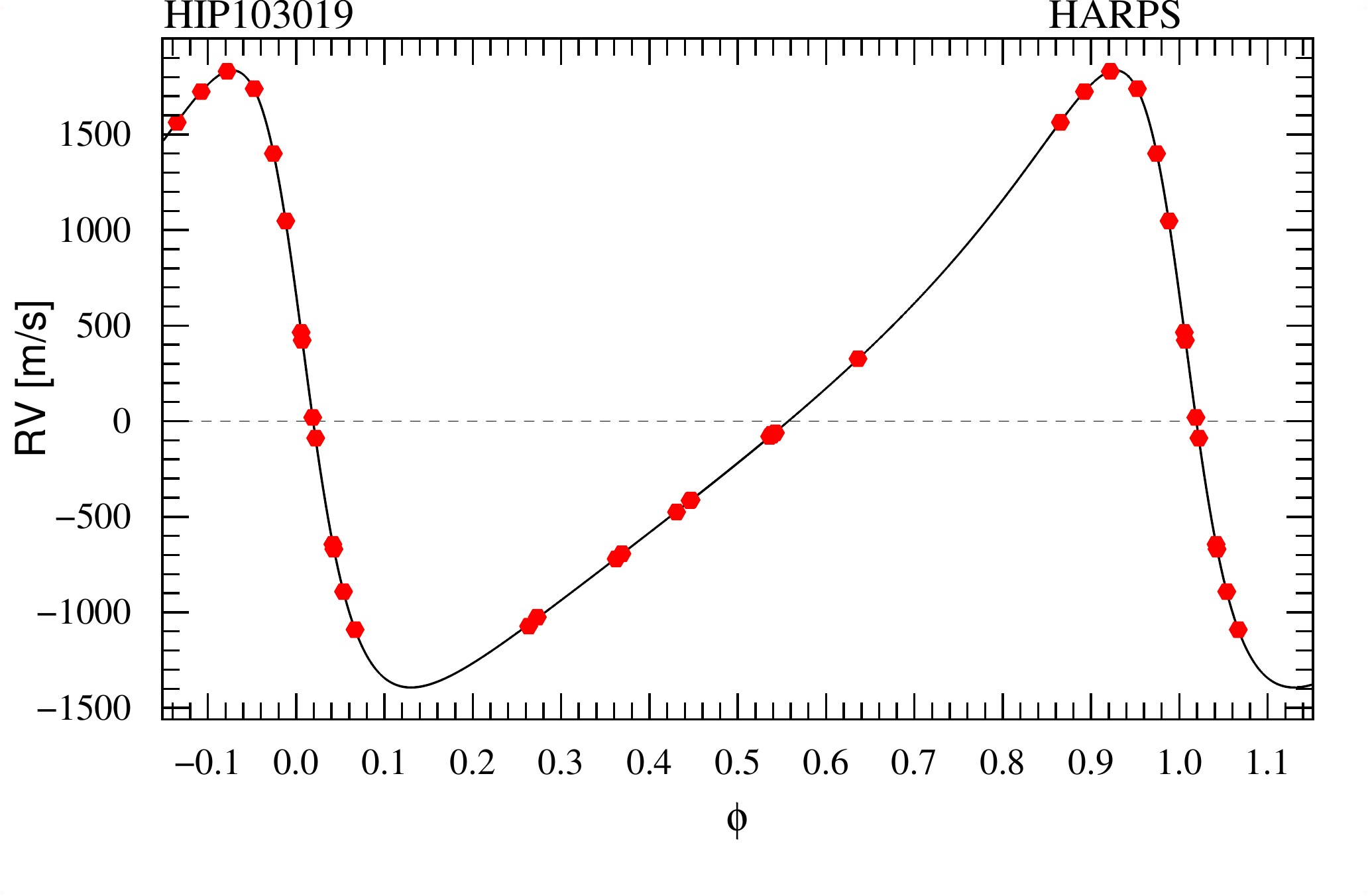} 
\caption{Phase-folded radial velocities of 5 stars with potential brown-dwarf companions. Error bars are smaller than the symbol size.}  \label{fig:rvlast} \end{center} \end{figure*}

\begin{figure}[h]\begin{center} 
\includegraphics[width= \linewidth]{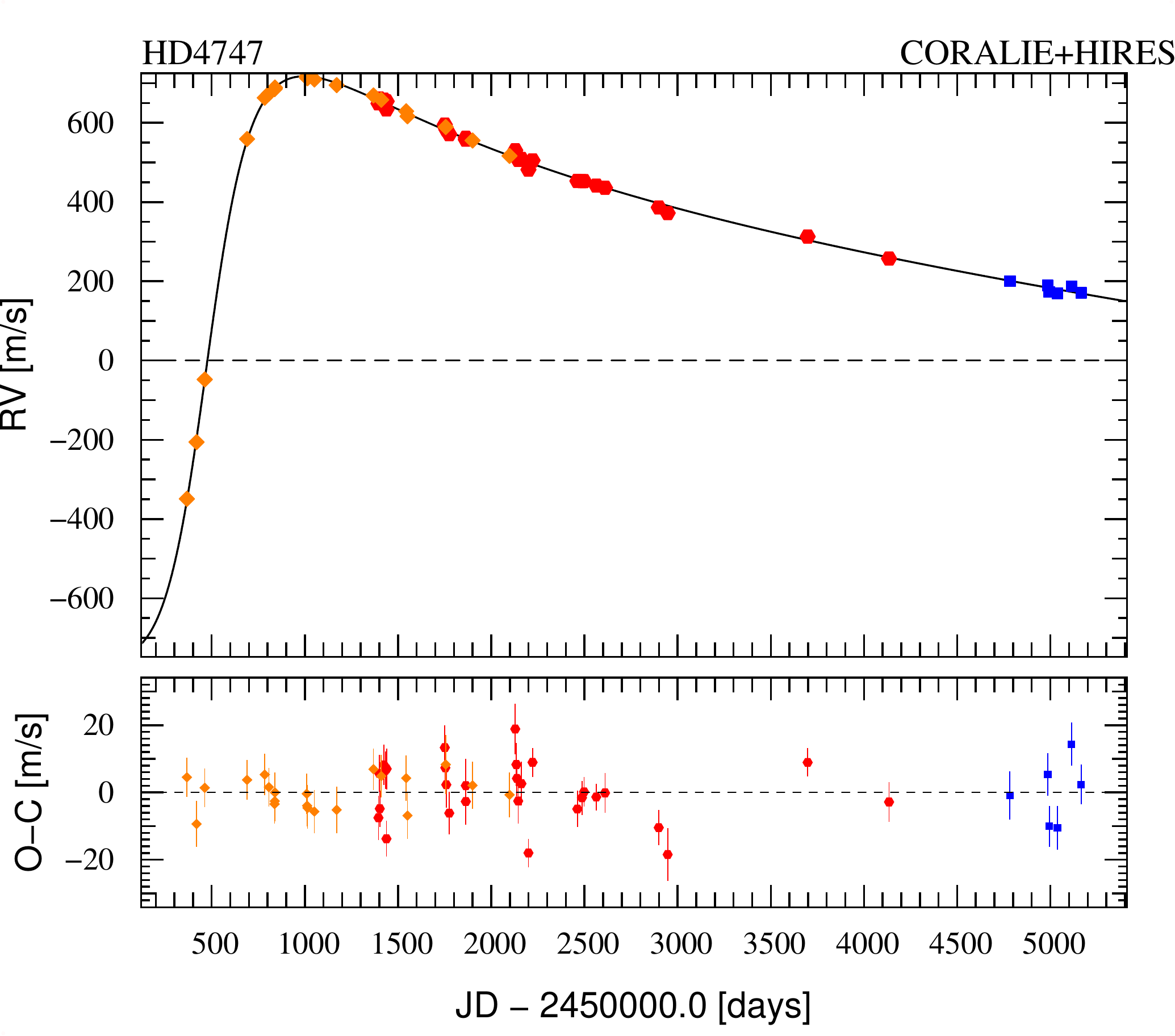}
\caption{Radial velocities (\emph{top}) and residuals (\emph{bottom}) to the best fit, represented by the solid line, of HD~4747 measured with {\footnotesize HIRES} (orange symbols), {\footnotesize CORALIE C98} (red symbols), and {\footnotesize CORALIE C07} (blue symbols).} \label{fig:HD4747} \end{center} \end{figure}

Table \ref{tab:KeplerOrbits} displays the orbital elements of the best fit solutions and Table \ref{tab:KeplerParams} lists the basic statistics and the derived orbit parameters (see also Fig. \ref{fig:orbchar}). $T_0$ is given as reduced Barycentric Julian Date (BJD), that is $\mathrm{JD}^{\star} = \mathrm{BJD} - 2\,400\,000$. $N_{\mathrm{RV}}$ is the total number of measurements, $\Delta T$ is the observation time-span, S/N is the median signal-to-noise ratio per pixel at 550 nm, $<\!\! \sigma_{\mathrm{RV}} \!\!>$ is the mean measurement uncertainty, $\chi_r$ is the reduced-chi value of the fit, G.o.F. is the goodness of fit, $\sigma_\mathrm{(O-C)}$ gives the root-mean-square dispersion of the fit residuals, $a_1 \sin{i}$ is the minimum astrometric semimajor axis expressed in unit of length, $f(m)$ is the mass function, $a_\mathrm{rel}$ is the  semimajor axis of the relative orbit, and finally $M_2 \sin i$ is the minimum companion mass. The quoted errors correspond to 1-$\sigma$ confidence intervals derived from 5000 Monte-Carlo simulations. Figures \ref{fig:rvfirst}, \ref{fig:rvlast}, and \ref{fig:HD4747} show the phase-folded radial velocities and the adopted orbital solution. All measurements are available in electronic form at CDS as indicated on the title page. 

\begin{figure}[h]\begin{center} 
\includegraphics[width= \linewidth]{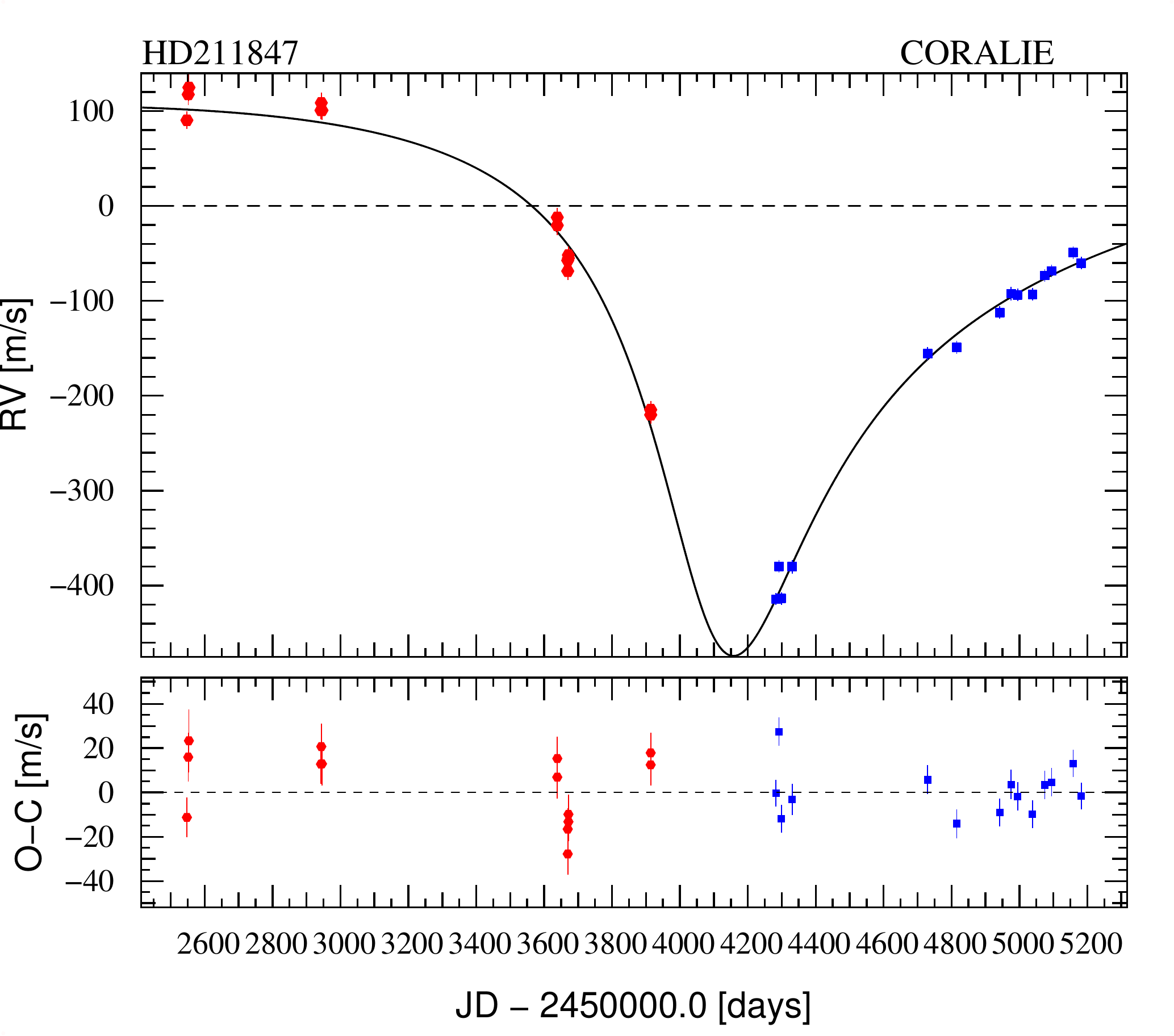} 
\caption{Radial velocities (\emph{top}) and residuals (\emph{bottom}) to the best fit, represented by the solid line, of HD~211847 measured with {\footnotesize CORALIE C98} (red circles), and {\footnotesize CORALIE C07} (blue squares).} \label{fig:HD211847} \end{center} \end{figure}

\begin{table*} [t]
\caption{Single-companion Keplerian orbital solutions of the surveyed stars}
\label{tab:KeplerOrbits} 
\centering  
\begin{tabular}{r r r r r r r r} 
\hline\hline %
Object & $\gamma_{\mathrm{C98}}$ & $\gamma_{\mathrm{C07}}$  & $P$ & $e$ & $K_1$ & $T_0$ & $\omega$ \\  
           &           (km\,s$^{-1}$)                 &  (km\,s$^{-1}$)                           &  (day)&        & (ms$^{-1}$)  & ($\mathrm{JD}^{\star}$) &   (deg)   \\
\hline 
HD~3277 & $-13.792^{+ 0.006}_{-0.006}$  &$-13.796^{+ 0.004}_{-0.004}$  &$46.1512^{+ 0.0002}_{-0.0002}$  &$0.285^{+ 0.001}_{-0.001}$  &$4076.4^{+ 3.5}_{-3.4}$  &$54324.14^{+    0.03}_{  -0.03}$  &$-39.4^{+ 0.3}_{-0.3}$ \vspace{1mm} \\
HD~17289 & $35.400^{+ 0.009}_{-0.009}$  &$35.420^{+ 0.017}_{-0.017}$  &$562.1^{+ 0.4}_{-0.4}$  &$0.532^{+ 0.004}_{-0.004}$  &$1414.5^{+ 10.0}_{-10.0}$  &$53762.5^{+    0.9}_{  -0.9}$  &$52.4^{+ 0.6}_{-0.6}$ \vspace{1mm} \\
HD~30501 & $23.725^{+ 0.026}_{-0.026}$  &$23.710^{+ 0.028}_{-0.028}$  &$2073.6^{+ 3.0}_{-2.9}$  &$0.741^{+ 0.004}_{-0.004}$  &$1703.1^{+ 26.0}_{-26.0}$  &$53851.5^{+    3.0}_{  -3.0}$  &$70.4^{+ 0.7}_{-0.7}$ \vspace{1mm} \\ 
HD~43848\tablefootmark{g} & $68.130^{+ 0.045}_{-0.046}$  &$68.118^{+ 0.042}_{-0.045}$    &$2354.3^{+ 8.9}_{-8.9}$  &$0.703^{+ 0.020}_{-0.019}$  &$570.1^{+ 48.1}_{-45.7}$  &$55585.8^{+   13.4}_{ -13.2}$  &$-131.7^{+ 1.3}_{-1.4}$ \vspace{1mm} \\ 
HD~43848\tablefootmark{h} & $66.845^{+ 0.492}_{-0.577}$  &$66.837^{+ 0.496}_{-0.575}$    &$2349.5^{+ 8.8}_{-8.5}$  &$0.885^{+ 0.030}_{-0.027}$  &$1859.3^{+ 570.1}_{-492.4}$  &$55634.6^{+   10.8}_{ -10.8}$  &$-151.5^{+ 4.0}_{-4.4}$ \vspace{1mm} \\ 
HD~52756 & $56.907^{+ 0.003}_{-0.003}$  &$56.904^{+ 0.005}_{-0.005}$  &$52.8657^{+ 0.0001}_{-0.0001}$  &$0.6780^{+ 0.0003}_{-0.0003}$  &$4948.8^{+ 3.0}_{-3.1}$  &$52828.617^{+    0.003}_{  -0.003}$  &$139.9^{+ 0.1}_{-0.1}$ \vspace{1mm} \\ 
HD~53680 & $89.065^{+ 0.005}_{-0.005}$  &$89.048^{+ 0.004}_{-0.004}$  &$1688.6^{+ 1.1}_{-1.1}$  &$0.475^{+ 0.002}_{-0.002}$  &$1239.8^{+ 4.1}_{-4.0}$  &$54182.9^{+    1.4}_{  -1.4}$  &$-133.2^{+ 0.3}_{-0.3}$ \vspace{1mm} \\ 
HD~74014 & $-16.371^{+ 0.002}_{-0.002}$  &$-16.387^{+ 0.008}_{-0.008}$  &$6936.3^{+ 134.9}_{-134.5}$  &$0.532^{+ 0.006}_{-0.006}$  &$617.5^{+ 2.3}_{-2.3}$  &$53431.9^{+    5.1}_{  -5.2}$  &$-34.7^{+ 0.6}_{-0.6}$ \vspace{1mm} \\ 
HD~89707 & $84.360^{+ 1.686}_{-1.581}$  &$84.390^{+ 1.699}_{-1.591}$  &$298.5^{+ 0.1}_{-0.1}$  &$0.900^{+ 0.039}_{-0.035}$  &$4189.5^{+ 1476.7}_{-1246.3}$  &$53296.4^{+    5.4}_{  -5.5}$  &$74.8^{+ 18.7}_{-19.7}$ \vspace{1mm} \\ 
HD~154697 & $-34.870^{+ 0.007}_{-0.007}$  &$-34.858^{+ 0.011}_{-0.011}$  &$1835.9^{+ 2.0}_{-2.0}$  &$0.165^{+ 0.002}_{-0.002}$  &$1230.1^{+ 8.1}_{-8.1}$  &$52579.7^{+ 9.5}_{ -9.5}$  &$179.1^{+1.6}_{-1.6}$ \vspace{1mm} \\ 
HD~164427A & $5.506^{+ 0.015}_{-0.014}$  &$5.512^{+ 0.015}_{-0.015}$  &$108.537^{+ 0.001}_{-0.001}$  &$0.551^{+ 0.002}_{-0.002}$  &$2231.1^{+ 12.6}_{-12.5}$  &$53026.8^{+    0.1}_{  -0.1}$  &$-3.3^{+ 0.4}_{-0.4}$ \vspace{1mm} \\ 
HD~167665 & $7.984^{+ 0.004}_{-0.004}$  &$8.003^{+ 0.008}_{-0.009}$  &$4451.8^{+ 27.6}_{-27.3}$  &$0.340^{+ 0.005}_{-0.005}$  &$609.5^{+ 3.3}_{-3.3}$  &$56987.6^{+   29.7}_{ -29.0}$  &$-134.3^{+ 0.9}_{-0.9}$ \vspace{1mm} \\ 
HD~189310 & $-10.258^{+ 0.003}_{-0.003}$  &$-10.250^{+ 0.005}_{-0.005}$  &$14.18643^{+ 0.00002}_{-0.00002}$  &$0.359^{+ 0.001}_{-0.001}$  &$2606.0^{+ 2.8}_{-2.8}$  &$53542.399^{+    0.004}_{  -0.004}$  &$50.1^{+ 0.1}_{-0.1}$ \vspace{1mm} \\ 
HIP~103019 & $-4.296^{+ 0.002}_{-0.002}$\tablefootmark{b}   &  \multicolumn{1}{c}{$\cdots$}  &$917.3^{+ 1.1}_{-1.1}$  &$0.502^{+ 0.001}_{-0.001}$  &$1614.2^{+ 1.8}_{-1.7}$  &$54681.5^{+    0.2}_{  -0.2}$  &$74.2^{+ 0.1}_{-0.1}$ \vspace{1mm} \\ 
\hline
HD~4747\tablefootmark{d}  &$9.949^{+ 0.007}_{-0.007}$  &$10.018^{+ 0.008}_{-0.008}$  &$7900$\tablefootmark{a}  &$0.676^{+ 0.003}_{-0.003}$  &$641.9^{+ 7.1}_{-7.3}$  &$58361.8^{+    3.9}_{  -4.0}$  &$-101.3^{+ 0.6}_{-0.6}$ \vspace{1mm} \\ 
HD~4747\tablefootmark{e} &$9.893^{+ 0.016}_{-0.016}$\tablefootmark{c}  &$9.904^{+ 0.024}_{-0.025}$  &$11593.2^{+ 1118.6}_{-1117.6}$  &$0.723^{+ 0.013}_{-0.013}$  &$703.3^{+ 16.7}_{-16.4}$  &$62059.1^{+ 1120.2}_{-1118.3}$  &$-94.2^{+ 1.6}_{-1.6}$ \vspace{1mm} \\
HD~4747\tablefootmark{f}  &$9.793^{+ 0.014}_{-0.014}$  &$9.757^{+ 0.011}_{-0.011}$  &$29000$\tablefootmark{a}  &$0.831^{+ 0.002}_{-0.002}$  &$807.9^{+ 14.3}_{-14.1}$  &$79470.8^{+    5.7}_{  -5.7}$  &$-84.9^{+ 0.6}_{-0.5}$ \vspace{1mm} \\ 
HD~211847\tablefootmark{d}  & $6.689$\tablefootmark{a}  &$6.689$\tablefootmark{a}  &$3750$\tablefootmark{a}  &$0.493^{+ 0.010}_{-0.010}$  &$267.2^{+ 6.5}_{-6.5}$  &$57830.9^{+   15.5}_{ -15.9}$  &$152.3^{+ 1.9}_{-1.9}$ \vspace{1mm} \\ 
HD~211847\tablefootmark{e}  & $6.689$\tablefootmark{a}   &$6.689$\tablefootmark{a}   &$7929.4^{+ 1999.1}_{-2500.2}$  &$0.685^{+ 0.068}_{-0.067}$  &$291.4^{+ 12.2}_{-12.0}$  &$62030.1^{+ 2010.4}_{-2501.4}$  &$159.2^{+ 2.0}_{-2.0}$ \vspace{1mm} \\ 
HD~211847\tablefootmark{f}   & $6.689$\tablefootmark{a}  &$6.689$\tablefootmark{a}  &$100000$\tablefootmark{a}  &$0.949^{+ 0.001}_{-0.001}$  &$338.1^{+ 7.1}_{-7.2}$  &$154108.7^{+    7.8}_{  -8.1}$  &$164.5^{+ 0.8}_{-0.8}$ \vspace{1mm} \\

\hline
\end{tabular} 
\tablefoot{\tablefoottext{a}  {Fixed.} \tablefoottext{b}  {{\footnotesize HARPS} systemic velocity.}  \tablefoottext{c} { {\footnotesize KECK HIRES} systemic velocity is $9.823^{+ 0.016}_{-0.016}$ km\,s$^{-1}$.}  \tablefoottext{d}  {Lower period limit.} \tablefoottext{e}  {Best-fit solution.} \tablefoottext{f}  {Upper period limit.} \tablefoottext{g} {Final solution at lower eccentricity. The MIKE systemic velocity is $-0.046^{+ 0.044}_{-0.046}$ km\,s$^{-1}$.} \tablefoottext{h} {Formal solution at higher eccentricity. The MIKE systemic velocity is $-1.331^{+ 0.493}_{-0.575}$ km\,s$^{-1}$.}} 
\end{table*}

\begin{table*} [t] \caption{Companion parameters for the surveyed stars}
\label{tab:KeplerParams} 
\centering  
\begin{tabular}{r r r r r r r r r r r r} 
\hline\hline %
Object & $N_{\mathrm{RV}}$ & $\Delta T$ & S/N & $<\!\! \sigma_{\mathrm{RV}} \!\!>$ &$\chi_\mathrm{r}$ & G.o.F. & $\sigma_\mathrm{(O-C)}$ & $a_1 \sin i$ & $f(m)$ & $a_\mathrm{rel}$ &$M_2 \sin i$  \\  
           &                                  &  (yr)           &       & (ms$^{-1}$)                       &             &            & (ms$^{-1}$)          &  ($10^{-3}$AU) & ($10^{-9}$M$_{\odot}$)& (AU)    &   ($M_J$)   \\
\hline 
HD~3277 & 14 & 10.4 & 35.4 & 8.4 &  \begin{math} 1.10 \pm 0.27 \end{math} &  0.55 & 5.38 &  $16.58^{+ 0.02}_{-0.02}$  &$ 285270^{+     796}_{   -791}$  &$0.249^{+ 0.004}_{-0.004}$  &$64.7^{+ 2.1}_{-2.2}$   \vspace{1mm} \\ 
HD~17289 & 29 & 11.3 & 46.9 & 6.6 &  \begin{math} 2.02 \pm 0.15 \end{math} &  6.07 & 12.46 & $61.9^{+ 0.3}_{-0.3}$  &$ 100178^{+    1619}_{  -1631}$  &$1.357^{+ 0.023}_{-0.023}$  &$48.9^{+ 1.7}_{-1.7}$   \vspace{1mm} \\ 
HD~30501 & 46 & 11.3 & 32.5 & 6.6 &  \begin{math} 0.91 \pm 0.11 \end{math} &  -0.73 & 6.33 & $217.9^{+ 2.0}_{-2.0}$  &$ 320923^{+    8957}_{  -9070}$  &$3.036^{+ 0.048}_{-0.049}$  &$62.3^{+ 2.1}_{-2.1}$   \vspace{1mm} \\
HD~43848\tablefootmark{f} & 38\tablefootmark{e} & 9.2 & 30.1 & 5.6 &  \begin{math} 1.31 \pm 0.13 \end{math} &  2.40 & 7.29 &  $87.5^{+ 4.8}_{-4.7}$  &$  16285^{+    2711}_{  -2638}$  &$3.360^{+ 0.057}_{-0.056}$  &$24.5^{+ 1.6}_{-1.6}$   \vspace{1mm} \\ 
HD~43848\tablefootmark{g} & 38\tablefootmark{e} & 9.2 & 30.1 & 5.6 &  \begin{math} 1.31 \pm 0.13 \end{math} &  2.40 & 7.29 &  $179.1^{+ 32.4}_{-28.1}$  &$ 150526^{+   78452}_{ -67284}$  &$3.385^{+ 0.057}_{-0.056}$  &$50.2^{+ 9.1}_{-8.1}$   \vspace{1mm} \\ 
HD~52756 & 33 & 10.1 & 28.8 & 6.1 &  \begin{math} 1.15 \pm 0.14 \end{math} &  1.17 & 7.72 & $17.68^{+ 0.01}_{-0.01}$  &$ 263682^{+     483}_{   -483}$  &$0.265^{+ 0.004}_{-0.004}$  &$59.3^{+ 2.0}_{-1.9}$   \vspace{1mm} \\ 
HD~53680 & 48 & 9.1 & 30.5 & 7.1 &  \begin{math} 1.03 \pm 0.11 \end{math} &  0.30 & 7.77 & $169.4^{+ 0.5}_{-0.5}$  &$ 227414^{+    2125}_{  -2059}$  &$2.620^{+ 0.043}_{-0.042}$  &$54.7^{+ 1.8}_{-1.8}$   \vspace{1mm} \\ 
HD~74014 & 118 & 11.3 & 41.8 & 5.3 &  \begin{math} 0.96 \pm 0.07 \end{math} &  -0.61 & 6.29 & $333.2^{+ 4.7}_{-4.8}$  &$ 102583^{+     955}_{   -947}$  &$7.222^{+ 0.147}_{-0.153}$  &$49.0^{+ 1.7}_{-1.7}$   \vspace{1mm} \\ 
HD~89707 & 37 & 9.0 & 57.5 & 6.1 &  \begin{math} 1.70 \pm 0.13 \end{math} &  5.03 & 11.63 &  $46.0^{+ 6.8}_{-5.8}$  &$ 158142^{+   62033}_{ -60856}$  &$0.877^{+ 0.015}_{-0.014}$  &$53.6^{+ 7.8}_{-6.9}$   \vspace{1mm} \\ 
HD~154697 & 49 & 10.8 & 31.6 & 7.2 &  \begin{math} 1.09 \pm 0.11 \end{math} &  0.85 & 7.82 &  $204.7^{+ 1.3}_{-1.3}$  &$ 339722^{+    6646}_{  -6630}$  &$2.960^{+ 0.048}_{-0.048}$  &$71.1^{+ 2.4}_{-2.4}$   \vspace{1mm} \\ 
HD~164427A & 21 & 10.7 & 43.1 & 6.7 &  \begin{math} 1.25 \pm 0.19 \end{math} &  1.39 & 7.12 & $18.6^{+ 0.1}_{-0.1}$  &$  72658^{+     994}_{   -991}$  &$0.472^{+ 0.007}_{-0.008}$  &$48.0^{+ 1.6}_{-1.6}$   \vspace{1mm} \\ 
HD~167665 & 38 & 10.9 & 76.3 & 5.0 &  \begin{math} 1.30 \pm 0.13 \end{math} &  2.36 & 8.28 &  $234.6^{+ 2.3}_{-2.3}$  &$  86931^{+    1800}_{  -1802}$  &$5.608^{+ 0.097}_{-0.096}$  &$50.6^{+ 1.7}_{-1.7}$   \vspace{1mm} \\ 
HD~189310 & 19 & 8.3 & 22.2 & 7.6 &  \begin{math} 0.70 \pm 0.20 \end{math} &  -1.43 & 4.39 & $3.172^{+ 0.003}_{-0.003}$  &$  21156^{+      56}_{    -57}$  &$0.109^{+ 0.002}_{-0.002}$  &$25.6^{+ 0.9}_{-0.8}$   \vspace{1mm} \\ 
HIP~103019 & 26 & 2.3 & 44.0 & 2.3 &  \begin{math} 1.92 \pm 0.16 \end{math} &  5.28 & 3.57 &  $117.7^{+ 0.1}_{-0.1}$  &$ 258370^{+     999}_{  -1019}$  &$1.678^{+ 0.027}_{-0.027}$  &$52.5^{+ 1.7}_{-1.7}$   \vspace{1mm} \\ 
\hline
HD4747\tablefootmark{b} & 56\tablefootmark{a} & 13.1 & 34.9 & 4.7 &  \begin{math} 1.56 \pm 0.10 \end{math} &  5.19 & 8.40 &  $343.4^{+ 4.7}_{-4.7}$  &$  86625^{+    3520}_{  -3588}$  &$7.286^{+ 0.119}_{-0.120}$  &$39.6^{+ 1.4}_{-1.4}$   \vspace{1mm} \\
HD4747\tablefootmark{c} & 56\tablefootmark{a}  & 13.1 & 34.9 & 4.7 &  \begin{math} 1.26 \pm 0.10 \end{math} &  2.50 & 6.47 &  $517.2^{+ 52.2}_{-50.8}$  &$ 137430^{+   15376}_{ -14928}$  &$9.425^{+ 0.636}_{-0.639}$  &$46.1^{+ 2.3}_{-2.3}$   \vspace{1mm} \\ 
HD4747\tablefootmark{d} & 56\tablefootmark{a} & 13.1 & 34.9 & 4.7 &  \begin{math} 1.56 \pm 0.10 \end{math} &  5.15 & 7.38 &$1199.3^{+ 25.2}_{-25.2}$  &$ 274046^{+   17203}_{ -17299}$  &$17.465^{+ 0.287}_{-0.288}$  &$58.1^{+ 2.4}_{-2.3}$   \vspace{1mm} \\ 
HD~211847\tablefootmark{b} & 28 & 7.2 & 41.2 & 5.9 &  \begin{math} 2.32 \pm 0.14 \end{math} &  7.94 & 13.58 &  $80.1^{+ 2.0}_{-2.0}$  &$   4885^{+     358}_{   -360}$  &$4.652^{+ 0.076}_{-0.075}$  &$17.0^{+ 0.7}_{-0.7}$   \vspace{1mm} \\ 
HD~211847\tablefootmark{c} & 28 & 7.2 & 41.2 & 5.9 &  \begin{math} 1.87 \pm 0.15 \end{math} &  5.39 & 11.44 & $148.2^{+ 33.3}_{-38.2}$  &$   7073^{+    1079}_{  -1084}$  &$7.536^{+ 1.367}_{-1.587}$  &$19.2^{+ 1.2}_{-1.2}$   \vspace{1mm} \\ 
HD~211847\tablefootmark{d} & 28 & 7.2 & 41.2 & 5.9 &  \begin{math} 2.02 \pm 0.14 \end{math} &  6.29 & 12.99 &  $977.8^{+ 23.1}_{-23.2}$  &$  12494^{+     883}_{   -888}$  &$41.624^{+ 0.695}_{-0.700}$  &$23.3^{+ 1.0}_{-1.0}$   \vspace{1mm} \\ 
\hline
\end{tabular} 
\tablefoot{\tablefoottext{a}  {21 velocity measurements are from \citet{Nidever:2002vn}.} \tablefoottext{b}  {Lower period limit.} \tablefoottext{c}  {Best-fit solution.} \tablefoottext{d}  {Upper period limit.} \tablefoottext{e} {10 velocity measurements are from \cite{Minniti:2009zr}.}  \tablefoottext{f} {Final solution at lower eccentricity.} \tablefoottext{g} {Formal solution at higher eccentricity.}}
\end{table*}

\subsection{Details on the surveyed stars}\label{sec:objnotes}
\begin{itemize}
  \item {HD~3277} (\object{HIP~2790}) is discovered to host a potential brown-dwarf companion. The star was listed as an uncertain spectroscopic binary in \cite{Tokovinin:2006rt}, who did not have sufficient radial-velocity measurements to constrain the orbit.
  \item {HD~4747} (\object{HIP~3850}) has a potential brown-dwarf companion discovered with radial velocities by \citet{Nidever:2002vn}. We combine {\footnotesize CORALIE} observations with the {\footnotesize KECK HIRES} radial velocities of \citet{Nidever:2002vn} retrieved via the SB9-catalogue \citep{Pourbaix:2004yq}. However, the orbital period is very long and not covered by the measurements, see Fig. \ref{fig:HD4747}. Tables \ref{tab:KeplerOrbits} and \ref{tab:KeplerParams} show the best-fit solution and the solutions with fixed period such that the $\chi$-value increases to $\chi_\mathrm{r} + 3 \sigma_{\chi} = 1.56$. The resulting period range is $P= 7900-29000$ days with corresponding minimum companion-mass $M_2 \sin i = 39.6 - 58.1\,M_J$.   
  \item {HD~17289} (\object{HIP~12726}) is discovered to host a potential brown-dwarf companion. \cite{Goldin:2007ly} independently found an orbital signature from Hipparcos astrometry alone with a period of $536 \pm 12$ days, which is close to our solution ($562.1 \pm 0.3$ days).  However, the orbital parameters derived by \cite{Goldin:2007ly} are less precise and less accurate than the radial-velocity solution. 
  
The dispersion of the fit residuals is abnormally large for {\footnotesize CORALIE}. The detailed analysis of the cross-correlation function (CCF) yields that its bisector velocity span and full-width-at-half-maximum (FWHM) are correlated with the orbital phase (Fig.~\ref{fig:bis}). As we will show in Sect.~\ref{sec:results}, the companion of HD~17289 is a star with mass of $0.52\,M_\odot$. The two stars are very close ($<50$~mas, Sect.~\ref{sec:3sigma}) and their estimated intensity ratio in the visible is about 1:180. Therefore, the companion's light is equally picked up by the science fibre of {\footnotesize CORALIE} and its signature appears in the collected spectra. The spectral lines of the companion distort the CCF depending on the orbital phase. As expected the CCF width is minimal when the radial-velocity curves of both components cross and equal the systemic velocity. The detailed modelling of this double-lined spectroscopic binary is outside the scope of this paper, but we have identified the binary as the cause of a phase-dependent radial-velocity bias, which explains the excess noise in our measurements.  
\begin{figure}\begin{center} 
\includegraphics[width= \linewidth, trim = 0cm 0.5cm 0cm 1cm, clip=true]{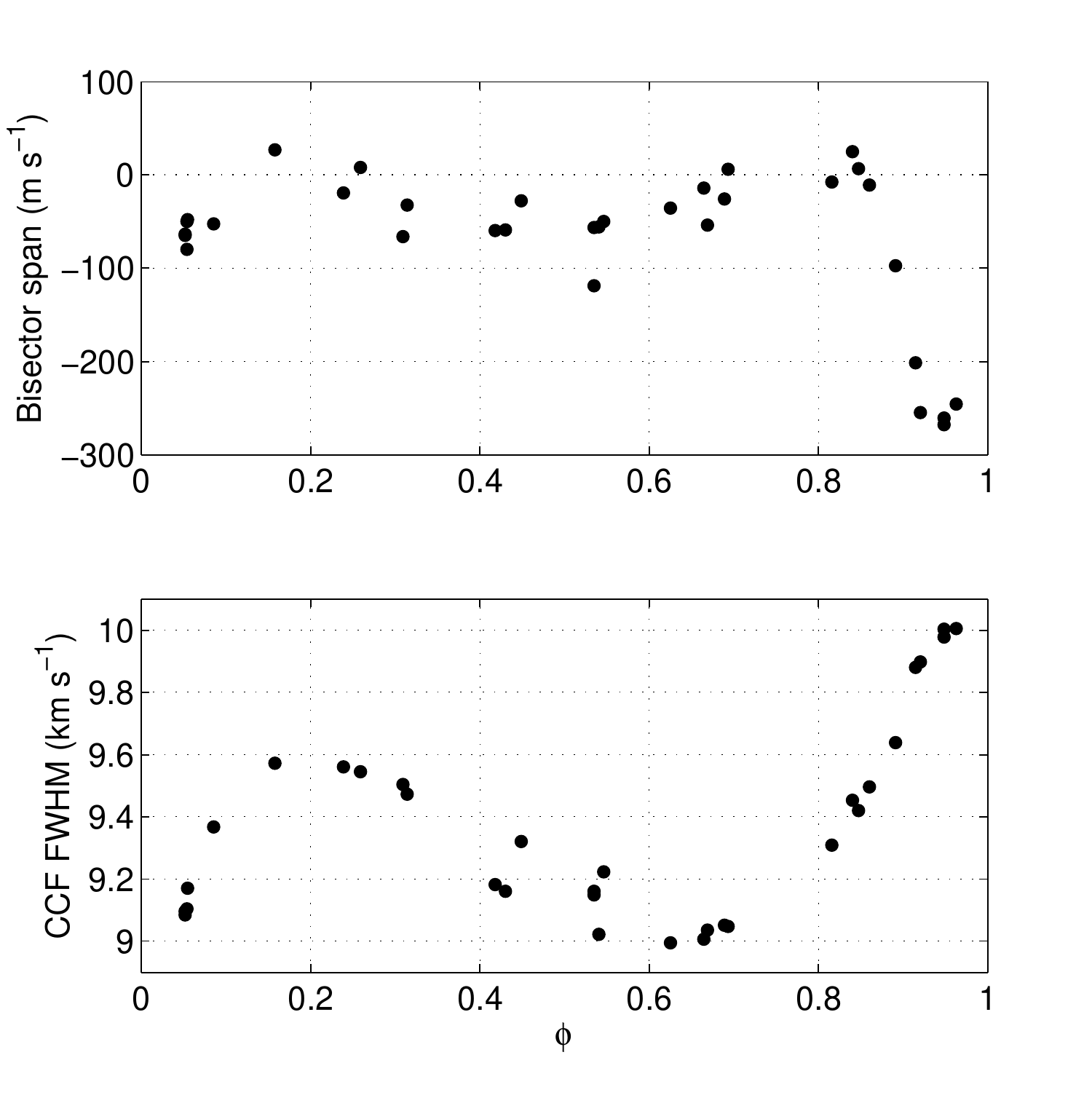} 
\caption{Bisector velocity span (\emph{top}) and FWHM of the cross-correlation-function (\emph{bottom}) as function of orbital phase for HD~17289. The corresponding radial velocities are shown in Fig.~\ref{fig:rvfirst}. Clearly, both observables show a phase-dependent modulation, which is likely caused by the companion.} 
\label{fig:bis} \end{center} \end{figure}
  \item \object{HD~30501} (\object{HIP~22122}) is discovered to host a potential brown-dwarf companion.   
  \item \object{HD~43848} (\object{HIP~29804}) has a potential brown-dwarf companion discovered with radial velocity by \cite{Minniti:2009zr}. An astrometric analysis using Hipparcos and the solution of \cite{Minniti:2009zr} was performed by \cite{Sozzetti:2010nx}. We combined the published measurements\footnote{By comparing the published radial velocities to our {\footnotesize CORALIE} measurements, we discovered that the published Julian dates corresponding to the radial velocities of \cite{Minniti:2009zr} are misprinted and offset by --2000 days.} with the {\footnotesize CORALIE} velocities to adjust the Keplerian model. While the orbital period is well defined, the distributions of $e$, $K_1$, $T_0$, and $\omega$ obtained from the Monte-Carlo simulations are bimodal. This is because the lower velocity turnover is not well sampled by the measurements. The fit-quality in terms of $\chi^2$ is comparable for all Monte-Carlo simulations and can not be used to identify the correct solution. In Fig.~\ref{fig:eccHD43848} we show the eccentricity and semi-amplitude distributions. Because the realisation of the lower-eccentricity orbit is more probable (77~\%) than the high-eccentricity orbit (23 \%), we consider only the orbits with $e\lesssim 0.78$ to find the final solution and to perform the astrometric analysis. For completeness, we also list the less probable and higher-eccentricity solution in Tables \ref{tab:KeplerOrbits} and~\ref{tab:KeplerParams}.
\begin{figure*}\begin{center} 
\sidecaption
\includegraphics[width=5.9cm]{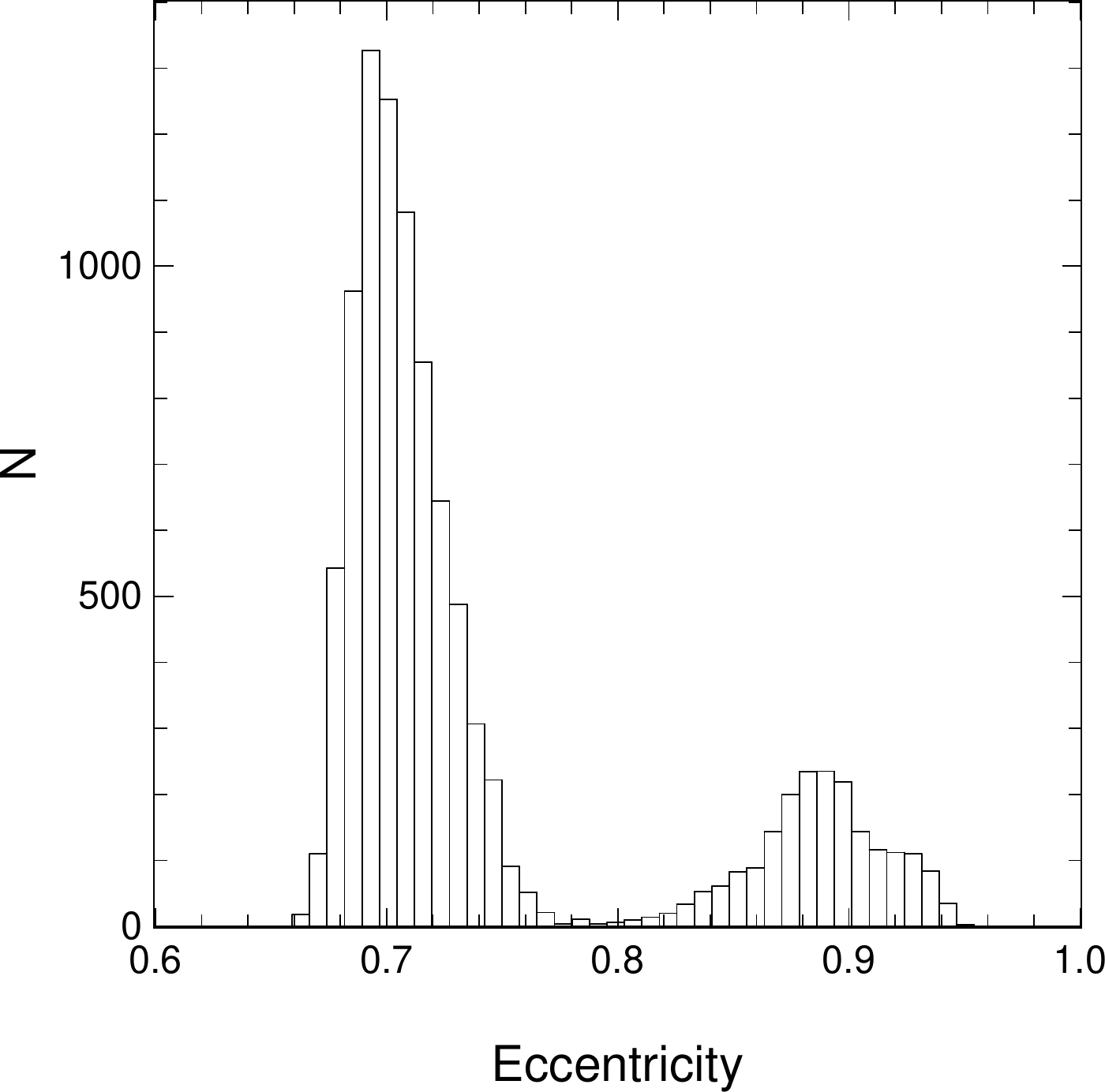} 
\includegraphics[width= 6cm]{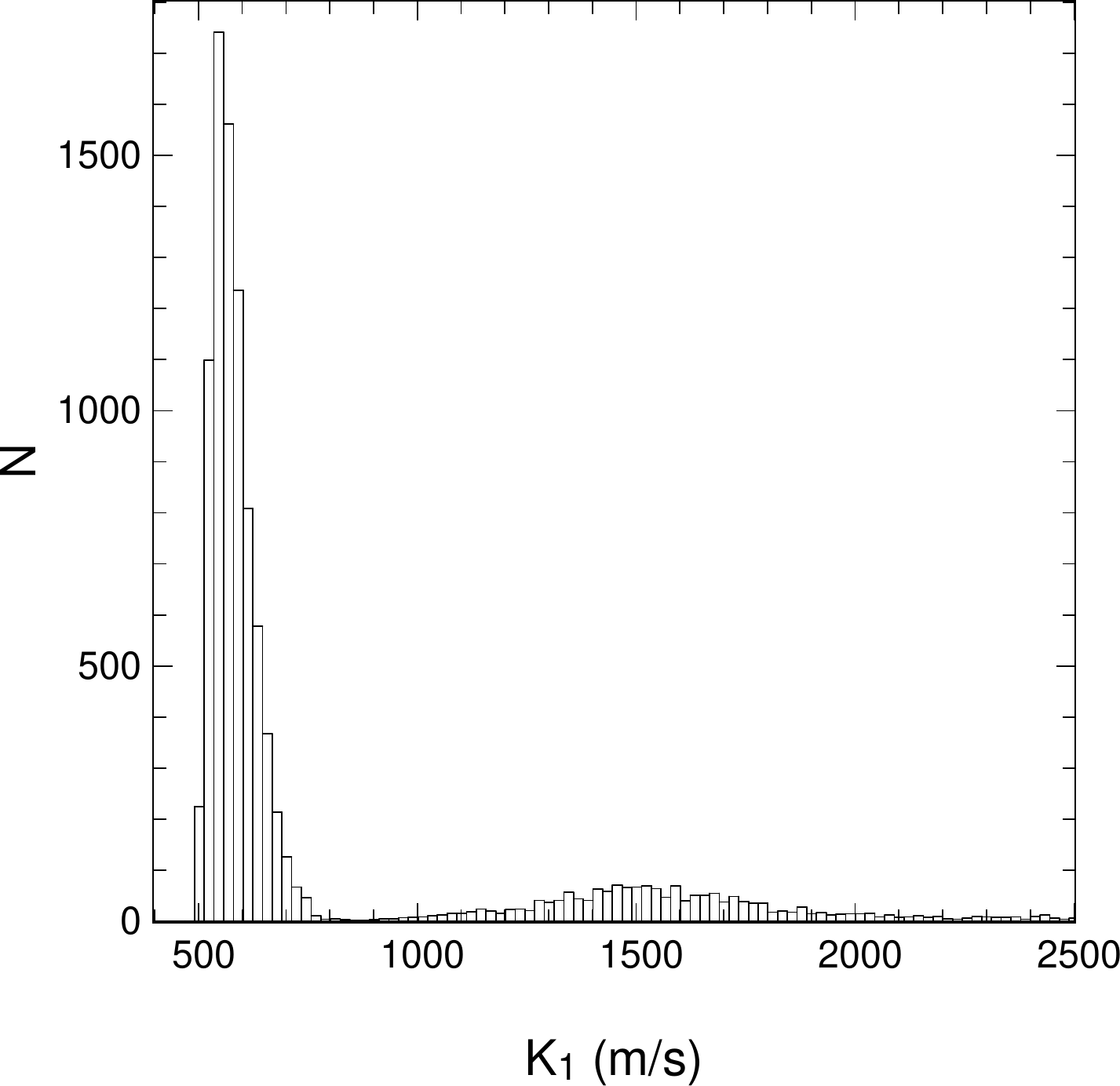} 
\caption{Eccentricities (\emph{left}) and velocity semi-amplitudes (\emph{right}) of 10000 Monte-Carlo simulations performed to find the radial-velocity orbit of HD~43848. The bi-modal structure is clearly seen in both histograms. The final solution is found by considering only the lower eccentricity realisations, which is done by separating the distributions at $\hat K_1 = 910\,$ms$^{-1}$ and correspondingly at $\hat e \sim 0.78$. The probability to obtain a high- or low-eccentricity solution is calculated by counting the number of realisations above and below this threshold value and amounts to 77 \% and 23 \%, respectively.} 
\label{fig:eccHD43848}
\end{center} \end{figure*}

\item \object{HD~52756} (\object{HIP~33736}) is discovered to host a potential brown-dwarf companion. 
\item \object{HD~53680} (\object{HIP~34052}) is discovered to host a potential brown-dwarf companion.
 \item \object{HD~74014} (\object{HIP~42634}) has a potential brown-dwarf companion discovered with radial velocities by \cite{Patel:2007ys}. The original orbit is compatible with our solution obtained from 118 {\footnotesize CORALIE} velocities covering the complete orbit.
  
  \item \object{HD~89707} (\object{HIP~50671}) has a potential brown-dwarf companion announced by \cite{Duquennoy:1991kx} and analysed by \cite{Halbwachs:2000rt} based on radial velocities collected with the {\footnotesize CORAVEL} and {\footnotesize ELODIE} instruments. The original orbit is compatible with our solution obtained from 37 {\footnotesize CORALIE} velocities. A combined astrometric analysis using Hipparcos was performed by \cite{Zucker:2001ve}. We find slight evidence for re-emission at the bottom of the \ion{Ca}{II H} absorption line (Fig. \ref{fig:HD89707activity}) and derive an activity index of $\log R'_\mathrm{HK} = -4.66 \pm 0.03$ from a high signal-to-noise {\footnotesize CORALIE} spectrum. Using the relation of \cite{Santos:2000vl}, we expect an additional activity-induced radial-velocity noise of $12\,$ms$^{-1}$, which explains the unusually-large fit residuals encountered for HD~89707. Accounting for the additional noise during the adjustment yields an acceptable fit quality with $\chi_\mathrm{r} = 0.91 \pm 0.13$. 
  
  \begin{figure}\begin{center} 
\includegraphics[width= \linewidth]{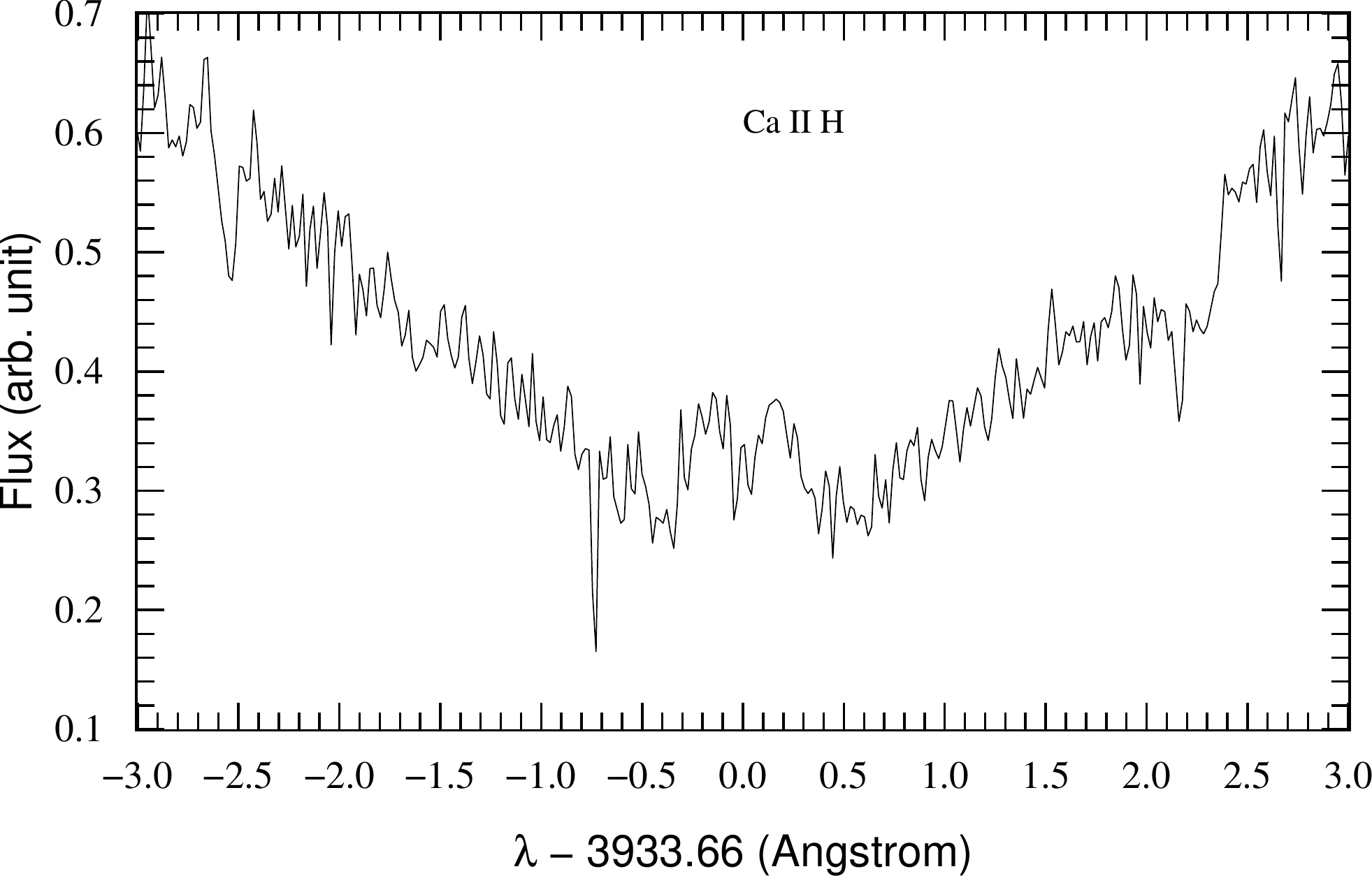} 
\caption{\ion{Ca}{II H} emission region for HD~89707 showing slight re-emission at the bottom of this absorption line.}  \label{fig:HD89707activity}\end{center} \end{figure}
  
  \item \object{HD~154697} (\object{HIP~83770}) is discovered to host a potential brown-dwarf companion.
  \item \object{HD~164427A} (\object{HIP~88531}) has a potential brown-dwarf companion announced by \cite{Tinney:2001yq}, whose original orbit is compatible with our solution. A combined astrometric analysis using Hipparcos was performed by \cite{Zucker:2001ve}. 
  \item \object{HD~189310}  (\object{HIP~99634}) is discovered to host a potential brown-dwarf companion in a short-period orbit of 14.2 days. 
  \item \object{HD~167665} (\object{HIP~89620}) has a potential brown-dwarf companion announced by \cite{Patel:2007ys}, whose original orbit is compatible with our solution. 
  \item \object{HD~211847} (\object{HIP~110340}) is discovered to host a potential brown dwarf companion. The long-period orbit is not completely covered by our measurements (Fig.~\ref{fig:HD211847}) and the {\footnotesize CORALIE} upgrade occured at minimum radial velocity. To obtain a reasonable solution, we therefore fix the offset between the parameters $\gamma_{\mathrm{C98}}$ and $\gamma_{\mathrm{C07}}$ to zero, which is the mean offset value found for three other stars of same spectral type. As for HD~4747, we derive realistic confidence intervals by fixing the period during the adjustment, such that the resulting $\chi$-value increases to $\chi_\mathrm{r} + 3 \sigma_{\chi} = 2.32$. The period lower limit is $P= 3750$ days and we set the upper limit to $P= 100000$ days, because we cannot derive an upper limit using the defined criterion. The corresponding minimum companion mass is $M_2 = 17.0 - 23.3\,M_J$. The dispersion of the best-fit residuals is abnormally large for {\footnotesize CORALIE}. We find evidence for re-emission at the bottom of \ion{Ca}{II H} and \ion{Ca}{II K}, see Fig.~\ref{fig:HD211847activity}, and derive an activity index of $\log R'_\mathrm{HK} = -4.74 \pm 0.04$ from a high signal-to-noise {\footnotesize CORALIE} spectrum. Using the relation of \cite{Santos:2000vl}, we expect an additional activity-induced radial-velocity noise of $11\,$ms$^{-1}$, which explains the fit residuals encountered for HD~211847. Accounting for the additional noise during the adjustment yields an acceptable fit quality with $\chi_\mathrm{r} = 1.15 \pm 0.15$.
  
\begin{figure}\begin{center} 
\includegraphics[width= \linewidth]{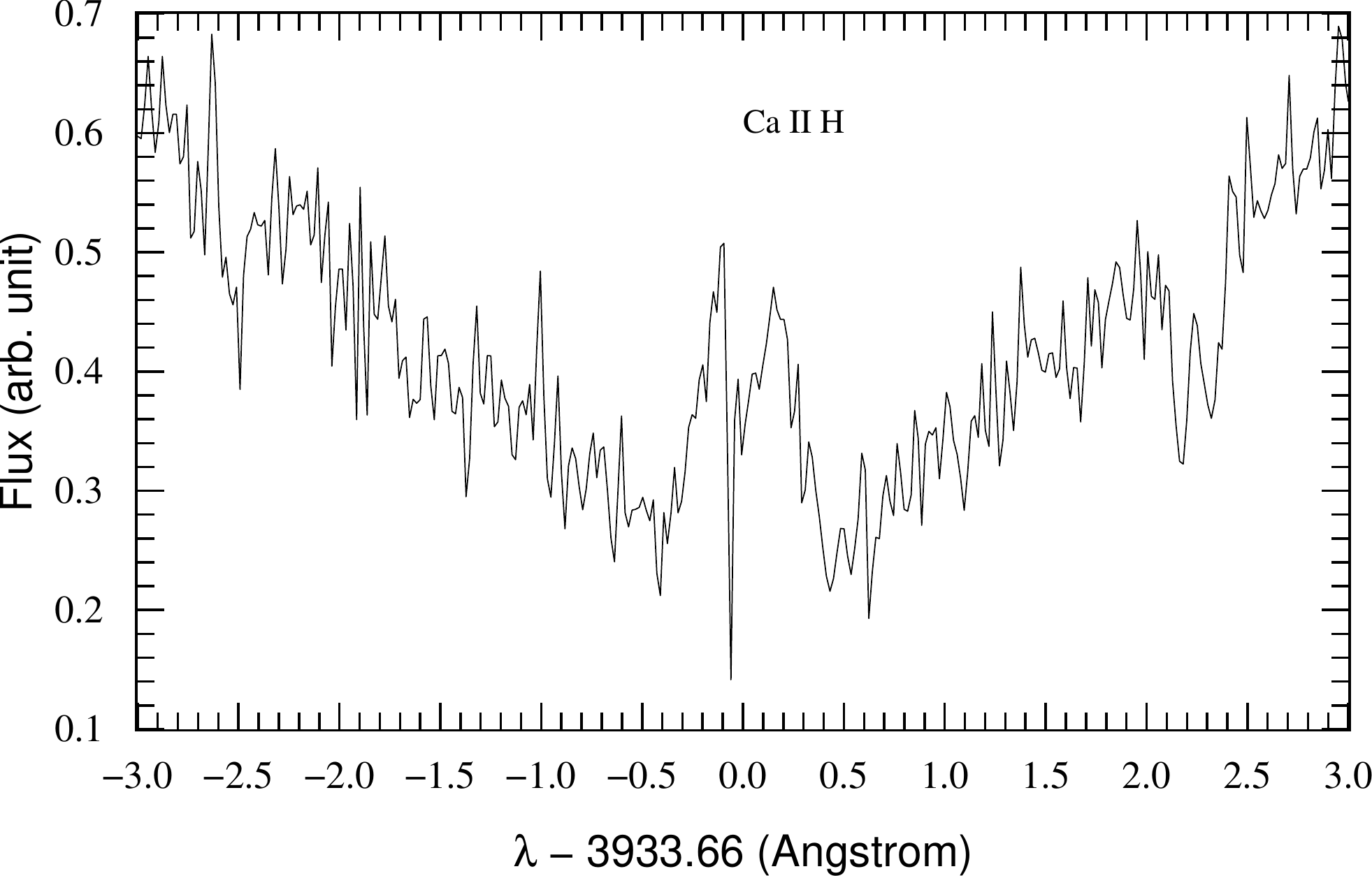} 
\caption{\ion{Ca}{II H} emission region for HD~211847 showing re-emission at the bottom of the absorption line.}
  \label{fig:HD211847activity}\end{center} \end{figure}

  \item \object{HIP~103019} is discovered to host a potential brown-dwarf companion using {\footnotesize HARPS}. The star is included in the search for planets in a volume-limited sample (e.g. Naef et al. 2010). Twenty-six radial velocities have been collected with {\footnotesize HARPS} with a mean uncertainty of 2.3 ms$^{-1}$. The dispersion of the fit residuals is abnormally large for {\footnotesize HARPS} ($\sigma_{O-C} = 3.57\,$ms$^{-1}$). Because of its faintness, we were not able to measure the activity level of \object{HIP~103019}. We therefore cannot exclude an additional radial-velocity jitter induced by the presence of spots on the surface of the star. The fit is satisfactory ($\chi_r=1.04$) if we include additional noise of 3.5 ms$^{-1}$. 
  \end{itemize}

\begin{table}\caption{List of targets selected from the literature}
\label{tab:littargets} 
\centering  
\begin{tabular}{r r r | r r r } 	
\hline\hline %
Nr. & \multicolumn{2}{c|}{Object} & Nr. & \multicolumn{2}{c}{Object}    \\  
      &HD        &  HIP                    &       &    HD &    HIP                                      \\
\hline 
1&	13189	&	10085 & 10 & GJ~595 &    76901 \\
2&	30339	&	22429 & 11 & 140913	&     77152 \\
3&	38529	&	27253 & 12 & 162020&	87330\\
4&	65430	&	39064 & 13 & 168443&	89844\\
5&	91669	&	51789 & 14 & 174457&	92418\\
6& 107383	&     60202 & 15 & 180777& 	94083\\
7& 119445	&     66892 & 16 & 190228	& 	98714\\
8& 131664	&     73408 & 17 & 191760	&     99661\\
9& 137510	& 	75535 & 18 & 202206	&   104903\\
\hline
\end{tabular} 
\end{table}

\subsection{Brown-dwarf candidates from the literature}\label{sec:littargets}
In addition to the stars from the {\footnotesize CORALIE} and {\footnotesize HARPS} surveys, we selected 18 stars from the literature using the following criteria: They have announced radial-velocity companions with minimum mass of 13-80 $M_J$, were published after 2002, and are not flagged as binary stars in the new Hipparcos reduction. The selected stars are listed in Table \ref{tab:littargets} and details are given below. 
\begin{itemize}
 \item \object{HD~13189} is a K2 II star with very uncertain mass of $2-7\,M_{\odot}$ and a companion announced by \cite{Hatzes:2005jw} with $M_2 \sin i= 8-20 \,M_J$.   
 \item \object{HD~30339} (F8, $1.1\,M_{\odot}$), \object{HD~65430} (K0V, $0.78\,M_{\odot}$), \object{HD~140913} (G0V, $0.98\,M_{\odot}$), and \object{HD~174457} (F8, $1.19\,M_{\odot}$) have companions announced by \cite{Nidever:2002vn} with $M_2 \sin i$ ranging between $42 - 78\,M_J$. 
 \item \object{HD~38529} is a G4 IV star with mass $1.48\,M_{\odot}$. Companion announced by \cite{Fischer:2003pi} with $M_2 \sin i= 12.7 \,M_J$. Refined orbits are given by \cite{Wright:2009fe} and \cite{Benedict:2010ph}. We use the latest orbital elements for analysis.
 \item \object{HD~91669} is a K0 dwarf with mass $0.88\pm0.04\,M_{\odot}$. Companion announced by \cite{Wittenmyer:2009mz} with $M_2 \sin i= 30.6 \,M_J$. 
 \item \object{HD~107383} is a G8 III giant star with mass $2.7\pm0.3\,M_{\odot}$ and a companion announced by \cite{Liu:2008uo} with $M_2 \sin i= 19.4 \,M_J$. 
 \item \object{HD~119445} is a G6 III giant star with mass $3.9\pm0.4\,M_{\odot}$ and a companion announced by \cite{Omiya:2009dn} with $M_2 \sin i= 37.6 \,M_J$. 
 \item \object{HD~131664} is a G3V star with mass $1.10\pm0.03\,M_{\odot}$. Companion announced by \cite{Moutou:2009fr} with $M_2 \sin i= 18.15\,M_J$.
 \item \object{HD~137510} is a G0 IV star with mass $1.42\,M_{\odot}$ and a companion announced by \cite{Endl:2004uq}. The updated data by \cite{Butler:2006pi} lists $M_2 \sin i= 22.7 \,M_J$. 
 \item \object{GJ~595} is a M3 dwarf with mass $0.28\,M_{\odot}$. Companion announced by \cite{Nidever:2002vn} with $M_2 \sin i = 60\,M_J$.
 \item \object{HD~162020} is a K3 V star with mass $0.75\,M_{\odot}$ and a companion announced by \citet{Udry2002} with $M_2 \sin i= 14.4\,M_J$. 
 \item \object{HD~168443} has two companions with minimum masses in the planetary (\object{HD~168443b}) and brown dwarf domain (\object{HD~168443c}, $M_2 \sin i= 18 \,M_J$) discovered by \citet{Marcy:2001rt} and \cite{Udry:2004oq} (see also \citealt{Udry2002}). Updated orbital elements are given by \citet{Wright:2009fe}, which we use for the astrometric analysis for the outer companion HD~168443c.
 \item \object{HD~180777} is a A9V star with mass $1.7\pm0.1\,M_{\odot}$ and a companion announced by \cite{Galland:2006qq} with $M_2 \sin i= 25 \,M_J$. 
 \item \object{HD~190228} is a G5 IV star with mass $0.83\,M_{\odot}$ and a planetary-mass companion with $M_2 \sin i= 3.58\,M_J$ \citep{Perrier:2003dw}. \cite{Zucker:2001ve} performed an astrometric analysis using Hipparcos and derived $M_2 = 67 \pm 29\,M_J$ at 95~\% confidence based on the spectroscopic elements of \cite{Sivan:2004rm} (see also Sect.~\ref{sec:comptargets}).
 \item \object{HD~191760} is a G3 IV star with mass $1.26\pm0.06\,M_{\odot}$. Companion announced by \cite{Jenkins:2009kx} with $M_2 \sin i= 38.17 \,M_J$. 
 \item \object{HD~202206} is a G6 V star with mass $1.15\,M_{\odot}$ and has two companions with minimum masses in the planetary (\object{HD~202206c}) and brown dwarf domain (\object{HD~202206b}, $M_2 \sin i= 17.4 \,M_J$) discovered by \cite{Udry2002} and \citet{Correia:2005bs}
\end{itemize}

We have now identified our target sample of stars with potential brown-dwarf companions and have obtained the orbital parameters accessible through radial-velocity measurements. As the next step, we will search the intermediate astrometric data of the new Hipparcos reduction for signatures of the orbital motion corresponding to the radial-velocity orbit.

\section{Combined astrometric analysis}\label{sec:method}
We describe a method to search the Intermediate Astrometric Data (IAD) of the new Hipparcos reduction for the orbital signatures of stars, whose spectroscopic elements are known from a reliable radial-velocity solution. The significance of the derived orbit is determined through the distribution-free permutation test \citep{Good:1994,Zucker:2001ve}.

Techniques to perform the simultaneous fitting to radial-velocity and Hipparcos astrometry data for substellar companions and to verify the statistical confidence of the solution have been developed by \cite{Zucker:2000vn, Pourbaix:2001qe, Halbwachs:2000rt, Zucker:2001ve}. More recently, \cite{Reffert:2006ly} and \cite{Sozzetti:2010nx} used similar approaches to claim the detection of three brown-dwarf companions, however with significance not exceeding 2-$\sigma$. Independently, successful detections of the astrometric signature of exoplanet candidates using the {\footnotesize HST} fine guidance sensor ({\footnotesize HST FGS}) optical interferometer are reported for a few stars with spectroscopic elements known from radial velocity (e.g. \citealt{Bean2007, Martioli:2010kx}). Even though the orbital phase is not always fully covered by {\footnotesize HST} observations, for instance the orbital coverage for \object{HD 33636} is 20~\% \citep{Bean2007}, their superior precision makes the detection possible. 

Our analysis is applied to stars with an orbital solution obtained from radial velocities that provides the spectroscopic elements $P$, $e$, $T_0$, $\omega$, and $K_1$. The system is modeled as a primary mass $M_1$ orbited by one invisible companion of mass $M_2$. Using Hipparcos astrometric measurements, we seek to determine the two remaining unknown parameters that characterise the orbit, which are the inclination $i$ and the longitude of the ascending node $\Omega$. The analysis is performed in three steps. First, the Hipparcos abscissa, possibly containing the companion signature, is constructed from the catalogue data. This is important, because the IAD contains only \emph{abscissa residuals}, which are the astrometric residuals after subtraction of the model adopted by the new Hipparcos reduction. Second, the best-fit values for $i$ and $\Omega$ are found by determining the global $\chi^2$ minimum on a two-dimensional search grid. Finally, the significance of the obtained solution is evaluated.

\subsection{Constructing the Hipparcos abscissa}
Based on the Hipparcos identifier of a given star, the astrometric data is retrieved from the new reduction catalogue. It includes the astrometric parameters of the solution, the solution type and the goodness of fit. The IAD is read from the \emph{resrec} folder on the catalogue DVD of \cite{:2007kx} and contains the satellite orbit number, the epoch $t$, the parallax factor $\Pi$, the scan angle orientation $\psi$, the abscissa residual $\delta \Lambda$, and the abscissa error $\sigma_{\Lambda}$ for every satellite scan.

For the stars in our sample, we encounter the solution types '1', '5', '7', and '9'. The solution type '5' indicates that the five standard astrometric parameters fit the data reasonably well, whereas a solution type '1' is termed a stochastic solution and is adopted when the five-parameter fit is not satisfactory and neither orbital nor acceleration models improve the solution in terms of $\chi^2$. The solution types '7' and '9' are given, when the solution has to include acceleration parameters (proper-motion derivatives of first and second order) to obtain a reasonable fit. The solution types '1', '7', and '9' can be the first indications of a possible astrometric perturbation by an unseen companion. The Hipparcos abscissa $\Lambda_{HIP}$ is then given by:
\begin{equation}\label{eq:abscrecon}
\begin{split}
\Lambda_{HIP} = \, &( \alpha^{\star} + \mu_{\alpha^\star} \, t + \Delta_7 \alpha^{\star} +\Delta_9 \alpha^{\star}) \, \cos \psi \\
&+ \left( \delta + \mu_\delta \, t + \Delta_7 \delta + \Delta_9 \delta \right) \, \sin \psi + \varpi \, \Pi + \delta \Lambda, 
\end{split}
\end{equation}      
with the modified right ascension $\alpha^\star = \alpha \cos \delta$, the declination $\delta$, the associated proper motions $\mu_{\alpha^\star}$ and $\mu_\delta$, and the parallax $\varpi$. The epoch $t$ is given as time in years since $t_{HIP} = 1991.25 \, \mathrm{A.D.} = 2448348.8125$ JD and the correction factors for solution types '7' and '9' are given by
\begin{equation}
\Delta_7 p = \dot{\mu}_p \left( t^2-0.81 \right)/2
\end{equation}      
\begin{equation}
\Delta_9 p = \ddot{\mu}_p \left( t^2-1.69\right)\,t/6,
\end{equation}      
where $p$ is either $\alpha^{\star}$ or $\delta$ \citep{:2007kx}. $\dot{\mu}_p$ and $\ddot{\mu}_p$ are acceleration and change in acceleration in the coordinate direction $p$. For solution types '1' and '5', $\Delta_7 p$ and $\Delta_9 p$ vanish and the five parameter model holds. 

Because of the linear nature of Eq. \ref{eq:abscrecon}, it is usually convenient to decompose each of the five standard parameters into a constant term and an offset, e.g. $\alpha^\star = \alpha^\star_0 + \Delta \alpha^\star$, and to set the constant terms to zero to compute the abscissa. As explained in Sect.~\ref{sec:linear}, however, our formalism requires the value of the parallax $\varpi$ to be present in Eq. \ref{eq:abscrecon}, whereas we can safely set $\alpha^\star_0 = \delta_0 = \mu_{\alpha^\star,0} = \mu_{\delta,0} = 0$ for the calculation of the abscissa.  
\subsection{Model function}
The model function describes the astrometric signature along the scan axis $\psi$ of an isolated star, orbited by one invisible companion. To develop our approach, we list the standard formulae (e.g. \citealt{Hilditch:2001kx}) applicable to such a system. The astrometric motion of the star is characterised by $P$, $e$, $T_0$, $i$, $\omega$, $\Omega$, and the astrometric semimajor axis $a$, expressed in angular units. We write the Thiele-Innes constants as: 
\begin{equation}\label{eq:thiele}
\begin{split}
A = a \, c_A  \hspace{1cm} & c_A = \cos \Omega \cos \omega - \sin \Omega \sin \omega \cos i \\
B = a \,c_B  \hspace{1cm} & c_B = \sin \Omega \cos \omega + \cos \Omega \sin \omega \cos i \\
F = a \,c_F  \hspace{1cm} & c_F = - \cos \Omega \sin \omega - \sin \Omega \cos \omega \cos i \\
G = a \,c_G \hspace{1cm} & c_G = - \sin \Omega \sin \omega + \cos \Omega \cos \omega \cos i 
\end{split}
\end{equation}
where $c_{A,B,F,G}$ are functions of $i$, $\Omega$, and $\omega$ and $a$ is expressed in milli-arcseconds (mas). The elliptical rectangular coordinates $X$ and $Y$ are functions of eccentric anomaly $E$ and eccentricity: 
\begin{eqnarray}
E - e \sin E &=& \frac{2\pi}{P} (t-T_0)\\
X &=& \cos E - e\\
Y &=& \sqrt{1-e^2} \sin E
\end{eqnarray}
If the single-companion assumption is true, the Hipparcos abscissa is described by a linear equation with 9 parameters, which are the 5 standard astrometric parameters plus the 4 Thiele-Innes constants:
\begin{equation}\label{eq:abscissa1}
\begin{split}
\Lambda_{HIP} = \,& ( \alpha^{\star} + \mu_{\alpha^\star} \, t ) \, \cos \psi + ( \delta + \mu_\delta \, t ) \, \sin \psi + \varpi \, \Pi \\
& + (B \, X + G \, Y) \cos \psi + (A \, X + F \, Y) \sin \psi.
\end{split}\end{equation}
The relationship between astrometrically and spectroscopically derived parameters is \citep{Pourbaix:2001qe}:
\begin{equation}\label{eq:pourbaix}
a \sin{i} = c_P  \, K_1 \, P \, \sqrt{1-e^2} \, \varpi \hspace{1cm}
c_P = 3.35729138 \cdot 10^{-5},
\end{equation}
where $K_1$, $P$, and $\varpi$ are expressed in ms$^{-1}$, years, and mas, respectively. Hence, the astrometric model function can also be written as non-linear function of 12 independent parameters:
\begin{equation}\label{eq:abscissa2}
\Lambda_{HIP} = \Lambda_{HIP}\left( \alpha^{\star},  \delta, \varpi,\mu_{\alpha^\star} , \mu_\delta,e,P,T_0,\omega,\Omega,i, K_1 \right)
\end{equation}
The observed radial velocity $v_{\mathrm{rad}}$ of such an orbit is given by
\begin{equation}
\label{eq:vrad}
v_{\mathrm{rad}} = K_1 [\cos (\theta + \omega) + e \cos \omega ] + \gamma,
\end{equation}
where $\gamma$ is the systemic velocity. The true anomaly $\theta$ and the semi-amplitude $K_1$ are constrained by 
\begin{equation}
\label{eq:trueAno}
\tan \frac{\theta}{2} = \sqrt{(1+e)/(1-e)} \, \tan \frac{E}{2},
\end{equation}
\begin{equation}
\label{eq:K1}
K_1 = (2 \pi \, a_1 \sin i)/[P(1-e)^{1/2} ],
\end{equation}

\subsection{Definition of the search grid}\label{sec:iOscan}
One technique to find a possible orbital signature consists in defining a two-dimensional search grid in inclination $i$ and ascending node $\Omega$ and to solve for the remaining parameters of the model function \citep{Zucker:2000vn, Reffert:2006ly}. In this way, the goodness-of-fit is obtained for each point on the grid and the best solution can be identified as the global $\chi^2$ minimum. The search grid is defined by $i = 0-180^{\circ}$ and $\Omega = 0-360^{\circ}$, to avoid negative semimajor axes and sample the complete range of values of the cosine-function (cf. Eqs. \ref{eq:thiele} and \ref{eq:pourbaix}). We use a grid size of $30\times30$, which provides a reasonable compromise between computation time and grid resolution. 

\subsection{Joint confidence intervals}  \label{sec:linear}
For each point on the $i$-$\Omega$-grid we find the least-square solution by solving a linear equation using matrix inversion, where $e$, $P$, $T_0$, $\omega$, $K_1$, $i$, and $\Omega$ are fixed. The $\chi^2$ value hence obtained for each grid-point allows us to derive joint confidence intervals on the $i$-$\Omega$-grid. The parallax dependence of the Thiele-Innes constants is removed by writing 
\begin{equation}
a = a_1 \, \varpi, \hspace{1cm} a_1 = c_P K_1 P \sqrt{1-e^2} \frac{1}{\sin i}
\end{equation}
and defining the new constant
\begin{equation}
\Upsilon = a_1 \left[ ( c_B X + c_G Y ) \cos \psi + ( c_A X + c_F Y ) \sin \psi \right].
\end{equation}
Note that $a_1$ is expressed in AU and corresponds to the angular semimajor axis $a$ of the astrometric orbit. Now we can rewrite Eq. \ref{eq:abscissa1} as 
\begin{equation}\label{eq:upsilon}
\Lambda_{HIP} = ( \alpha^{\star} + \mu_{\alpha^\star} \, t ) \cos \psi + (\delta + \mu_\delta \, t) \sin \psi + \varpi \, \left(\Pi + \Upsilon \right).
\end{equation}
This relation is linear in the five remaining free parameters ($\alpha^{\star}$, $\delta$, $\varpi$, $\mu_{\alpha^\star}$, $\mu_\delta$) and can easily be solved analytically. The companion signature is solely contained in an additive modulation $\Upsilon(i,\Omega,e,P,T_0,\omega,K_1)$ of the parallax factor. It is important to realise that the presence of $\Upsilon$ forces the parallax, rather than a parallax offset, to be present in the model function, which originates in the abscissa reconstruction from Eq. \ref{eq:abscrecon}. For the remaining parameters it is sufficient to consider the offsets $\Delta \alpha^{\star}$, $\Delta \delta$, $\Delta \mu_{\alpha^\star}$, and $\Delta \mu_{\delta}$ to the catalogue values, because the abscissa has been constructed with $\alpha^\star_0 = \delta_0 = \mu_{\alpha^\star,0} = \mu_{\delta,0} = 0$.

\begin{figure*}\begin{center} 
\sidecaption
\includegraphics[width= 6cm]{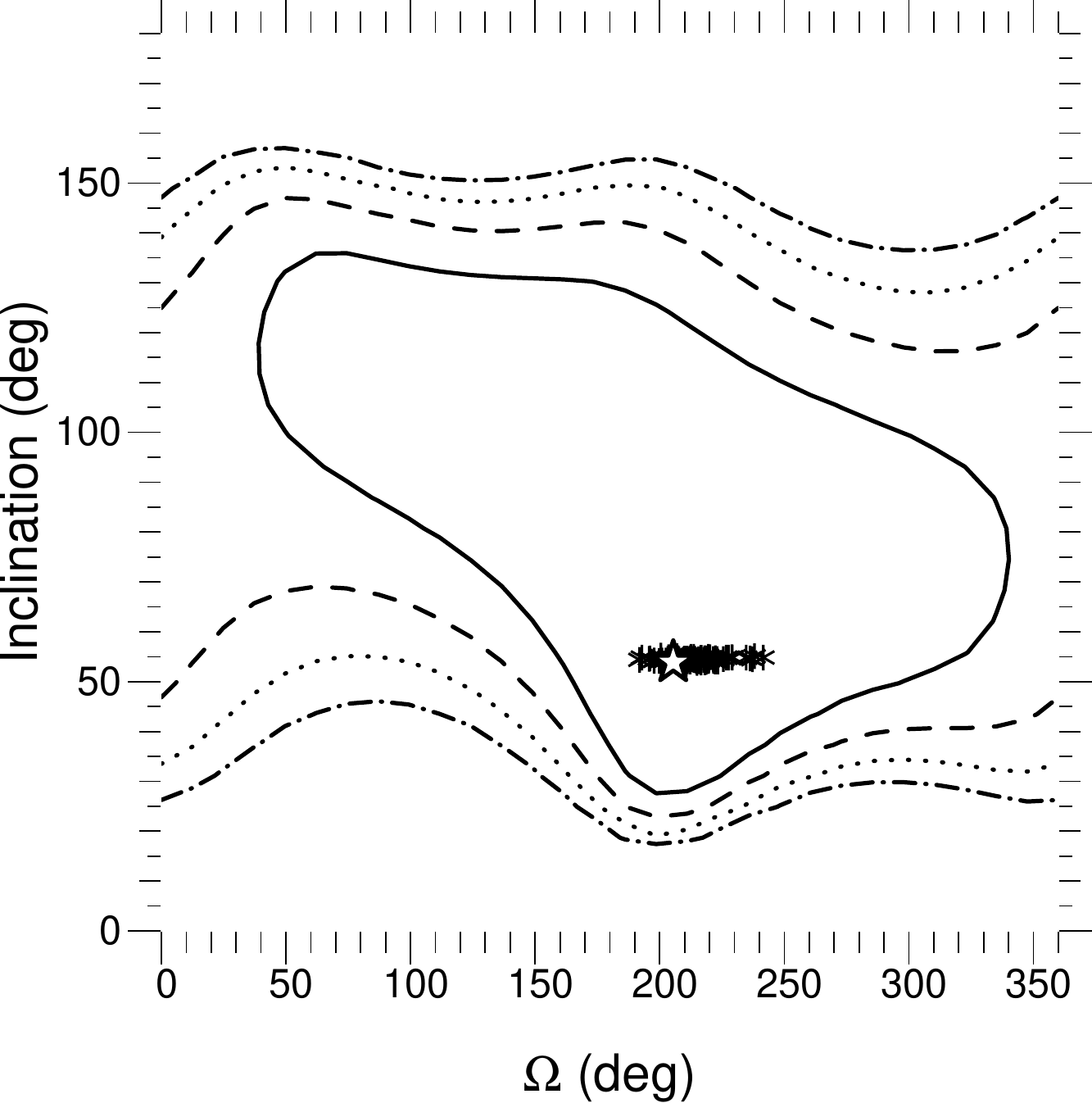} 
\includegraphics[width= 6cm]{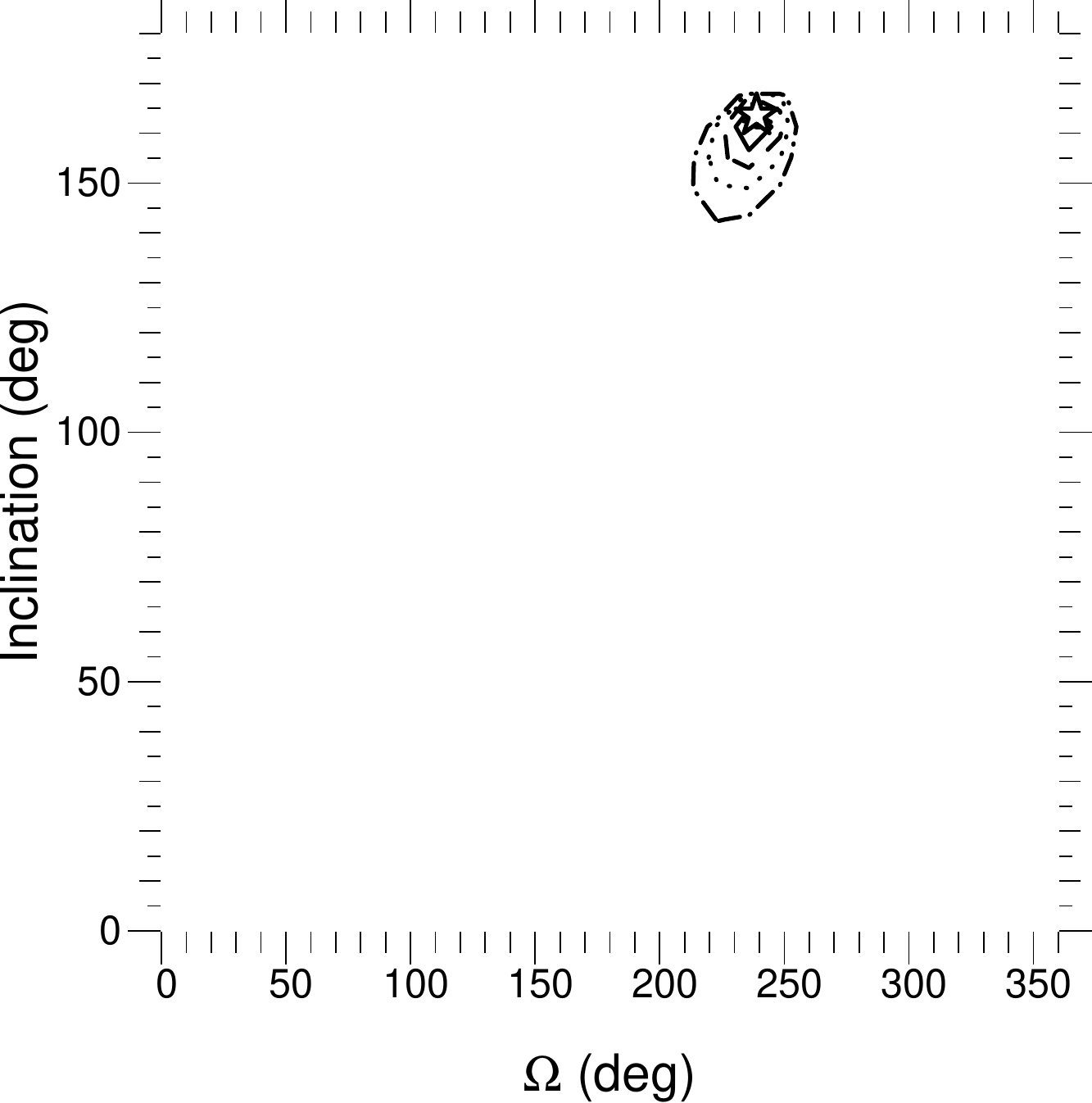} 
\caption{Joint confidence contours on the $i$-$\Omega$-grid for the low-significance orbit of HD~167665 (\emph{left}) and the high-significance orbit of HD~53680 (\emph{right}). The contour lines correspond to confidences at 1-$\sigma$ (solid), 2-$\sigma$ (dashed), 3-$\sigma$ (dotted), and 4-$\sigma$ (dash-dotted) level. The crosses indicate the position of the best non-linear adjustment solution for each of the 100 draws and the star corresponds to the adopted orbit.} 
\label{fig:HD167665contour}
\end{center} \end{figure*}

For every combination of $i$ and $\Omega$ the linear equation \ref{eq:upsilon} is solved, yielding the five astrometric parameters and the corresponding $\chi^2$. The best-fit parameters are identified by the minimum $\chi^2$ value on the $i$-$\Omega$-grid. The results of the linear adjustment are not directly used in the quoted final solution, but they serve as a consistency check for the results from the non-linear fitting and are used for graphical illustration of the joint confidence intervals as shown in Fig.~\ref{fig:HD167665contour}. Low-significance orbits can show several local $\chi^2$-minima corresponding to approximately opposite orbit orientations (see also e.g. \citealt{Zucker:2001ve, Reffert:2006ly}). We find that moderate- and high-significance orbits have one global $\chi^2$-minimum and the confidence contours cover a small area of the $i$-$\Omega$-space (similarly to HD~53680 in Fig.~\ref{fig:HD167665contour}). The solution parameters corresponding to an opposite-orientation orbit are always beyond the 4-$\sigma$ contour, i.e. all orbital parameters of significant solutions are unambiguously determined.

\subsection{Determining the final solution} \label{sec:nonlinear} 
While the solution method discussed in the previous section is favourable because the model function is linear, its accuracy is limited by the resolution of the $i$-$\Omega$-grid. For instance, the resolution of a square grid with 900 points is $6\degr$ and $12\degr$ in $i$ and $\Omega$, respectively. To avoid this limitation, we determine the best solution via $\chi^2$-minimisation of a model function, where $i$ and $\Omega$ are free parameters and thus can adopt continuous values. Therefor, the $i$-$\Omega$-grid is used to define the starting values of a non-linear least-squares fit by the Levenberg-Marquardt method. The model function is given by Eq. \ref{eq:abscissa2}, however limited to seven free parameters: the 5 astrometric parameters $\alpha^{\star}$, $\delta$, $\varpi$, $\mu_{\alpha^\star}$, $\mu_\delta$ plus $i$ and $\Omega$. We select the non-linear solution yielding the smallest $\chi^2$ and perform 1000 Monte-Carlo simulations. Every Monte-Carlo realisation consists of generating a set of Hipparcos abscissa measurements and consequent $\chi^2$-minimisation of the non-linear model and therefore provides 1000 sets of solution elements.   

To incorporate the uncertainties of the spectroscopic parameters we run the complete analysis for 100 sets of spectroscopic parameters, where each set is randomly drawn from gaussian distributions with mean and standard deviation given by the radial-velocity solution and its error, respectively (similarly to \citealt{Sozzetti:2010nx}). The final solution accounts for all 1000 Monte-Carlo solutions, which are obtained for each of the 100 sets of spectroscopic parameters. To achieve this, we combine all solutions to yield distributions of 100\,000 values for each parameter in Eq. \ref{eq:abscissa2}. The final parameter value and its error is derived from the mean and the 1-$\sigma$ confidence interval of the associated distribution, respectively. The values and errors of derived quantities, such as the companion mass, are obtained in the same way and thus take into account possible correlations between individual parameters.

\subsection{Outliers in the astrometric data}
Outlying astrometric measurements are examined if they deviate by more than 4-$\sigma$ from the orbital solution. Here, $\sigma$ is calculated as the root-mean-square of the fit residuals. The datapoint is discarded if there are at least three measurements at same satellite orbit number and if it is more than 3-$\sigma$ away from these datapoints. If less than three datapoints at same orbit number are present, they are all removed. The analysis is then iterated. This procedure is similar to the one applied by \cite{Pourbaix:2000sf} and in some cases also agrees with the outlier rejection, that is applied by the new Hipparcos reduction itself. Outliers are removed for the following objects: GJ~595, HD~3277, HD~43848, HD~74842, HD~154697, HD~164427A, HD~167665, HD~191760, and HIP~103019. This never requires more than one iteration. 

\subsection{Statistical significance of the orbit}
Orbital solutions can be found for any set of astrometric data and the crucial step in the analysis is to determine the credibility of the derived orbits. We accomplish this by applying two statistical tests: the F-test for its simplicity and the permutation test for its lack of underlying assumptions.

The F-test is extensively used for the statistical analysis of astrometric orbit signatures in Hipparcos data \citep{Pourbaix:2000sf, Pourbaix:2001rt, Pourbaix:2001qe, Reffert:2006ly}, although it relies on the assumption that the measurement errors are Gaussian. That this assumption is not necessarily fulfilled for Hipparcos data is mentioned by \cite{Zucker:2001ve}, who do not apply the F-test for their analysis. The F-test yields the probability that the null assumption \emph{no orbital motion present} is true, by comparing the $\chi^2$-values of two models with differing number of parameters. In our case, we consider the $\chi^2_7$-value of the 7-parameter solution by non-linear minimisation for each of the 100 draws from the spectroscopic elements. The F-test is evaluated on the median $\chi^2_7$-value with respect to the $\chi^2_5$ value, obtained from the standard 5-parameter solution, and results in the null probability, which is tabulated in Table \ref{tab:mass2sigma} and gives the probability that the simpler model is valid. We use the F-test as an additional indicator for the statistical confidence of our solution, but the conclusive argument is derived from the permutation test.

The permutation test has the important advantage to be part of the distribution-free tests \citep{Zucker:2001ve} and therefore does not rely on the assumption of Gaussian error distribution. For our purpose, it can be summarised by the idea that a periodic signal present in a given data set will be destroyed by random permutation of the individual datapoints in all but a few cases. In the present analysis, the periodicity is defined by the radial velocity solution. To derive a significance, the semimajor axis of the solution orbit, derived with the original astrometric data, is compared to the semimajor axes of pseudo-orbits, which are derived from randomly permuted astrometric data. If a signal is present in the astrometric data, it will be destroyed by the random permutations and the solution orbit will stick out of the pseudo-orbits with its large amplitude.

\begin{figure*}\begin{center} 
\sidecaption
\includegraphics[width= 6cm]{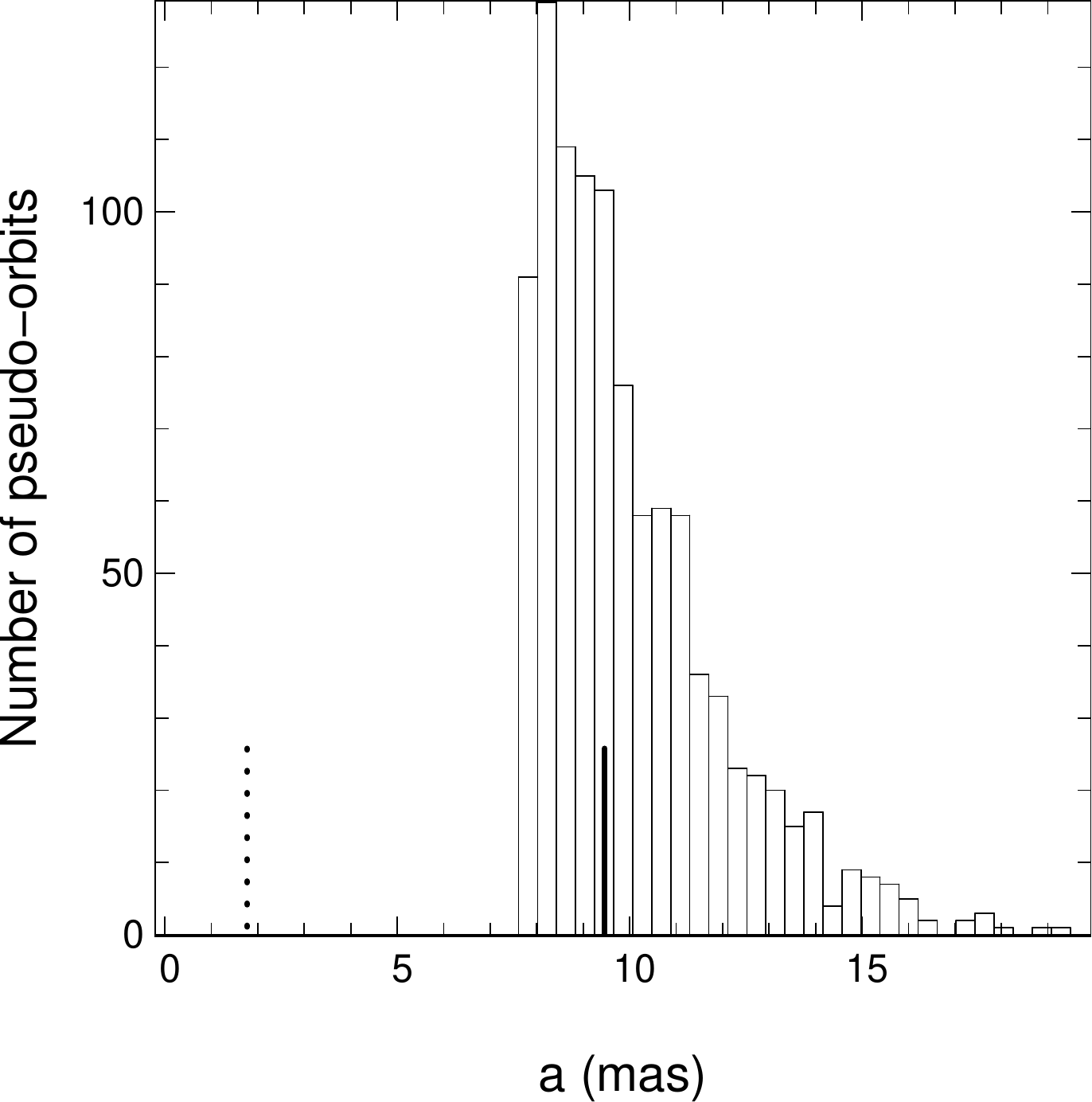} 
\includegraphics[width= 6cm]{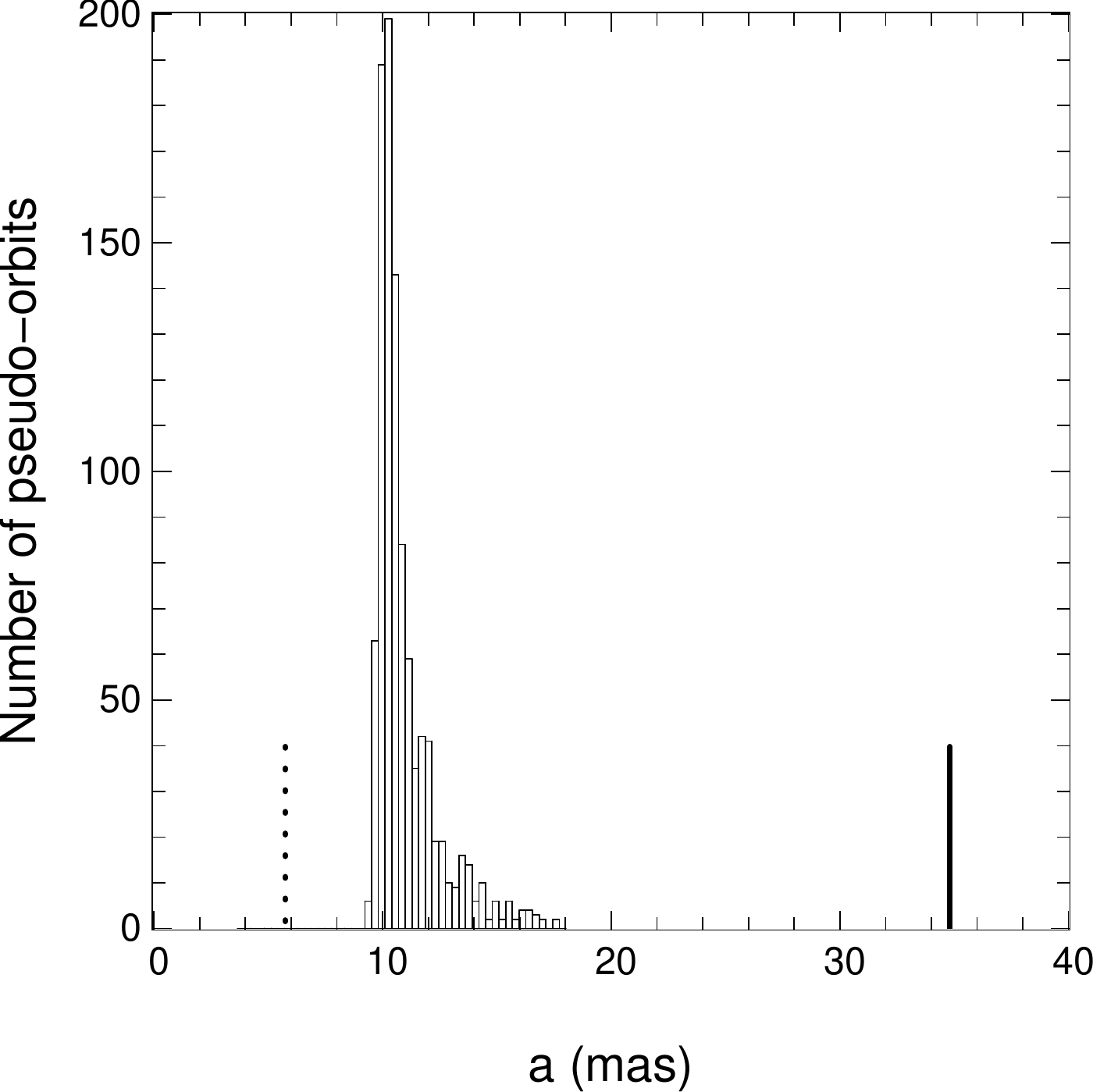} 
\caption{Histogram showing the semimajor axes of the 1000 pseudo-orbits for the low-significance orbit of HD~167665 (\emph{left}) and the high-significance orbit of HD~53680 (\emph{right}). The vertical solid line indicates the semimajor axis of the non-permuted, best-fit solution and the dotted line indicates the median Hipparcos single-measurement precision $\sigma_\Lambda$ for the respective star.} 
\label{fig:HD53680a1permute}
\end{center} \end{figure*}

Because of the pre-reduction present in the original Hipparcos IAD \citep{ESA:1997vn}, its permutation had to be done with great care \citep{Zucker:2001ve}. Fortunately, it became considerably easier with the availability of the new Hipparcos reduction IAD. The permutation is applied only to the list of abscissa residuals $\delta \Lambda$, while taking the possible contribution of the 7- and 9-parameter solution into account. The remaining data, i.e. epochs, parallax factors, and scan angle orientations are left unchanged. To obtain the pseudo-orbits, 1000 permutations are performed and the analysis is run on the synthesised data, using the nominal spectroscopic elements. The statistical significance of the solution orbit is equal to the percentage of pseudo-orbits, that have a smaller semimajor axis than the solution orbit \citep{Zucker:2001ve}. Examples for high- and low-significance orbits are shown in Fig.~\ref{fig:HD53680a1permute}. In contrast to \cite{Halbwachs:2000rt}, the analysis of the pseudo-orbits is done with spectroscopic parameters identical to the nominal ones, hence the method of comparing the distribution of $a/\sigma_a$ to the expected Rayleigh-Rice law is not applicable.

\subsection{Validation of the method}
The combined analysis of radial-velocity and astrometric data can lead to erroneous conclusions, if the results are not carefully checked \citep{Han:2001kx, Pourbaix:2001rt}. Especially the relationship between the inclination and the semimajor axis has to be investigated during the interpretation, since over- and underestimation of these parameters during the fitting process affect the derived companion masses. To develop good confidence in our results, we validate our method both by simulation and by comparison with published work. 

\subsubsection{Simulation} \label{sec:simulation}
Simulations are employed to investigate the effect of Hipparcos measurement precision on the derived orbital parameters and the orbit significance. Of particular interest is the behaviour of the occurring biases as function of signal-to-noise given by the ratio of orbital semimajor axis and measurement precision.

Instead of modeling a complete dataset, we chose the case of {HD~17289}, whose orbit is discussed in Sect.~\ref{sec:results} and where the Hipparcos observations cover twice the orbital period. To inject the astrometric signature of a companion, we use the satellite configuration for this star, i.e. the scan angles, the parallax factors, and the measurement precision. Noise is generated by adding a random bias term to each measurement, which is drawn from a Gaussian distribution with standard deviation given by the satellite measurement precision. Configurations with different signal-to-noise are generated by altering the radial velocity semi-amplitude $K_1$ of the simulated orbit, while keeping all other parameters fixed. Consequently, the emulated Hipparcos errors in this simulation are Gaussian, which is a simplifying assumption and may not be fulfilled in reality. However, the simulation is valid to investigate the biases of the method. 

\begin{figure}
\begin{center} 
\includegraphics[width= \linewidth, trim = 0cm 0.5cm 0cm 1.3cm, clip=true]{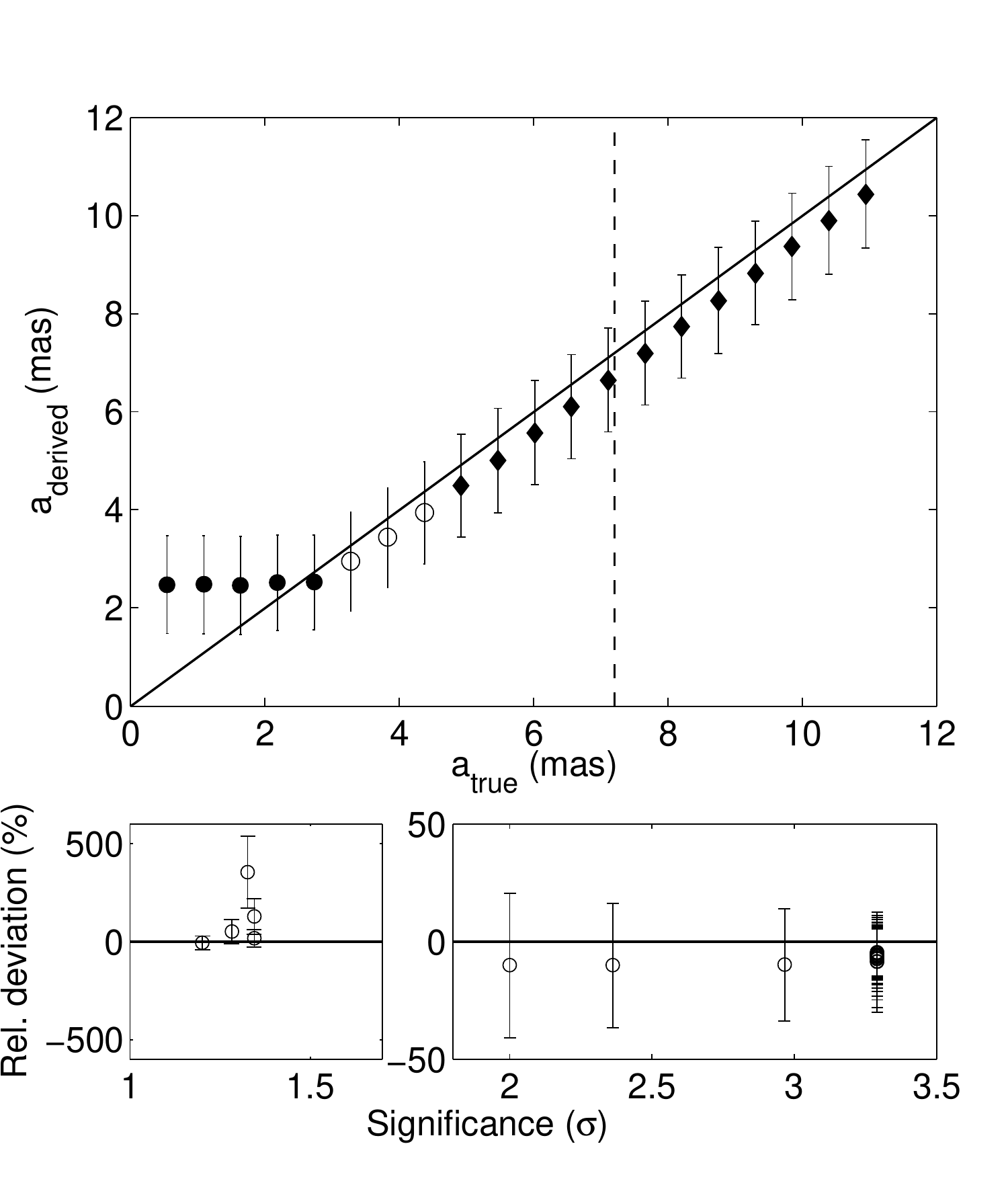} 
\caption{Simulated difference between the derived semimajor axis $a_\mathrm{derived}$ and its true value $a_\mathrm{true}$. \emph{Top}: $a_\mathrm{derived}$ as function of $a_\mathrm{true}$. Filled circles, open circles, and diamonds designate orbits of $1\!-\!2\,\sigma$, $2\!-\!3\,\sigma$, and $>\!3\,\sigma$ significance, respectively. The solid line has unity slope and the dashed line shows the single-measurement precision of $\sigma_\Lambda = 7.2$~mas. \emph{Bottom}: The relative deviation calculated as $(a_\mathrm{derived}- a_\mathrm{true})/a_\mathrm{true}$ as function of significance. The accuracy and relative precision of the solution improve with significance. At the detection limit of 3-$\sigma$, the deviation from the true value is below 0.5 mas.} \label{fig:simuresult_a}
\end{center} \end{figure}

\begin{figure}
\begin{center} 
\includegraphics[width= 0.7\linewidth]{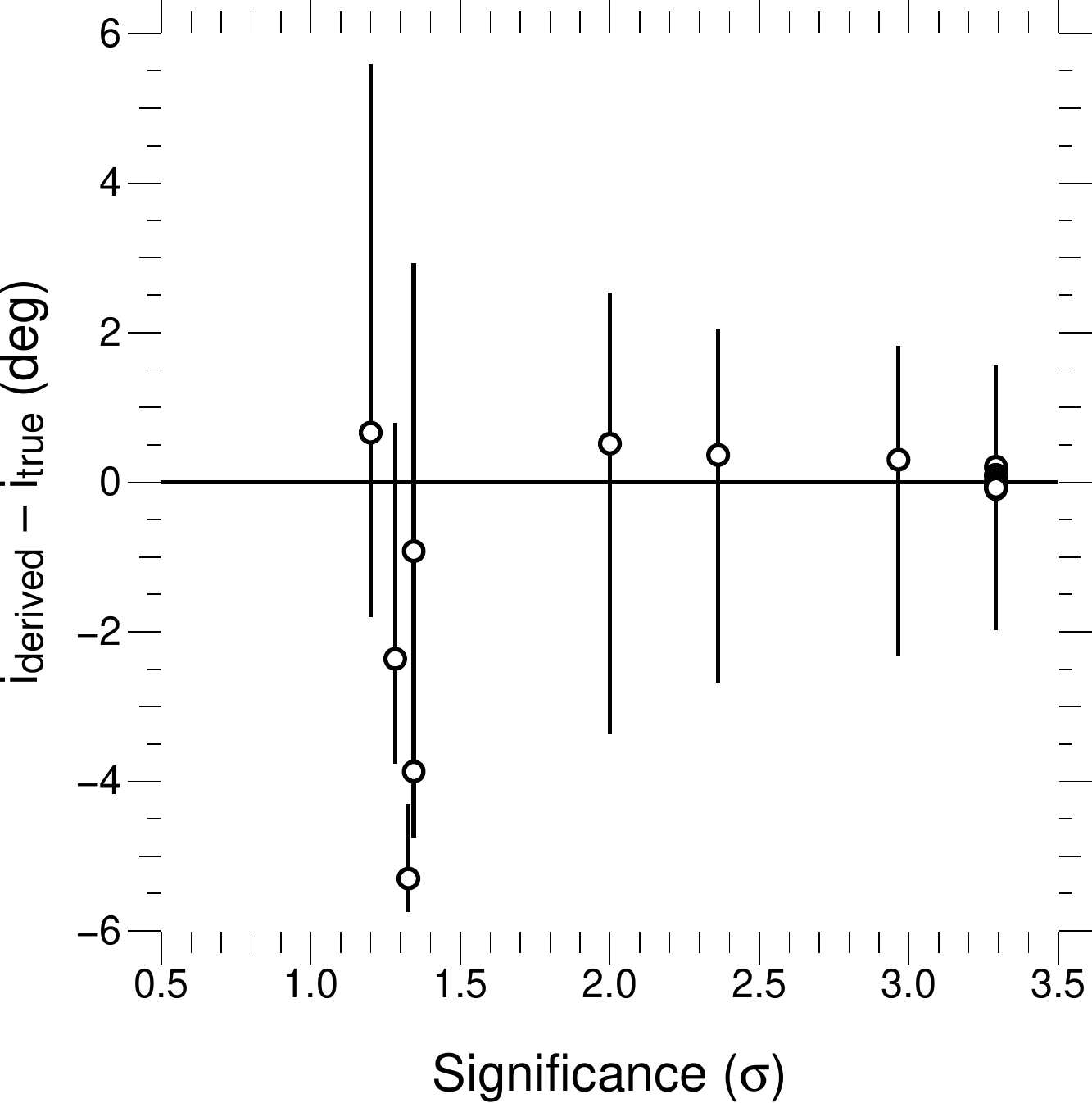} 
\caption{Simulated difference between the derived inclination $i$ and its true value. The accuracy and the precision of the solution improve with significance. At the detection limit of 3-$\sigma$, the deviation from the true value is below $0.02^{\circ}$.} \label{fig:simuresult_i}
\end{center} \end{figure}

The results are illustrated in Figs. \ref{fig:simuresult_a} and \ref{fig:simuresult_i}, where the accuracy of the determined values for inclination and semimajor axis are displayed as function of statistical significance, derived from the permutation test. The accuracy is expressed in terms of the difference between determined and true value, i.e. for the inclination it is computed as $i_{\mathrm{derived}} - i_{\mathrm{true}}$. Generally, the accuracy of the derived $a$ and $i$ increase with signal-to-noise and significance. For significance above 2-$\sigma$, the accuracy is comparable to the precision. However, the semimajor axis retains a small bias even at high signal-to-noise, whose magnitude and sign is influenced by the satellite configuration concerning orbit orientation and period. The determination of the orbit inclination greatly improves with significance. To investigate if our analysis is biased even when exceeding the 3-$\sigma$ significance, we repeated the simulation for 3 different satellite configuration, i.e. three different stars. We find no sign for a systematic bias, favouring large or small inclinations and semimajor axes. Above 3-$\sigma$ significance, derived and true values always agree within the error bars. This simulation shows that a large bias can be introduced in the solution, if an orbit significance of 1-$\sigma$ is used to qualify a detection. In contrast, the biases are small if higher significance is requested. 

We find that an orbit is detected at better than 3-$\sigma$ significance, when its semimajor axis amounts to $\sim$$70$~\% of the measurement precision. This result may depend on the assumption of Gaussian errors and the specific satellite configuration, but it confirms the warning of \cite{Pourbaix:2001qe}, that derived orbits with sizes comparable to the instrument precision have to be evaluated very carefully.   

\begin{table*} \caption{Results for the comparison sample.}
\label{tab:compResults} 
\centering  
\begin{tabular}{r r r l r l r l r} 
\hline\hline %
Object & HIP & \multicolumn{2}{c}{$a$} & \multicolumn{2}{c}{$M_2$} & \multicolumn{2}{c}{Significance} & Ref.\\  
           &             & \multicolumn{2}{c}{(mas)} & \multicolumn{2}{c}{($M_J$)}&\multicolumn{2}{c}{(\%)} & \\  
\hline 
HD~190228 & 98714 & $ 1.8^{+ 0.5}_{-0.5}$ & $[1.8\pm0.8]$        & $ 65.7^{+ 20.4}_{-20.4}$\tablefootmark{a}     & $[67.1^{+ 30.4}_{-28.3} ]$      & 96.3     & [95] & 1 \vspace{1mm}\\ 
GJ~1069      & 19832 & $ 17.7^{+ 0.8}_{-0.8}$ & $  [17.1\pm0.8]$  & $ 263.2^{+ 18.7}_{-18.7}$   & $[253.5^{+14.7}_{-15.7}]$     & $>99.9$ &[99.9] & 1 \vspace{1mm}\\ 
GJ~1069      & 19832 & $ 17.7^{+ 0.8}_{-0.8}$ & $ [17.2\pm0.8]$   & $ 263.2^{+ 18.7}_{-18.7}$   & $[256.7\pm14.7]$                   & $>99.9$ & [$\cdots$] & 2 \vspace{1mm}\\ 
HD~145849 & 79358 & $ 11.1^{+ 0.7}_{-0.7}$ & $ [9.69\pm0.85]$ & $ 1443.5^{+ 18.5}_{-20.0}$ & $ [1424.8^{+73.3}_{-31.4}]$ & $>99.9$ & [$\cdots$] & 3 \vspace{1mm}\\ 
\hline
\end{tabular}
\tablebib{(1) ~\cite{Zucker:2001ve}; (2) \citet{Halbwachs:2000rt}; (3) \citet{Torres:2007kx}.}
\tablefoot{The values in square brackets are the corresponding values from the reference paper. Our error on the companion mass does not include the uncertainty of the primary mass. \tablefoottext{a} Primary mass from \cite{Sivan:2004rm}}	
\end{table*}

\subsubsection{Comparison sample}\label{sec:comptargets}
For independent validation of our method and to investigate the impact of the new reduction, we screened the literature for objects for which a similar analysis was performed. We chose the three stars \object{HR~6046}, \object{GJ~1069}, and \object{HD~190228} with companions found by \cite{Torres:2007kx}, \cite{Halbwachs:2000rt}, and \cite{Zucker:2001ve}\footnote{HD~190228 has also been analysed by \cite{Pourbaix:2001rt}.}, respectively, and exhibiting diverse orbit inclinations and companion masses. All these works use the original IAD, whereas we use the new reduction IAD.

For all three stars, good agreement is found between our results and the published works, both in terms of value and 1-$\sigma$ uncertainty of semimajor axis and companion mass, as is shown in Table \ref{tab:compResults}. Also the orbit significances derived from pseudo-orbits given by \cite{Zucker:2001ve} for two objects match our results. For \object{HD~190228}, we find the same solution as \cite{Zucker:2001ve} and a 2-$\sigma$ significance (see Sect. \ref{sec:2sresults} for further discussion of this object). We confirm the detection of the \object{GJ~1069} orbit, caused by a low mass star \citep{Zucker:2001ve, Halbwachs:2000rt}, and the binary orbit of \object{HR~6046} \citep{Torres:2007kx} at better than 3-$\sigma$ significance. Especially the case of \object{GJ~1069}, that is now analysed by three independent teams finding essentially identical results, attests that our analysis is robust.\\

Both the simulation results and the analysis of a sample of comparison stars confirm that our method for the combined analysis of radial-velocity orbits and Hipparcos astrometry is correct. The consideration of the orbit significance ensures that the derived result is bias-free to reasonable extent.

\subsection{Companion mass limits}\label{sec:mlim}
The combination of astrometric and spectroscopic data makes it possible to find a complete orbital solution. However, the result can be misleading or wrong, if the significance of the solution is not verified. For significant orbits the derived parameters, e.g. the companion mass, and their errors are valid. For low-significance orbits, the derived parameters and their errors are very probably false. Therefore, the distribution function for the companion mass of a low-significance orbit cannot be used to derive an upper and lower mass limits. 

The question is then turned to the definition of a significant orbit. In the literature, various significance estimators (e.g. F-test or permutation test) and detection limits are employed.  We use a 3-$\sigma$ detection limit based on the permutation test and deem orbits at less than 2-$\sigma$ unreliable. For a 2-$\sigma$ orbit we adopt a reduced validity and expect that the mass limits do hold.

In principle, it is possible to derive the upper mass limit for any companion by employing simulations similar to the one described in Sect. \ref{sec:simulation}. Such simulation is extremely elaborate and its application to the list of host stars with undetected astrometric motion is beyond the scope of this paper. However, we showed that an orbit is detected at better than 3-$\sigma$ significance, when its semimajor axis amounts to $\sim$$70$~\% of the measurement precision, if the Hipparcos observations cover one complete orbit. Including an additional confidence margin, we thus claim that the signature of a fully covered orbit would have been detected, if its semimajor axis equalled the single-measurement precision. Hence, an upper limit to the companion mass can be set by enforcing the equation $a = \sigma_\Lambda$ for orbits with significance below $2$-$\sigma$ and $N_\mathrm{orb} > 1$.

\section{Results}\label{sec:results}
For all objects in our sample we are able to obtain an orbital solution from the combination of the radial-velocity orbit and the Hipparcos astrometric data, but not all of these orbits are credible. To distinguish a real orbit from meaningless output of the fitting procedure, we use the permutation test to evaluate the orbit significance. This results in the detection of 6 orbits having high significance above 3-$\sigma$ (Tables \ref{tab:sol2sigma} and \ref{tab:mass2sigma}). The orbits of 5 objects have moderate significance between 2-$\sigma$ and 3-$\sigma$. Three orbits have low significance between 1-$\sigma$ and 2-$\sigma$ and the orbits of 17 objects have very-low significance ($<\!1$-$\sigma$, Table~\ref{tab:insignificantOrbits})\footnote{As motivated in Sect. \ref{sec:objnotes}, we do not present an astrometric analysis for HD~4747 and HD~211847 and therefore the number of objects listed here adds up to 31 instead of 33.}.  

The Tables \ref{tab:KeplerOrbits}, \ref{tab:target2sigma}, and \ref{tab:insignificantOrbits} list the spectroscopic and astrometric elements of the adopted orbital solution together with the orbit significance and give additional information: $N_{\mathrm{orb}}$ is the number of orbits covered by the Hipparcos observations, $\sigma_\Lambda$ is the median Hipparcos single-measurement precision, $a \sin i$ is the minimum astrometric semimajor axis, $M_2$~(3-$\sigma$) are upper and lower companion-mass limits at 3-$\sigma$, $a_\mathrm{rel}$ is the semimajor axis of the relative orbit, $\chi^2_{7,\mathrm{red}}$ is the reduced chi-square value of the adopted 7-parameter solution, and the Null probability gives the result of the F-test. The quoted errors correspond to Monte-Carlo based 1-$\sigma$-confidence intervals.

For orbits with significance below $2$-$\sigma$ and provided that the orbital period is fully covered by Hipparcos observations, we derived the upper limit to the companions mass $M_{2,\mathrm{up-lim}}$ by enforcing the equation $a = \sigma_\Lambda$ (cf.~Sect.~\ref{sec:mlim}).

\begin{table*} 
\caption{Astrometric solution elements for targets with orbits that have significance above 1-$\sigma$.}
\label{tab:sol2sigma} 
\centering  
\begin{tabular}{r r r r r r r r r } 
\hline\hline %
Object  & $\Delta \alpha^{\star}$ & $\Delta \delta$ & $\varpi$ &$\Delta \varpi$  & $\Delta \mu_{\alpha^\star}$ & $\Delta \mu_{\delta}$ & $i$ & $\Omega$  \\  
        &  (mas)                     & (mas)             &  (mas)    &  (mas)            &  (mas $\mathrm{yr}^{-1}$)       & (mas $\mathrm{yr}^{-1}$) & (deg)	      & (deg)	    \\  
\hline 
HD~17289 & $-6.7^{+ 0.8}_{-0.8}$ & $6.0^{+ 0.8}_{-0.8}$  & $20.7^{+ 0.6}_{-0.6}$ & $-0.1$ & $1.2^{+ 0.6}_{-0.6}$ & $-2.2^{+ 0.7}_{-0.7}$ & $ 173.2^{+ 0.4}_{-0.5}$ & $ 2.5^{+ 3.9}_{-3.9}$  \vspace{1mm} \\ 
 HD~30501 & $9.1^{+ 3.0}_{-3.0}$ & $-12.6^{+ 3.5}_{-3.5}$  & $48.1^{+ 0.4}_{-0.4}$ & $0.2$ & $-3.0^{+ 0.8}_{-0.8}$ & $-4.0^{+ 1.8}_{-1.8}$ & $ 49.1^{+ 10.1}_{-7.8}$ & $ 99.7^{+ 12.0}_{-12.4}$  \vspace{1mm} \\ 
 HD~43848\tablefootmark{a} & $2.7^{+ 1.3}_{-1.3}$ & $4.0^{+ 1.3}_{-1.3}$  & $26.4^{+ 0.7}_{-0.7}$ & $-0.1$ & $-0.2^{+ 1.0}_{-1.0}$ & $-5.6^{+ 1.1}_{-1.1}$ & $ 165.0^{+ 2.2}_{-3.0}$ & $ 279.3^{+ 12.9}_{-13.0}$  \vspace{1mm} \\
HD~43848\tablefootmark{b} & $3.7^{+ 1.9}_{-1.9}$ & $7.9^{+ 2.0}_{-2.0}$  & $26.5^{+ 0.7}_{-0.7}$ & $0.1$ & $-0.7^{+ 0.9}_{-0.9}$ & $-4.4^{+ 1.1}_{-1.1}$ & $ 155.7^{+ 6.6}_{-14.7}$ & $ 237.9^{+ 11.7}_{-11.7}$  \vspace{1mm} \\ 
 HD~53680 & $6.2^{+ 2.1}_{-2.1}$ & $41.3^{+ 4.0}_{-4.0}$  & $58.2^{+ 0.8}_{-0.8}$ & $0.8$ & $-21.5^{+ 2.2}_{-2.2}$ & $2.0^{+ 1.7}_{-1.7}$ & $ 163.6^{+ 1.4}_{-1.7}$ & $ 238.9^{+ 2.9}_{-2.9}$  \vspace{1mm} \\ 
 HD~164427A & $0.4^{+ 0.7}_{-0.7}$ & $1.5^{+ 0.5}_{-0.5}$  & $25.8^{+ 0.6}_{-0.6}$ & $-0.5$ & $0.1^{+ 0.6}_{-0.6}$ & $0.2^{+ 0.4}_{-0.4}$ & $ 11.8^{+ 3.1}_{-2.1}$ & $ 338.9^{+ 12.3}_{-12.3}$  \vspace{1mm} \\ 
 HIP~103019 & $5.6^{+ 1.6}_{-1.6}$ & $4.2^{+ 1.3}_{-1.3}$  & $30.2^{+ 1.8}_{-1.8}$ & $1.9$ & $0.7^{+ 1.9}_{-1.9}$ & $0.2^{+ 1.6}_{-1.6}$ & $ 160.9^{+ 2.2}_{-2.7}$ & $ 143.1^{+ 8.6}_{-8.6}$  \vspace{1mm}  \\ 
 HD~3277 & $-1.0^{+ 0.5}_{-0.5}$ & $0.3^{+ 0.5}_{-0.5}$  & $34.9^{+ 0.6}_{-0.6}$ & $0.1$ & $-0.8^{+ 0.6}_{-0.6}$ & $-0.6^{+ 0.5}_{-0.5}$ & $ 166.5^{+ 2.1}_{-3.1}$ & $ 270.2^{+ 11.2}_{-11.2}$  \vspace{1mm} \\ 
 HD~131664 & $-0.3^{+ 1.6}_{-1.6}$ & $-1.9^{+ 0.8}_{-0.8}$  & $18.6^{+ 0.7}_{-0.7}$ & $0.9$ & $-1.6^{+ 1.1}_{-1.1}$ & $-2.3^{+ 0.9}_{-0.9}$ & $ 167.7^{+ 1.9}_{-2.8}$ & $ 321.0^{+ 31.6}_{-30.7}$  \vspace{1mm} \\ 
 HD~154697 & $4.2^{+ 1.7}_{-1.9}$ & $5.9^{+ 2.8}_{-3.2}$  & $29.4^{+ 1.0}_{-1.0}$ & $-1.1$ & $-7.7^{+ 1.8}_{-1.9}$ & $2.3^{+ 1.8}_{-1.8}$ & $ 148.8^{+ 4.6}_{-6.1}$ & $ 293.9^{+ 65.5}_{-78.6}$  \vspace{1mm} \\ 
  HD~174457 & $-0.7^{+ 0.7}_{-0.7}$ & $1.3^{+ 0.6}_{-0.6}$  & $18.5^{+ 0.7}_{-0.7}$ & $-0.2$ & $-0.3^{+ 0.8}_{-0.9}$ & $-0.8^{+ 0.9}_{-0.9}$ & $ 137.0^{+ 11.9}_{-14.1}$ & $ 133.0^{+ 17.2}_{-11.0}$  \vspace{1mm} \\ 
 HD~190228 & $0.6^{+ 0.6}_{-0.6}$ & $-1.3^{+ 0.7}_{-0.7}$  & $16.0^{+ 0.6}_{-0.6}$ & $-0.2$ & $-1.2^{+ 0.5}_{-0.5}$ & $0.3^{+ 0.5}_{-0.5}$ & $ 4.3^{+ 1.8}_{-1.0}$ & $ 61.0^{+ 22.7}_{-22.9}$  \vspace{1mm} \\ 
 HD~74014 & $-57.4^{+ 33.6}_{-33.5}$ & $49.4^{+ 39.6}_{-39.3}$  & $29.5^{+ 0.7}_{-0.7}$ & $-0.0$ & $-13.3^{+ 8.6}_{-8.6}$ & $-4.9^{+ 5.2}_{-5.3}$ & $ 171.1^{+ 3.0}_{-10.6}$ & $ 248.7^{+ 16.7}_{-27.2}$  \vspace{1mm} \\ 
 HD~168443c & $-0.7^{+ 0.7}_{-0.7}$ & $0.5^{+ 0.6}_{-0.6}$  & $26.3^{+ 0.7}_{-0.7}$ & $-0.4$ & $1.2^{+ 1.0}_{-0.9}$ & $-0.1^{+ 1.0}_{-1.0}$ & $ 35.7^{+ 14.5}_{-11.2}$ & $ 139.1^{+ 33.4}_{-34.8}$  \vspace{1mm} \\ 
 HD~191760 & $-0.5^{+ 1.0}_{-1.0}$ & $1.0^{+ 0.8}_{-0.8}$  & $11.5^{+ 1.1}_{-1.1}$ & $-0.8$ & $0.7^{+ 1.4}_{-1.4}$ & $-0.4^{+ 1.0}_{-1.0}$ & $ 15.4^{+ 21.0}_{-6.8}$ & $ 85.8^{+ 46.6}_{-16.0}$  \vspace{1mm} \\ 
 \hline
\end{tabular} 
\tablefoot{The top 6 targets have high-significance orbits ($\geqslant 99.7$ \%). The bottom 3 targets have low-significance orbits (1-$\sigma\!-2\!$-$\sigma$) and the parameters represent the formal solution. The remaining 5 orbits have moderate significance (2-$\sigma-\!3\!$-$\sigma$). \tablefoottext{a} {Final solution at lower eccentricity.} \tablefoottext{b} {Formal solution at higher eccentricity.} }
\end{table*}

\begin{table*} 
\caption{Solution parameters and companion masses for stars with orbits that have significance above 1-$\sigma$.}
\label{tab:mass2sigma} 
\centering  
\begin{tabular}{r r r r r r r r r r r r r r} 
\hline\hline %
Object & Sol. & $N_\mathrm{orb}$ & $\sigma_{\Lambda}$  & $a \sin i$ & $a$ & $M_2$  & $M_2$ (3-$\sigma$)& $a_{\mathrm{rel}}$  & $\chi^2_\mathrm{7,red}$ & Null prob. & Significance \\  
       & type &           & (mas)            &(mas)     & (mas) & ($M_J$) & ($M_J$) &(mas) &  & (\%)  & (\%) \\  
\hline  
 HD~17289 & 1 & 2.1 & 7.2 & 1.29 & $ 10.8^{+ 0.7}_{-0.7}$ & $ 547.4 ^{+47.7}_{-47.3} $ & ( 386, 700) & 32.0 & 0.32& 7e-29 &$>99.9$ \vspace{1mm} \\ 
 HD~30501 & 5 & 0.6 & 3.3 & 10.44 & $ 14.0^{+ 1.8}_{-1.9}$ & $ 89.6 ^{+12.3}_{-12.5} $ & (  65, 132) & 147.7 & 0.79& 0.005 &99.9 \vspace{1mm} \\ 
 HD~43848\tablefootmark{a} & 9 & 0.5 & 4.2 & 2.32 & $ 8.9^{+ 1.2}_{-1.2}$ & $ 101.8^{+ 15.0}_{-15.0}$ & $( 59,150)$ & 90.8 & 1.09& 4e-8 &$>99.9$ \vspace{1mm} \\
 HD~43848\tablefootmark{b} & 9 & 0.5 & 4.2 & 4.93 & $ 10.5^{+ 1.3}_{-1.3}$ & $ 120.2^{+ 17.1}_{-17.2}$ & $( 73,179)$ & 91.7 & 1.07& 1e-8 &$>99.9$ \vspace{1mm} \\ 
 HD~53680 & 1 & 0.7 & 5.7 & 9.73 & $ 34.9^{+ 3.2}_{-3.2}$ & $ 226.7 ^{+24.0}_{-23.9} $ & ( 159, 301) & 160.7 & 0.82& 3e-16 &$>99.9$ \vspace{1mm} \\ 
 HD~164427A & 5 & 11.1 & 2.4 & 0.49 & $ 2.3^{+ 0.5}_{-0.5}$ & $ 269.9 ^{+63.2}_{-63.2} $ & ( 101, 482) & 12.8  & 1.15& 0.003 &99.7 \vspace{1mm} \\ 
 HIP~103019 & 5 & 1.2 & 7.1 & 3.33 & $ 10.8^{+ 1.3}_{-1.3}$ & $ 188.1^{+ 26.5}_{-26.4}$ & $(118,280)$ & 53.4 & 1.13& 4e-10 &$>99.9$ \vspace{1mm} \\ 
 HD~3277 & 5 & 24.8 & 3.6 & 0.58 & $ 2.5^{+ 0.5}_{-0.5}$ & $ 344.2 ^{+76.1}_{-76.1} $ & ( 143, 600) & 9.4 & 1.00& 5e-4 &97.7 \vspace{1mm} \\ 
 HD~131664 & 5 & 0.6 & 4.7 & 0.88 & $ 4.3^{+ 0.8}_{-0.8}$ & $ 89.6^{+ 16.4}_{-16.6}$ & $( 44,153)$ & 60.3 & 0.77& 2e-5 &97.9 \vspace{1mm} \\ 
 HD~154697 & 7 & 0.6 & 3.1 & 6.26 & $ 11.7^{+ 1.7}_{-1.7}$ & $ 151.9 ^{+24.9}_{-25.0} $ & (  81, 235) & 89.2  & 1.60& 1e-6 &96.0 \vspace{1mm} \\ 
 HD~174457 & 5 & 1.3 & 3.2 & 1.77 & $ 2.8^{+ 0.6}_{-0.6}$ & $ 107.8^{+ 23.8}_{-24.1}$ & $( 65,190)$ & 35.2 & 1.22& 12.3 &95.9 \vspace{1mm} \\ 
 HD~190228 & 5 & 1.0 & 3.5 & 0.13 & $ 1.8^{+ 0.5}_{-0.5}$ & $ 49.4^{+ 14.8}_{-14.8}$ & $( 10, 99)$ & 32.9 & 1.08& 0.3 &95.4 \vspace{1mm} \\ 
 HD~74014 & 5 & 0.1 & 2.5 & 9.84 & $ 64.0^{+ 32.3}_{-33.4}$ & $ 421.3 ^{+243.8}_{-252.1} $ & (  51,1568) & 235.4  & 1.34& 19.9 &80.6   \vspace{1mm} \\ 
HD~168443 & 5 & 0.6 & 1.9 & 1.25 & $ 2.4^{+ 0.6}_{-0.6}$ & $ 34.3^{+ 9.0}_{-9.1}$ & $( 18, 67)$ & 75.9 & 1.53& 3.6 &75.0 \vspace{1mm} \\ 
HD~191760 & 5 & 1.6 & 3.3 & 0.46 & $ 1.9^{+ 0.8}_{-0.8}$ & $ 185.3^{+ 94.1}_{-93.4}$ & $( 39,659)$ & 16.0 & 1.39& 27.6 &77.5  \vspace{1mm} \\ 
\hline
\end{tabular} 
\tablefoot{The top 6 orbits have high significance $\geqslant 99.7$ \% and are therefore fully characterised. The bottom 3 orbits are of low significance (1-$\sigma\!-2\!$-$\sigma$) and represent the formal solution. The remaining 5 orbits have moderate significance (2-$\sigma-\!3\!$-$\sigma$) and provide lower and upper mass-limits for the companions. \tablefoottext{a} {Final solution at lower eccentricity.} \tablefoottext{b} {Formal solution at higher eccentricity.} }
\end{table*}

\begin{table*} \caption{Spectroscopic elements of literature targets with orbits of better than 1-$\sigma$ significance.}
\label{tab:target2sigma} 
\centering  
\begin{tabular}{r r r r c r c c r r } 
\hline\hline %
Object & HIP    & $M_1$ & $M_2 \sin i$  & $P$ & $e$ & $K_1$ & $T_0$ & $\omega$ & Ref.\\  
       &        &($M_{\odot}$) & ($M_J$) & (day)& & (ms$^{-1}$)  & ($\mathrm{JD}^{\star}$) & (deg) & \\  
\hline 
 HD~131664 & 73408 & $1.1$ & $ 18.2  $  & $1951.0 \pm  41.0$ & $0.64 \pm 0.02$ & $359.5 \pm  22.3$ & $52060.0 \pm   41.0$ & $149.7 \pm 1.0$ & 1 \\ 
 HD~174457 & 92418 & $1.2$ & $ 65.8 $  & $840.8 \pm   0.1$ & $0.23 \pm 0.01$ & $1250.0 \pm  10.0$ & $52020.0 \pm    4.0$ & $139.0 \pm 1.0$ & 2 \\ 
 HD~190228 & 98714 & $0.8$ & $ 3.6 $  & $1146.0 \pm  16.0$ & $0.50 \pm 0.04$ & $ 91.0 \pm   5.0$ & $51236.0 \pm   25.0$ & $100.7 \pm 3.2$ & 3 \\ 
 HD~168443c & 89844 & $1.0$ & $ 18.1 $  & $1748.2 \pm   1.0$ & $0.212 \pm 0.001$ & $298.1 \pm   0.6$ & $53769.8 \pm    0.0$ & $64.7 \pm 0.5$ & 4 \\ 
 HD~191760 & 99661 & $1.2$ & $38.0$  & $505.6 \pm   0.4$ & $0.63 \pm 0.01$ & $1047.8 \pm  38.7$ & $54835.7 \pm    2.1$ & $200.4 \pm 0.3$ & 5 \\ 
\hline
\end{tabular} 
\tablebib{(1)~\citet{Minniti:2009zr}; (2) \citet{Nidever:2002vn}; (3) \citet{Perrier:2003dw}; (4) \citet{Wright:2009fe}; (5) \citet{Jenkins:2009kx}.} 
\end{table*}

\begin{table*} \caption{Parameters for targets with very-low significance orbits ($<$1-$\sigma$)}
\label{tab:insignificantOrbits} 
\centering  
\begin{tabular}{r r r r r r r r r r r r } 
\hline\hline  
Object & HIP & Sol. & $N_\mathrm{orb}$ & $\sigma_{\Lambda}$ &  $M_1$        & $M_2 \sin i$   & $a \sin i$ & Ref. & Null prob. & Significance & $M_{2,\mathrm{up-lim}}$  \\  
       &     & type &           & (mas)                & ($M_{\odot}$) & ($M_J$)         & (mas)     &  &  (\%)     & (\%)   & ($M_{\odot}$)     \\  
\hline 
 HD~52756 & 33736 & 5 & 22.5 & 4.2 &  $  0.8 $ & $59.0  $        & $ 0.53 $ & 1  & 26.1      & 54.0 & 0.72\\
 HD~89707 & 50671 & 5 & 3.5 & 2.7 &  $  1.0 $ & $53.0  $        & $ 1.54 $ & 1  & 19.5      & 52.0  & 0.11\\
  HD~167665 & 89620 & 5 & 0.2 & 1.8 &  $  1.1$ & $ 50.6  $        & $ 7.70 $ & 1  & 25.1      & 54.0  & $\cdots$ \\
 HD~189310 & 99634 & 5 & 82.0 & 4.2 &  $  0.8 $ & $ 26.0$        & $ 0.10 $ & 1  & 2.6      & 58.3  & 2.93\\
 GJ~595       & 76901 & 1 & 12.2 & 19.9 &  $  0.3 $ & $ 60.0 $        & $ 3.50 $ & 2  & 89.5      & 8.0  & 0.63\\
 HD~13189 & 10085 & 5 & 1.9 & 3.5 &  $  4.5 $ & $ 14.0 $        & $ 0.01 $ & 3  & 71.3      & 11.0  & 9.62\\
 HD~30339 & 22429 & 5 & 60.0 & 3.8 &  $  1.1 $ & $ 77.8 $        & $ 0.12 $ & 2  & 35.6      & 23.0 & 10.46\\
 HD~38529 & 27253 & 5 & 0.4 & 1.6 &  $  1.5 $ & $ 13.4  $        & $ 0.80 $ & 4  & 55.9      & 9.0 & $\cdots$ \\ 
 HD~65430 & 39064 & 5 & 0.3 & 3.1 &  $  0.8 $ & $ 67.8$        & $ 12.79 $ & 2  & 21.0      & 61.0  &$\cdots$\\ 
 HD~91669 & 51789 & 5 & 2.1 & 6.0 &  $  0.9 $ & $ 30.6  $        & $ 0.44 $ & 5  & 17.9      & 52.0  & 0.53\\
 HD~107383 & 60202 & 5 & 3.5 & 1.1 &  $  2.7 $ & $19.4 $        & $ 0.10 $ & 6  & 64.2      & 9.0 & 0.22\\
 HD~119445 & 66892 & 5 & 2.7 & 2.2 &  $  3.9 $ & $37.6  $        & $ 0.06 $ & 7  & 77.3      & 0.0 & 1.86\\
 HD~137510 & 75535 & 5 & 1.4 & 1.9 &  $  1.4 $ & $22.7  $        & $ 0.70 $ & 8  & 45.7      & 34.0 & 0.061\\
 HD~140913 & 77152 & 5 & 7.3 & 5.1 &  $  1.0 $ & $ 43.2 $        & $ 0.49 $ & 2 & 20.7      & 20.0 &0.56\\
 HD~162020 & 87330 & 5 & 92.7 & 5.0 &  $  0.8 $ & $ 14.4 $        & $ 0.05 $ & 9  & 20.8      & 50.0  & 7.32\\  
 HD~180777 & 94083 & 5 & 40.8 & 1.2 &  $  1.7 $ & $25.0  $        & $ 0.11 $ & 10  & 51.9      & 7.0 & 0.27\\
 HD~202206b & 104903 & 5 & 3.9 & 3.8 &  $  1.1 $ & $ 17.4 $        & $ 0.26 $ & 11  & 48.7      & 10.0 &  0.27\\ 
\hline
\end{tabular} 
\tablebib{(1) This work; (2) \cite{Nidever:2002vn}; (3) \cite{Hatzes:2005jw}; (4) \cite{Benedict:2010ph}; (5) \cite{Wittenmyer:2009mz}; (6) ~\cite{Liu:2008uo}; (7) \cite{Omiya:2009dn}; (8) \cite{Butler:2006pi}; (9) \cite{Udry2002}; (10) \cite{Galland:2006qq}; (11) \cite{Correia:2005bs}.} 
\end{table*}

\subsection{Orbits with high significance exceeding 3-$\sigma$}\label{sec:3sigma}
\begin{figure*}\begin{center} 
\includegraphics[width= 0.33\linewidth, trim= 0 -1cm 0 0]{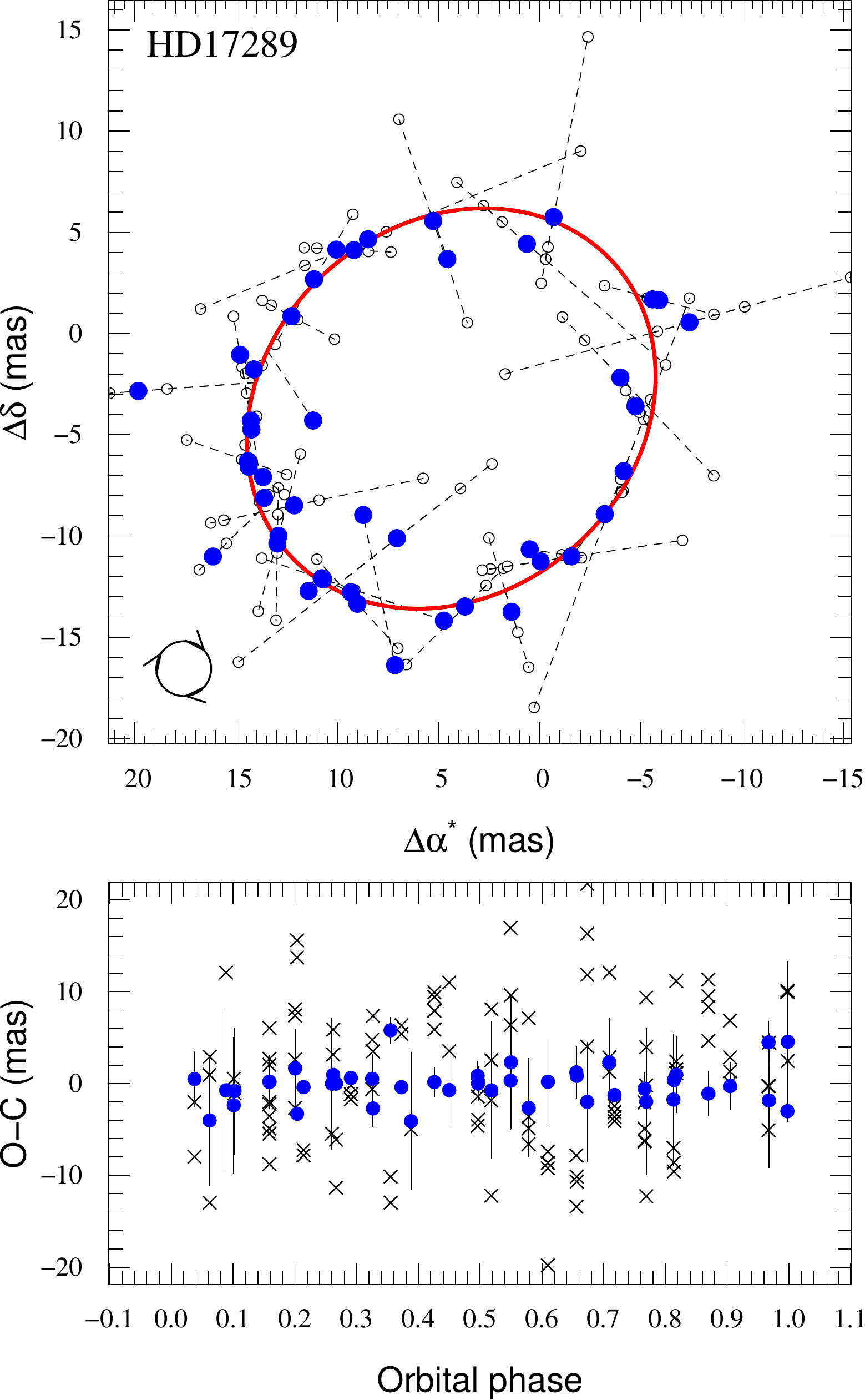}
\includegraphics[width= 0.33\linewidth, trim= 0 -1cm 0 0]{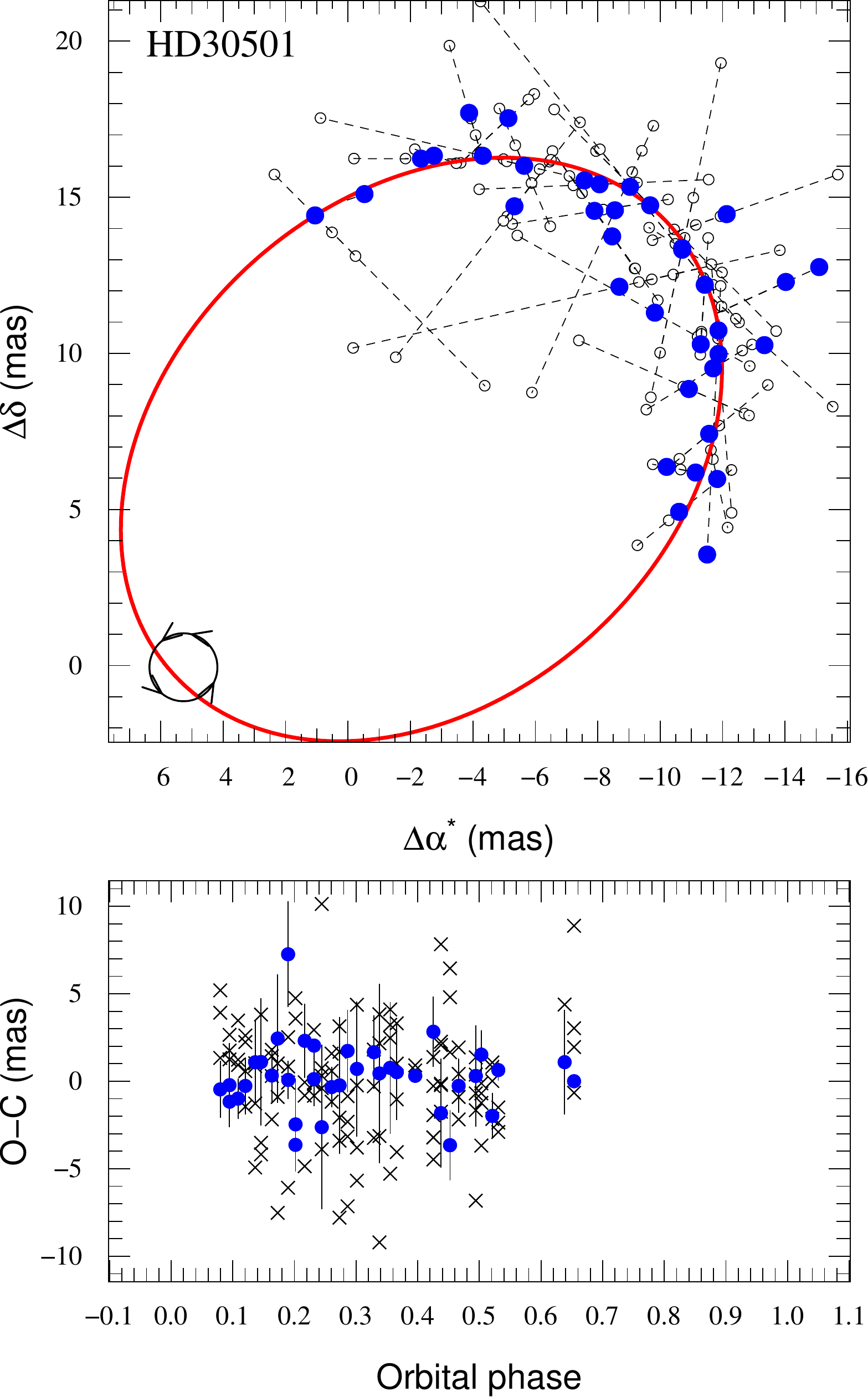} 
\includegraphics[width= 0.33\linewidth, trim= 0 -1cm 0 0]{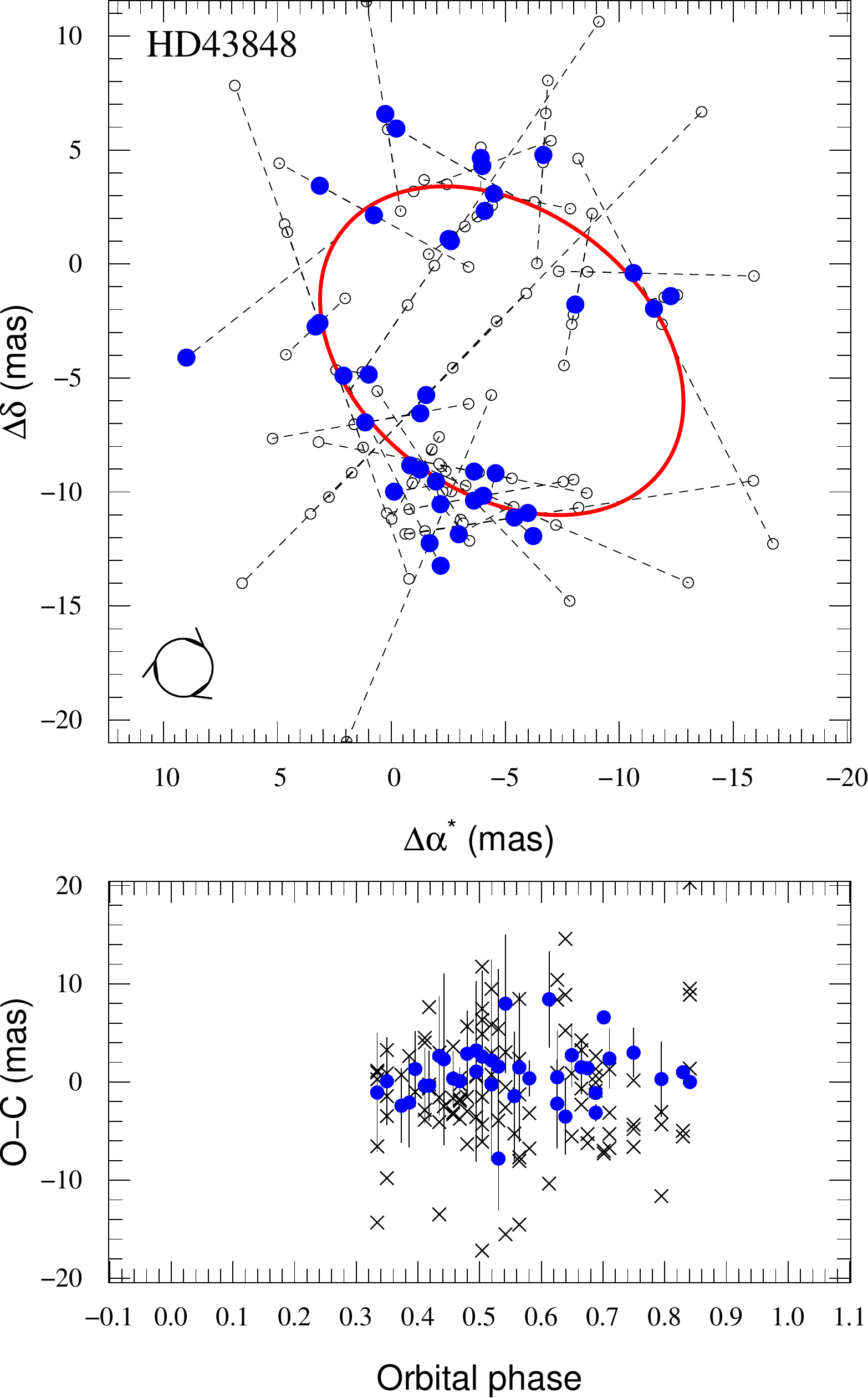} 
\includegraphics[width= 0.33\linewidth]{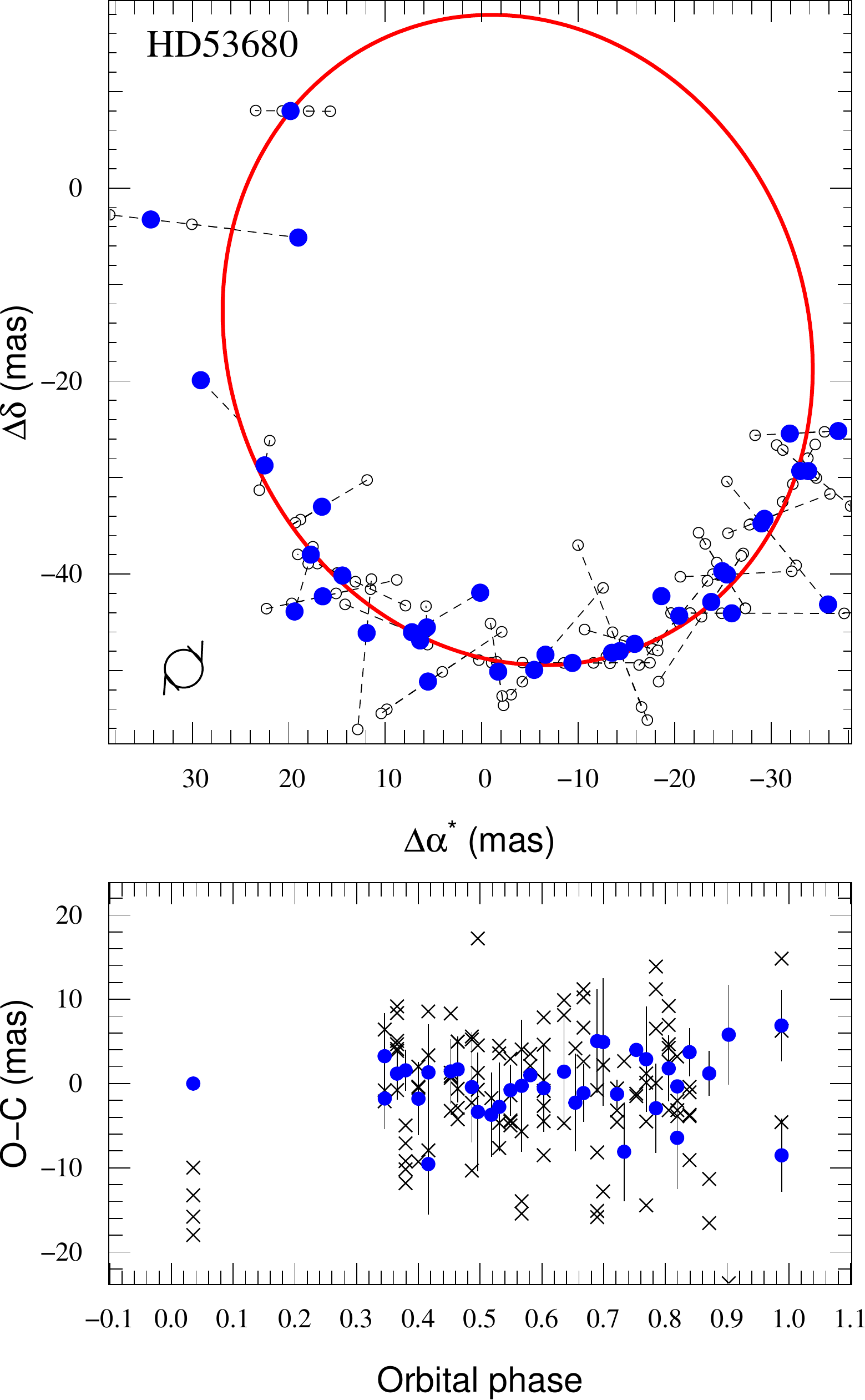} 
\includegraphics[width= 0.33\linewidth]{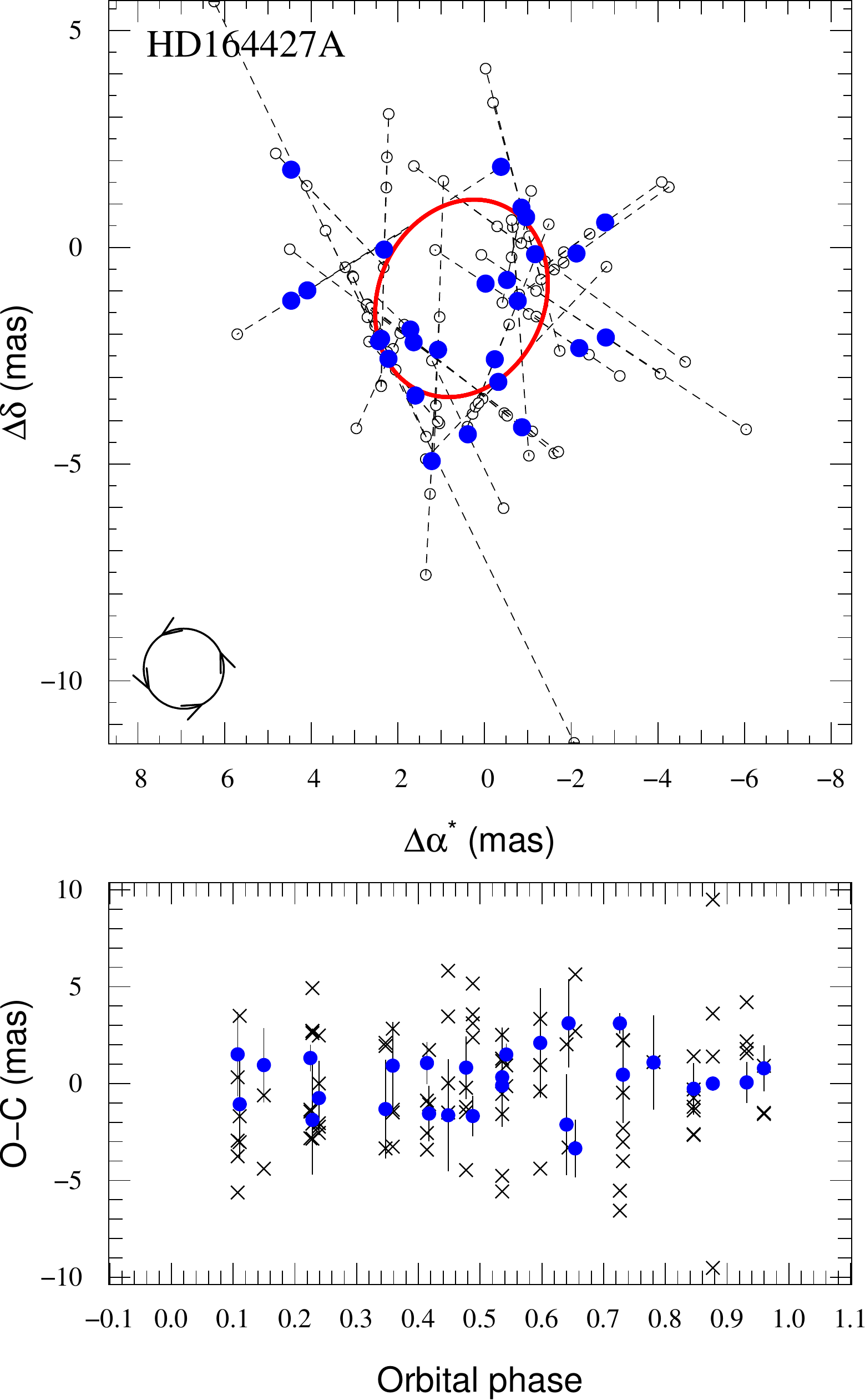} 
\includegraphics[width= 0.33\linewidth]{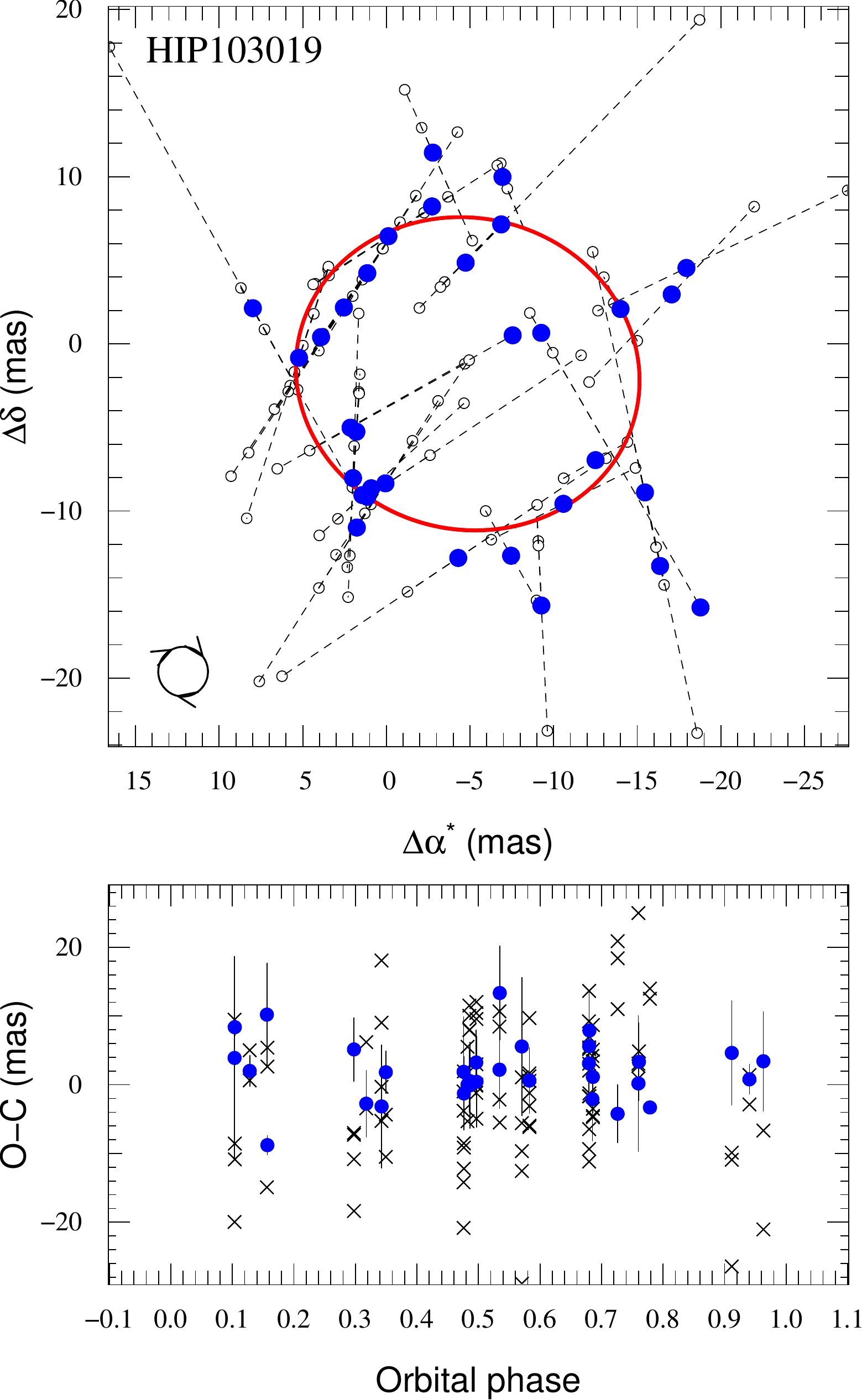} 
 \caption{Visualisation of the high-significance orbits. \emph{Top panels:} Astrometric stellar orbits projected on the sky. North is up and East is left. The solid red line shows the orbital solution and open circles mark the individual Hipparcos measurements. Dashed lines with orientation along the scan angle $\psi$ and length given by the O-C residual of the orbital solution connect the measurements with the predicted location from our model. The blue solid circles show the normal points for each satellite orbit number. The curl at the lower left corner indicates the orientation of orbital motion. \emph{Bottom panels:} O-C residuals for the normal points of the orbital solution (filled blue circles) and of the standard 5-parameter model without companion (black crosses). The error bars of the normal points correspond to the dispersion of Hipparcos measurements if there are several per satellite orbit and to the individual Hipparcos abscissa error if there is only one measurement.} 
\label{fig:orbits}
 \end{center} \end{figure*}

Clear orbital signatures are detected for six target stars. The orbit significances exceed $99.7 \%$, which we adopt as criterion for undoubtful detection. The orbits show a large variety in size ($a=2.3-35$ mas) and inclination ($i = 12-173 ^{\circ}$) and reveal companion masses ranging from $90 \,M_J$ to $0.52 \, M_{\odot}$. The obtained relative precision on the companion mass is typically of the order of 10 \% and does not include any contribution of the primary-mass uncertainty. The visualisation of Hipparcos astrometric data is not very intuitive, because positions are measured in one dimension only, i.e. along the scan-angle orientation $\psi$. We choose a representation similar to \cite{Torres:2007kx} and the resulting stellar astrometric orbits are shown in Fig.~\ref{fig:orbits}. For better display, we compute normal points for each satellite orbit number. Their values and errors are given by the mean and standard-deviation of the Hipparcos abscissae obtained during one given satellite orbit, respectively. Normal points are used only for the visualisation and not during the orbit adjustment. We prove that orbit detections are also possible in apparently difficult conditions, such as incomplete orbit coverage by the satellite ($N_{orb}=0.5$ for HD~43848) and an instrument precision comparable to the the orbit size (\object{HD~164427A}). The final 7-parameter fit for the detected orbits has an acceptable reduced chi-square value ($\chi^2_{7,\mathrm{red}} = 0.8-1.2)$ in all but one case. For HD~17289 this value is very small at 0.3 and the Hipparcos precision is unusually large at 7.2~mas, which may indicate that the astrometric errors  for this star are overestimated. As for the radial-velocity measurements, the light emitted by the actual stellar companion of this star may have disturbed the Hipparcos observations. 
  
Three out of these six stars have non-standard Hipparcos solutions. Thus, first indications for the astrometric disturbance induced by the companion were already detected during the standard Hipparcos data analysis. An additional outcome of the companion solution are refined positions, distances, and proper motions of the objects (Table \ref{tab:sol2sigma}). For 5 stars the parallax estimate becomes more precise, whereas its value is compatible within the error bars. Only the parallax of HIP~103019 is 1.9 mas larger than the value given by the new reduction. Changes in position and proper motion are of the order of 1~mas and 1~mas~yr$^{-1}$, with extrema of 41~mas and 22~mas~yr$^{-1}$, respectively.  

\begin{itemize}
  \item \object{HD~17289}: The orbit inclination is $173.2 \pm 0.5^{\circ}$ and the companion mass is $0.52\pm0.05\,M_{\odot}$, which makes it the most massive companion detected in this study. \object{HD17289} is thus identified as a binary star with mass ratio $q=2$. \cite{Goldin:2007ly} used solely Hipparcos astrometry to find an astrometric orbit with semimajor axis $a=11.7^{+3.8}_{-1.6}$, which is compatible with our more precise value of $a=10.8\pm0.7$. Because the orbit is completely characterised, we can calculate the projected angular separation between the two components. Along the orbit, their separation on the sky ranges between $15-49$~mas. The mass-luminosity relations of \cite{Delfosse:2000xq} estimates an apparent $V$-magnitude of the secondary of $V_2 = 13.18$. The primary magnitude obtained from Hipparcos is $V = 7.56$ (Table \ref{tab:stellarCharHIP}) and the expected intensity contrast between the secondary and primary is $\sim$1:180 in the visible. The combination of close separation and moderate intensity contrast causes a small and periodic bias in the radial-velocity measurement, see Sect.~\ref{sec:objnotes}.
  \item \object{HD~30501}: The orbit inclination is $49^{+10}_{-8} $ deg and the companion mass is $90 \pm 12 \,M_J$, which is the lightest companion detected in this study. This companion is just at the threshold between a very low-mass star and a brown dwarf. 
  \item \object{HD~43848}: The orbit inclination is $165^{+2}_{-3} $ deg and the companion is a very low mass star with mass $0.10 \pm 0.01\, M_{\odot}$. This value is slightly lower and more precise than the one derived by \cite{Sozzetti:2010nx}, who obtain $0.11^{+0.16}_{-0.04} \, M_{\odot}$ based on the radial velocities of \citealt{Minniti:2009zr}. This solution is found with the lower-eccentricity orbit (cf. Sect.~\ref{sec:rvsol}). For completeness, we include the formal astrometric solution for the less probable and higher-eccentricity radial-velocity orbit in Tables~\ref{tab:sol2sigma} and \ref{tab:mass2sigma}. The resulting companion mass obtained in this case would be higher at $M_2 = 0.11 \pm 0.02\, M_{\odot}$. Hipparcos astrometry cannot be used to determine the actual eccentricity, because the fit quality in terms of $\chi^2$ is comparable in both cases. 
  \item \object{HD~53680}: The orbit inclination is $163.6^{+1.4}_{-1.7} $ deg and the companion is a low mass star with mass $0.22 \pm 0.02\, M_{\odot}$.
  \item \object{HD~164427A}: The orbit inclination is $11.8^{+3.1}_{-2.1} $ deg and the companion is a low mass star with mass $0.26 \pm 0.06\, M_{\odot}$. The stellar companion has also been detected by \cite{Zucker:2001ve}, who derived a slightly less inclined orbit and therefore obtained the larger mass of $0.35 \pm 0.09\, M_{\odot}$.
  \item \object{HIP~103019}: The orbit inclination is $160.9^{+2.2}_{-2.7} $ deg and the companion is a very low mass star with mass $0.18 \pm 0.03\, M_{\odot}$. 
\end{itemize}

\subsection{Orbits with moderate significance between 2-$\sigma$ and 3-$\sigma$}\label{sec:2sresults}
\begin{figure*}\begin{center} 
\includegraphics[width= 0.33\linewidth, trim= 0 -1cm 0 0]{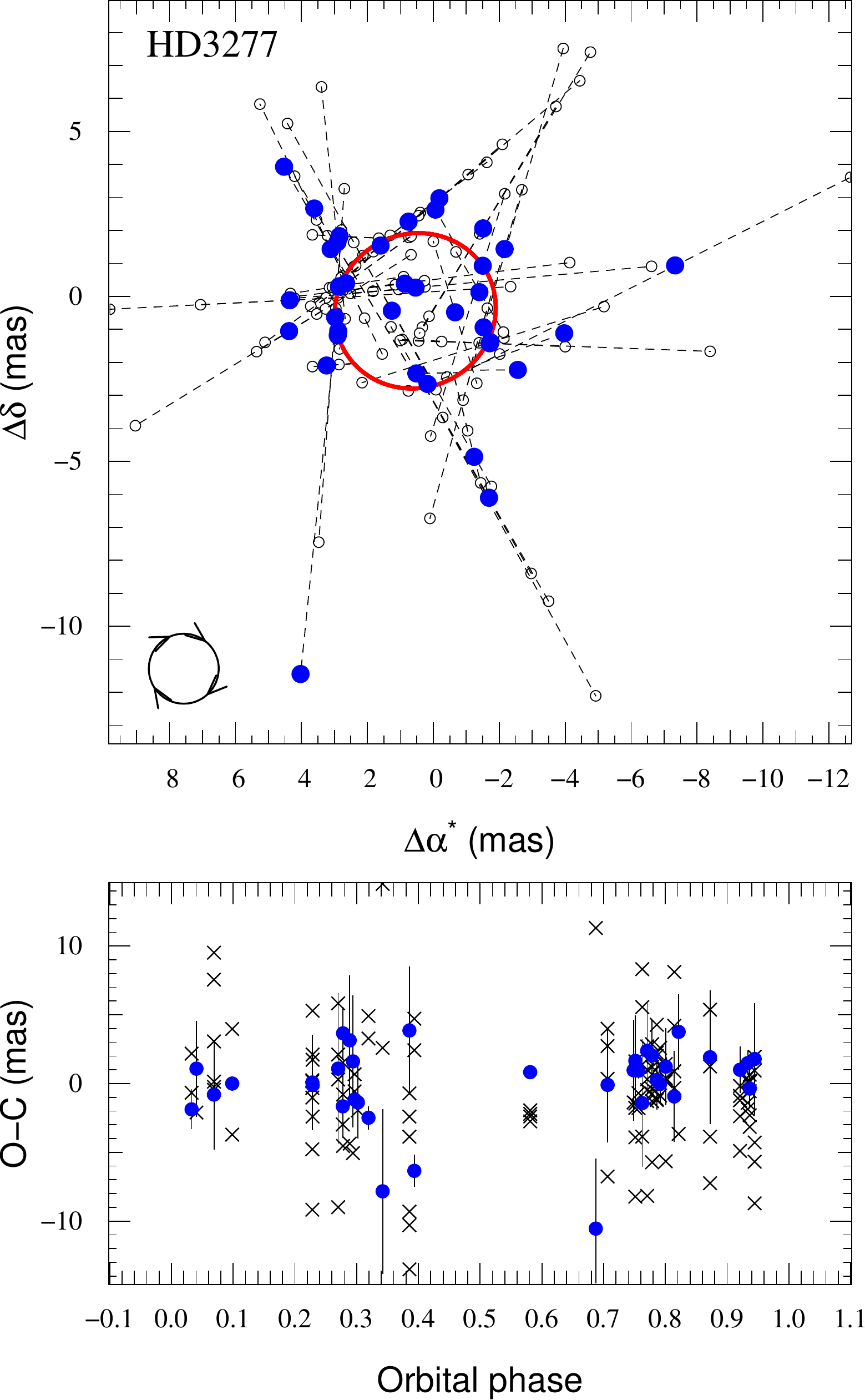} 
\includegraphics[width= 0.33\linewidth, trim= 0 -1cm 0 0]{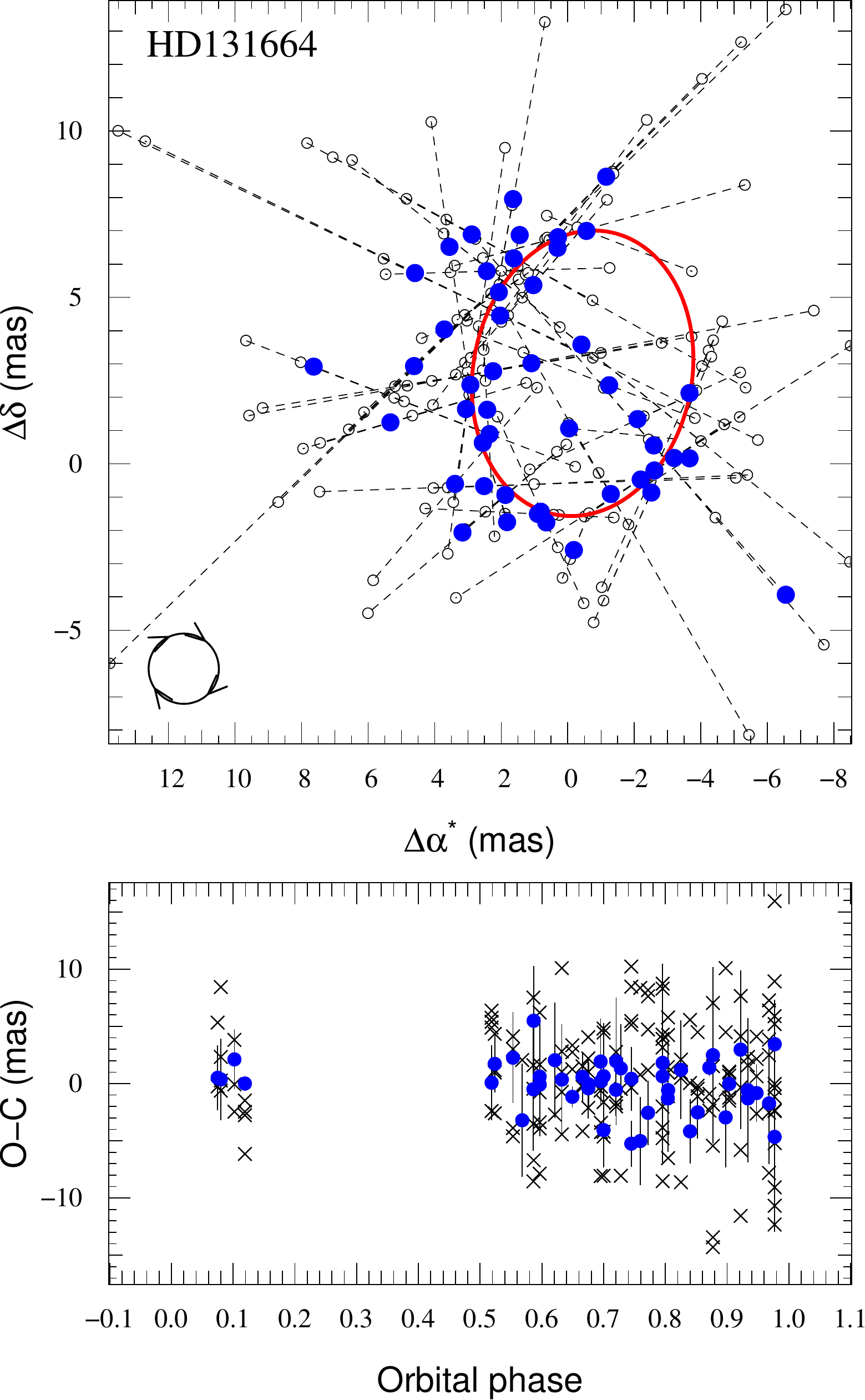} 
\includegraphics[width= 0.33\linewidth, trim= 0 -1cm 0 0]{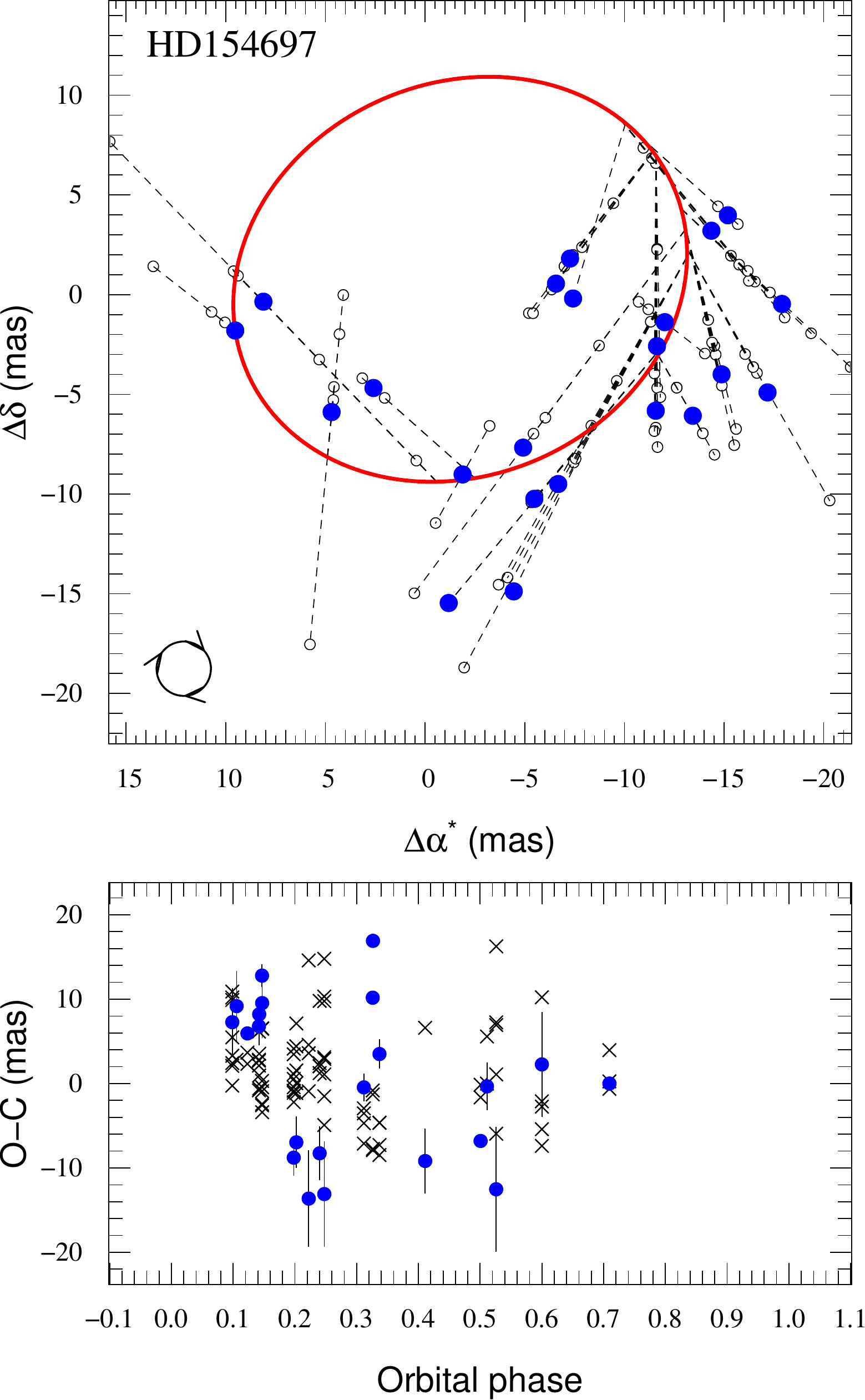} 
\includegraphics[width= 0.33\linewidth]{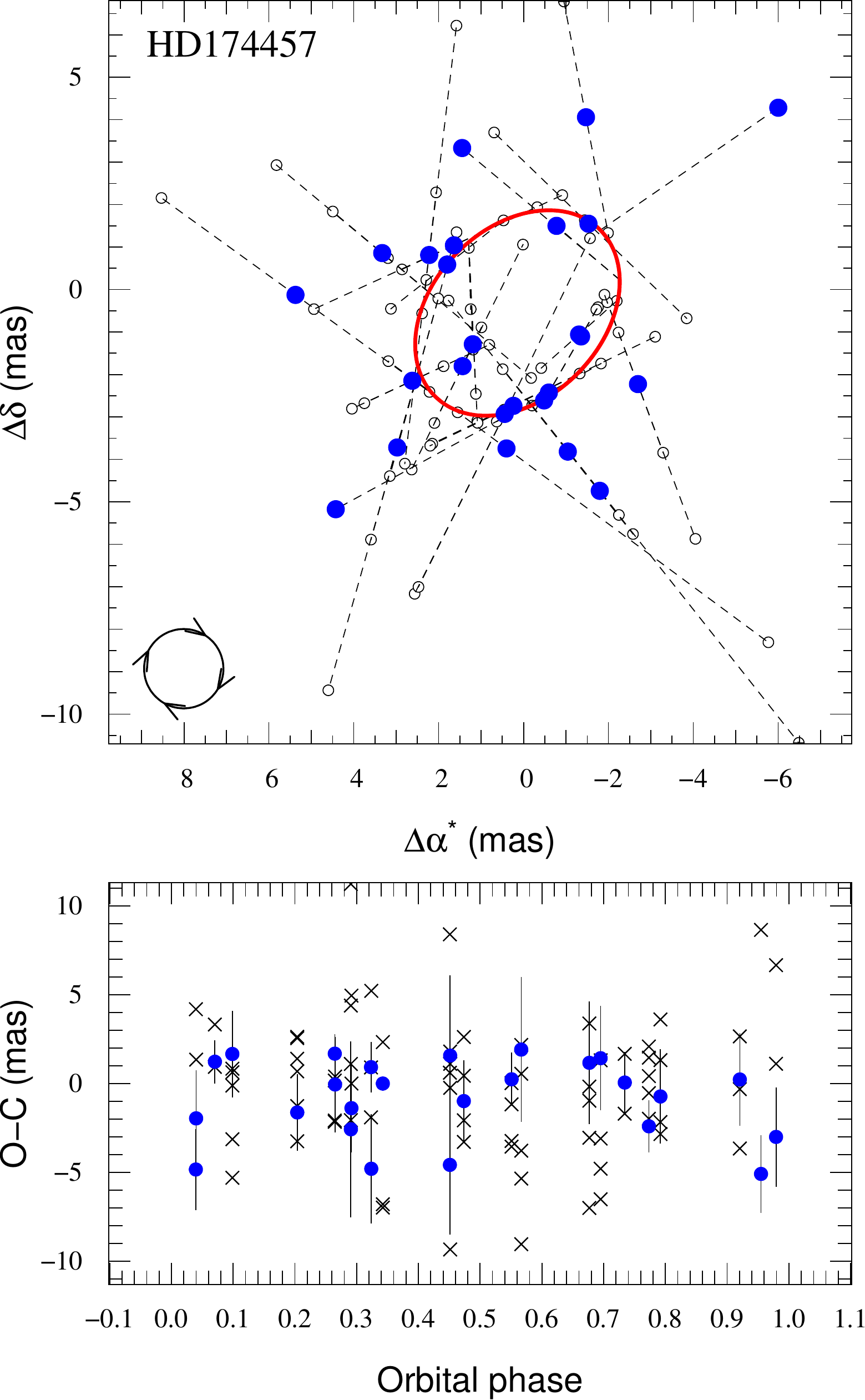} 
\includegraphics[width= 0.33\linewidth]{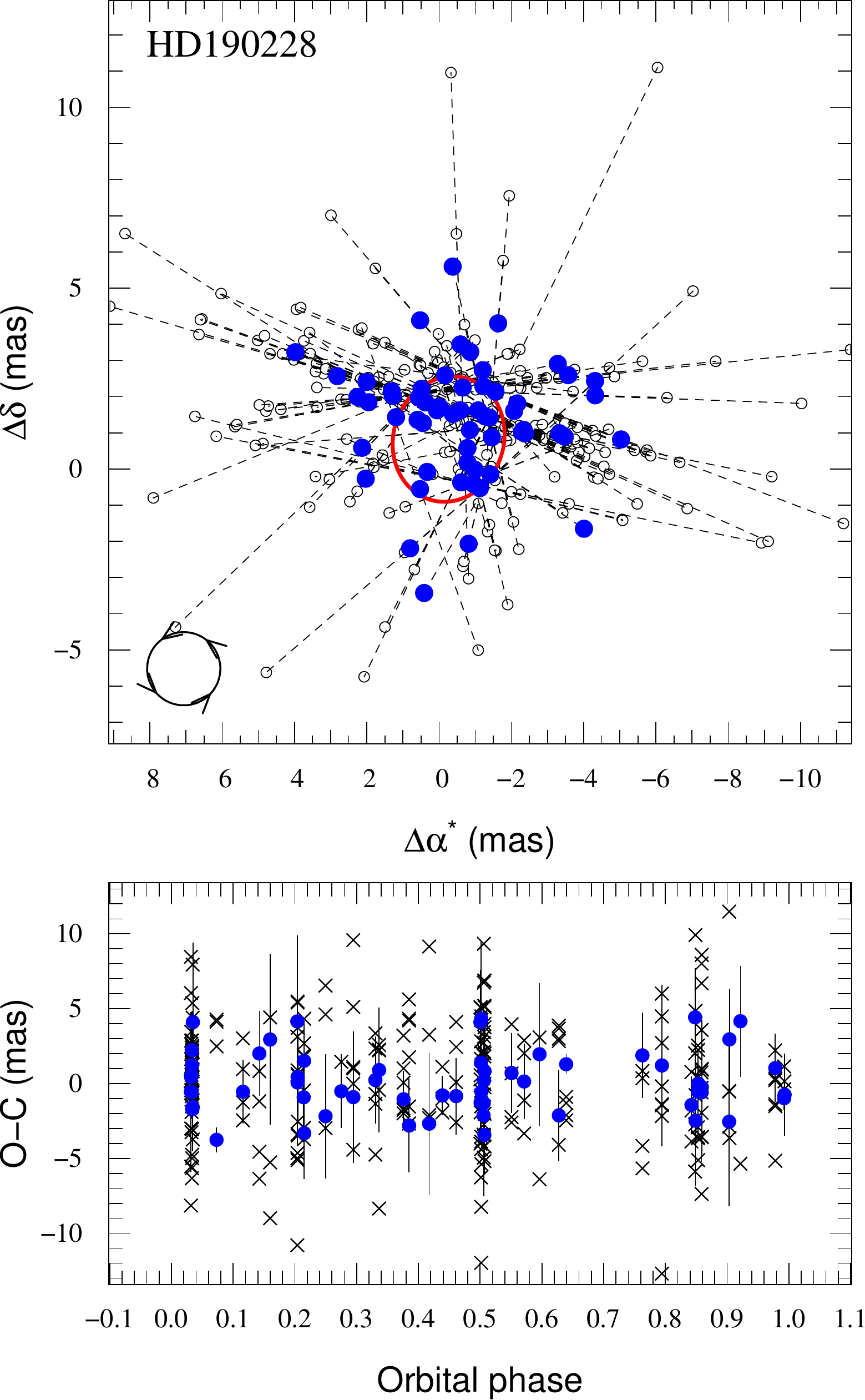} 
 \caption{Visualisation of the formal solutions for the moderate-significance orbits. The panels are analogous to Fig.~\ref{fig:orbits}. The formal orbit of HD~154697 is not valid and exhibits large residuals (see the text for explanation). It is shown here to illustrate the possible problems that can occur when visualising astrometric orbits. Despite their moderate significance, the four remaining orbits visually appear well constrained.}
\label{fig:orbits2}
 \end{center} \end{figure*}

The orbits of 5 stars are found with significance of $95.4-99.7$~\%. Although we do not consider that these solutions accurately characterise the stellar orbits, the moderate level of significance indicates that the orbital motion is present in the astrometric data. The formal orbits are shown in Fig.~\ref{fig:orbits2} and allow us to obtain a visual impression of the fit quality. Four out of five orbits have sizes comparable to or smaller than the instrument precision and the astrometric signal is therefore hidden. The permutation test allows us to assess the credibility of the derived orbit especially in this difficult regime. Since there is signal in the astrometric data, the orbit inclination is constrained and we expect the companion mass to be confined within the 3-$\sigma$ limits given in Table \ref{tab:mass2sigma}.

\begin{figure}
\begin{center} 
\includegraphics[width= 0.7\linewidth]{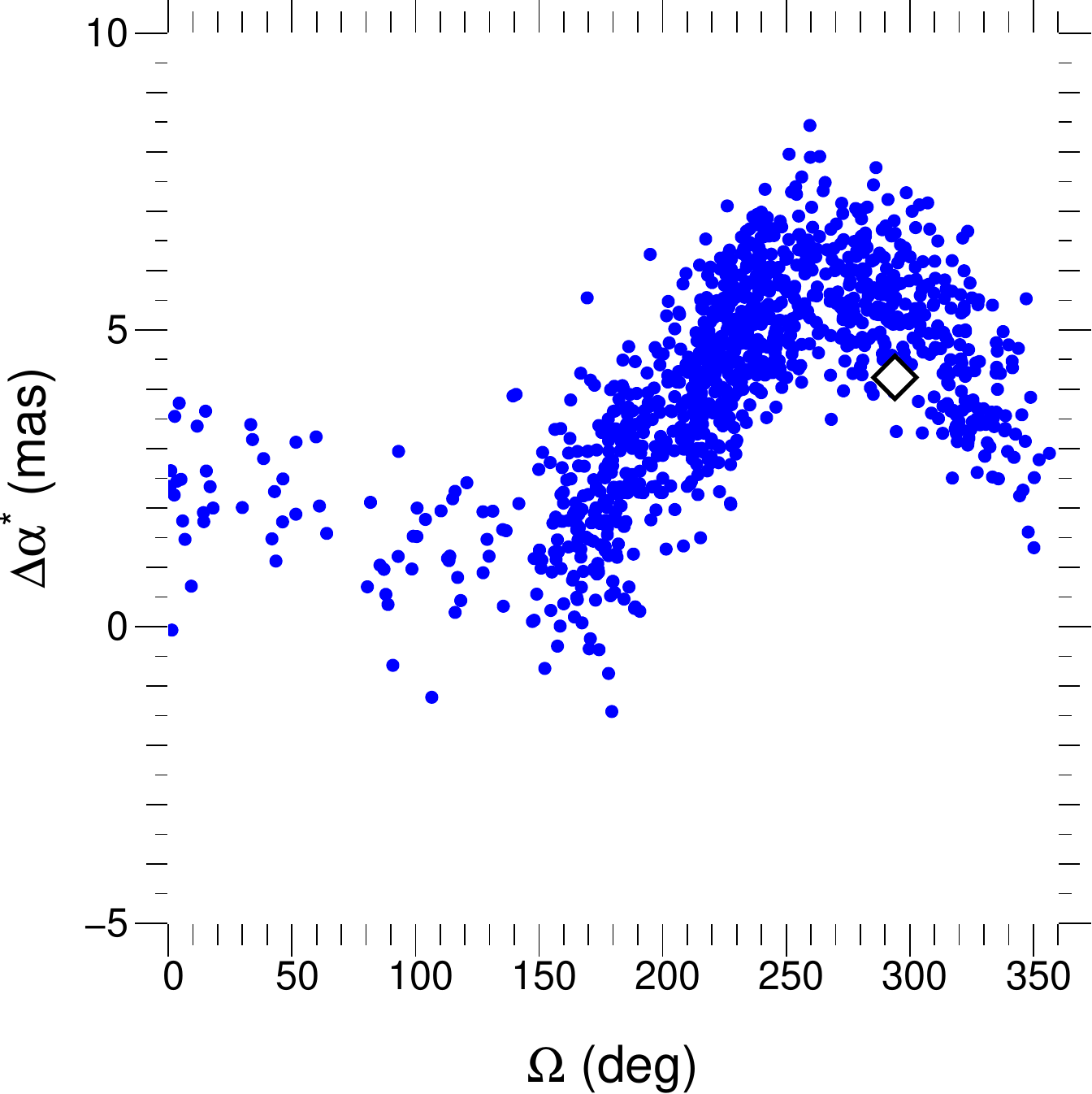} 
\caption{Correlation between the adjusted position and longitude $\Omega$ present in the distribution of the combined 100\,000 Monte-Carlo realisations for HD~154697. The adopted solution is marked with the diamond.} 
\label{fig:OMEa_corr}
\end{center} \end{figure}

\begin{itemize}
  \item \object{HD~3277}: The companion is of stellar nature with its mass confined within $0.14-0.57\,M_{\odot}$. The formal companion mass at moderate significance is $0.33 \pm 0.07\, M_{\odot}$. The upper mass-limit which we independently derived from $a = \sigma_\Lambda$ is $M_{2,\mathrm{up-lim}} = 0.51\,M_{\odot}$ and compatible with our solution. 
  \item \object{HD~131664}: The companion is a very low-mass star or a brown dwarf with a mass of $44-153 \, M_J$. The available data is not sufficient to distinguish between these two cases. The formal companion mass at moderate significance is $90 \pm 16 \,M_J$. Although the orbit significance agrees with the result of \cite{Sozzetti:2010nx}, who obtain a companion mass of $23^{+26}_{-5} \, M_J$ with 95 \% confidence, the mass range derived from our analysis is significantly higher.  
  \item \object{HD~154697}: The companion is a low-mass star with a mass confined within $0.08-0.22\,M_{\odot}$. The formal companion mass at moderate significance is $0.15 \pm 0.02\, M_{\odot}$. The longitude of the ascending node $\Omega$ of this orbit is hardly constrained and is correlated with the position offsets $\Delta \alpha^\star$ and $\Delta \delta$. The $\Omega - \Delta \alpha^\star$-correlation is shown in Fig.\ref{fig:OMEa_corr}. Because the adopted solution is derived from the mean of the complete distribution, it does not represent a valid solution with a valid orbit, as is also visible in Fig.~\ref{fig:orbits2}, where the fit residuals are large. However, this does not affect the validity of the derived companion mass, because that is independent of $\Omega$.  
  \item \object{HD~174457}: The companion is a very low-mass star or a brown dwarf. Its mass is confined within $0.06-0.18\,M_{\odot}$. The formal companion mass at moderate significance is $0.10 \pm 0.02\, M_{\odot}$. The upper mass-limit which we independently derived from $a = \sigma_\Lambda$ is $M_{2,\mathrm{up-lim}} = 0.12\,M_{\odot}$ and compatible with our solution. 
  \item \object{HD~190228}: The companion is probably a brown dwarf, although we cannot exclude a very low-mass star. Assuming a primary mass of $0.83 \,M_{\odot}$ \citep{Perrier:2003dw}, we find the formal companion mass of $49 \pm 15\, M_J$ with lower and upper limits of $10-99\,M_J$, cf. Table \ref{tab:compResults}. The upper mass-limit derived from $a = \sigma_\Lambda$ independently reproduces this result and yields $M_{2,\mathrm{up-lim}} = 99\,M_J$.
    \end{itemize}

\subsection{Orbits with low significance between 1-$\sigma$ and 2-$\sigma$}
The orbits of three stars are found at low significance of $68.3-95.4$~\%, i.e. the astrometric data contains very little or no orbital signal. As we have shown by simulation (cf. Sect. \ref{sec:simulation}), orbital solutions at this significance level are prone to large biases. Therefore, the solution parameters shall not be considered valid in these cases. The formal solution of the fitting procedure is given for completeness in Tables \ref{tab:sol2sigma} and \ref{tab:mass2sigma}.  

\begin{itemize}
  \item {HD~74014}: The poor Hipparcos coverage of 10~\% of this long-period orbit impedes the detection of the astrometric orbit. With a minimum semimajor axis of $a \sin i = 9.8$ mas, the orbit would have been detected on a longer timespan, given the instrument precision of 2.5 mas. Because the orbit is not fully covered, we cannot set an upper companion-mass limit $M_{2,\mathrm{up-lim}}$.
  \item {HD~168443}: The formal solution for the companion mass of $34\pm9\,M_J$ is in agreement with the result of \cite{Reffert:2006ly}, who derive $34\pm12\,M_J$, using the radial velocities of \cite{Marcy:2001rt} and the original Hipparcos data. From our analysis, using the {\footnotesize CORALIE} radial velocities and the new Hipparcos reduction IAD, we conclude that the mass obtained for the outer companion {HD~168443c} is of low confidence. Because the orbit is not fully covered, we cannot set an upper companion-mass limit $M_{2,\mathrm{up-lim}}$. 
  \item \object{HD~191760}: The minimum orbit size of this star is very small ($a \sin i = 0.46$~mas) and Hipparcos astrometry does not reveal the signature of the companion discovered by \cite{Jenkins:2009kx}. However, we can set an upper mass-limit for the companion of $M_{2,\mathrm{up-lim}} = 0.28\,M_{\odot}$.
\end{itemize}

\subsection{Orbits with very-low significance and the detection of one brown-dwarf companion}
The analysis of 17 objects yields orbits with significance below 1-$\sigma$. Because the astrometric data does not contain orbital information, the derived result is physically meaningless and only a mathematical solution. Therefore we do not list the solution parameters. Table \ref{tab:insignificantOrbits} shows the targets and their basic parameters. Two characteristics can be observed that impede the astrometric orbit detection. In the one case, applicable to HD~65430 and HD~167665, the orbital coverage is poor and the orbit is not detected although the expected signal is large compared to the instrument precision. In the other case, for instance for GJ~595, HD~13189, and HD~140913, the minimum orbital signature given by $a \sin i$ is very small compared to the instrument precision. 

The companions of six stars have maximum mass $M_{2,\mathrm{up-lim}}$ above $0.63\,M_{\odot}$, which does not represent a considerable constraint of the system. Nine companions have upper mass-limits of $M_{2,\mathrm{up-lim}} = 0.11 - 0.63\,M_{\odot}$ and thus in the M-dwarf mass range. Finally, the mass of the companion of HD~137510 is confined between $M_2 \sin i = 22.7\,M_J$ and  $M_{2,\mathrm{up-lim}} = 64.4\,M_J$. Therefore, we find that \object{HD~137510b} has to be a brown dwarf. This detection became possible, because of the confidence we gained on the capability of Hipparcos to detect astrometric orbital motion. Using a simpler argument, \citet{Endl:2004uq} derived an upper mass-limit of $94\,M_J$, which was not stringent enough to prove the substellar nature of the companion.

This sample also contains \object{HD~38529}, for which \cite{Benedict:2010ph} found the orbit inclination of the outer companion \object{HD~38529c} and derived its mass of $M_2 = 17.6\,M_J$ using {\footnotesize HST FGS} astrometry. Our analysis made use of the radial velocity orbit given by \cite{Benedict:2010ph}, but fails to detect a significant orbit, although the Hipparcos precision of 1.6 mas is in the range of the {\footnotesize HST} orbit size ($a=1.05 \pm 0.06$ mas). However, the Hipparcos orbit coverage is poor at 0.4, which can explain the non-detection. 

\section{Discussion}\label{sec:discussion}
The recent compilation of close companions to Sun-like stars with minimum masses of $10-80\,M_J$ by \cite{Sozzetti:2010nx} lists 39 objects. With the discovery of 9 new companions, we increase the known number of such objects by about 20~\%. Ten companions in the \cite{Sozzetti:2010nx} list are probably stars. We confirm the stellar nature in two cases (\object{HD~43848B} and \object{HD~164427B}) and detect 6 new stellar companions (\object{HD~3277B}, \object{HD~17289B}, \object{HD~30501B}, \object{HD~53680B}, \object{HD~154697B}, and \object{HIP~103019B}) at the 3-$\sigma$ significance level. All of these companions are M dwarfs with masses between $90\,M_J$ and $0.52\,M_\odot$. For 3 companions, we derive mass limits that enclose the boundary between low-mass stars and brown dwarfs: HD~131664 ($M_2 = 44-153\,M_J$), HD~174457 ($M_2 = 65-190\,M_J$), and HD~190228 ($M_2 = 10-99\,M_J$). HD~190228 initially was a planet-host star discovered by \cite{Sivan:2004rm}. We analysed it to validate our method and consequently recognised its importance as a promising brown dwarf candidate. 

Furthermore, we set upper mass limits below $0.63\,M_{\odot}$ to the companions of nine stars. These companions are therefore M-dwarfs or brown dwarfs.  The upper mass-limits of 12 companions within our sample remain unconstrained, either because the derived limit is very high ($>\!0.63\,M_{\odot}$) or because the orbital period is not covered by Hipparcos measurements. Finally, we found that the companion of HD~137510 is a brown-dwarf with a mass range of $22.7-64.4\,M_J$.

\begin{figure}\begin{center} 
\includegraphics[width= \linewidth, trim = 0cm 2cm 0cm 2cm, clip=true]{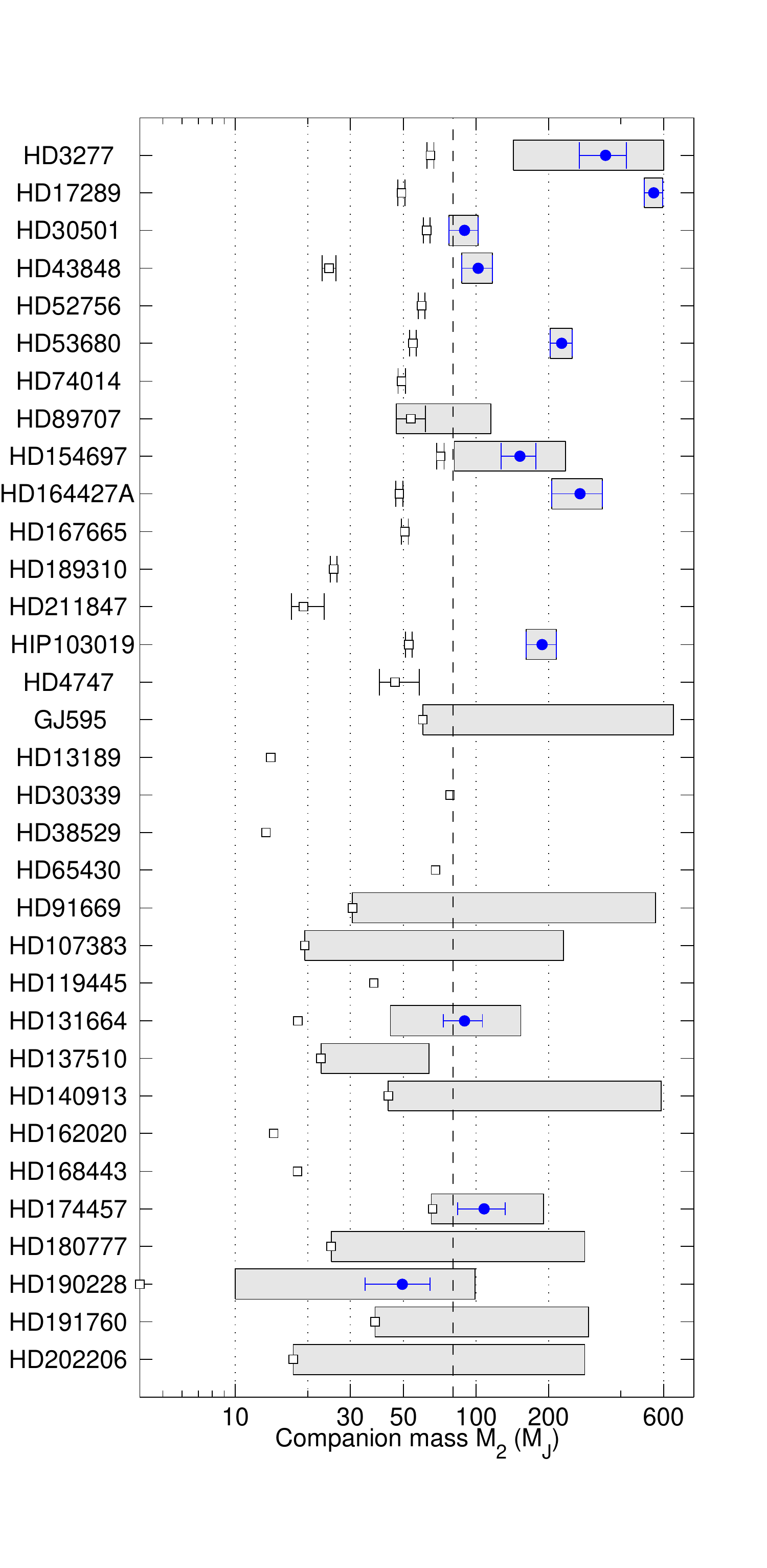}  
\caption{Constraints on the companion masses of the target sample. Open squares show the minimum companion mass $M_2 \sin i$ derived from radial velocities. Blue filled circles indicate the formal companion masses $M_2$ derived from astrometry for orbits with $>\!2\sigma$-significance. Grey-shaded areas show the companion mass interval constrained by astrometry. The vertical dashed line indicates the adopted threshold mass of $80\,M_J$ between brown dwarfs and stars. Twelve systems could not be constrained by Hipparcos astrometry, either because the derived mass limit was very high ($>\!0.63\,M_{\odot}$) or because the orbital period was not covered by Hipparcos measurements.}
\label{fig:m2all} 
  \end{center}\end{figure}

Our results are summarised in Fig.~\ref{fig:m2all}, where we show the companion-mass constraints obtained from Hipparcos astrometry for all targets in our analysis sample.

\subsection{Combining radial velocities and Hipparcos astrometry}
We used the astrometric data of the new Hipparcos reduction to set constraints on the masses of potential brown-dwarf companions, detected in radial-velocity surveys. Radial-velocity measurements of a stellar orbit yield 5 of the 7 Keplerian elements and do not constrain the orbit's inclination and longitude of the ascending node. We demonstrated how these two remaining elements can be determined, if the orbital signature is present in the Hipparcos astrometry, and how to distinguish a significant orbit. 

We used the permutation test to define the criterion for orbit detection. The implementation of this powerful test is considerably simplified by the use of the new Hipparcos reduction, because the individual measurements of the IAD are independent. A source of possible errors and complications that was present in the original Hipparcos data is therefore avoided. We emphasize the importance to eliminate outliers of the IAD, because otherwise the final result can be falsified. On a couple of comparison targets drawn from the literature, we have verified that our analysis method is robust. On these targets, the performance of the new and the original Hipparcos IAD is comparable and no major difference of the results was observed.

\begin{figure}\begin{center} 
\includegraphics[width= \linewidth]{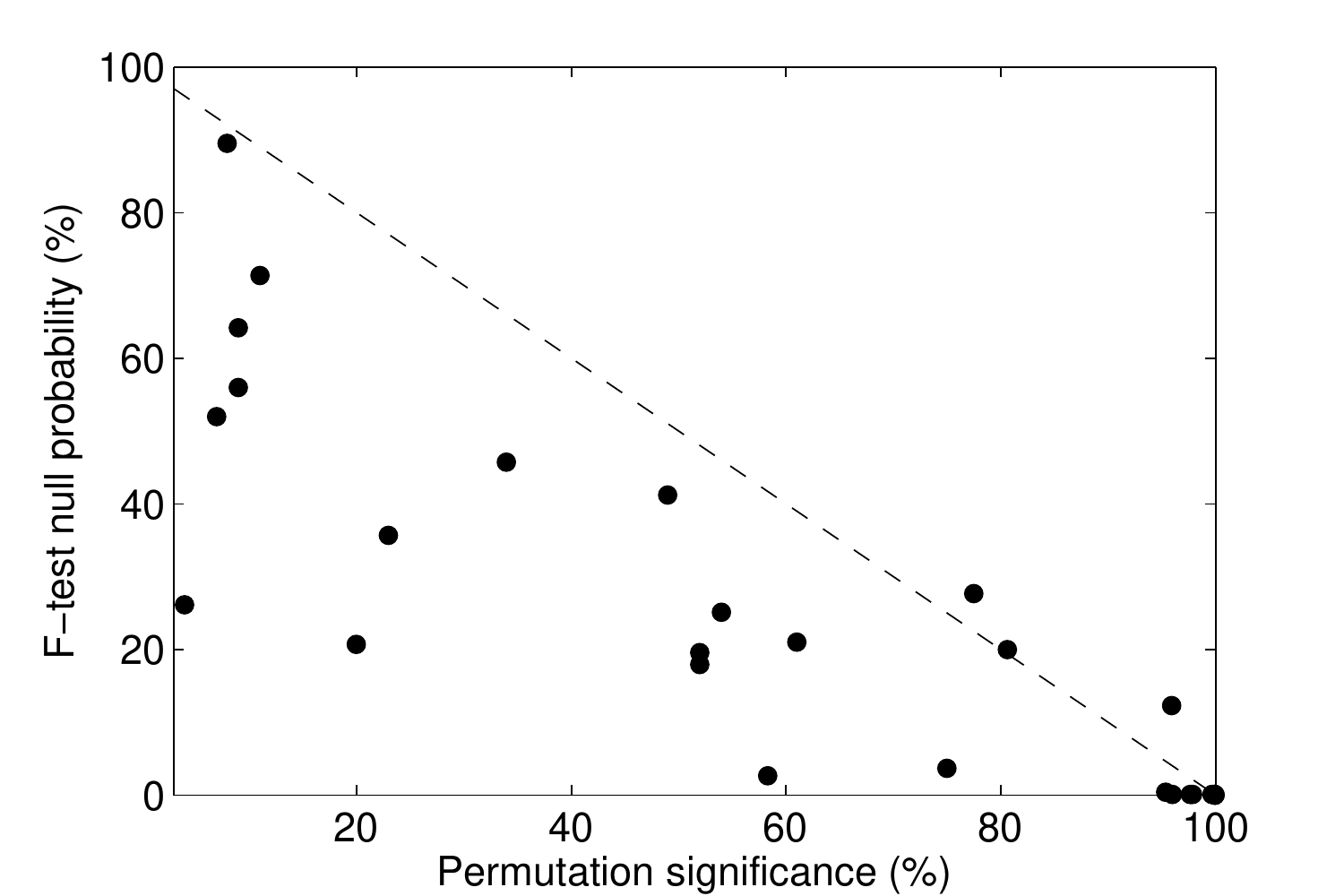} 
\caption{Comparison of the orbit significance indicators derived from the permutation test and from the F-test.}
\label{fig:significance} 
  \end{center}\end{figure}

In addition to the permutation test, we simultaneously performed the F-test and can thus compare the performance of both significance indicators, keeping in mind that the F-test is valid under the assumption of Gaussian errors only. Figure \ref{fig:significance} displays the respective values listed in Tables \ref{tab:mass2sigma} and \ref{tab:insignificantOrbits}. We find a general agreement between the two indicators, i.e. the null probability decreases with increasing permutation significance. However, the permutation test is more stringent. If we had chosen to rely on the F-test and imposed a null probability of 0.3 \% as detection limit, we would have detected 10 orbits. These overlap with the 6 orbits, which we detected with the criterion based on the permutation test (cf. Table~\ref{tab:mass2sigma}).

\begin{figure}\begin{center} 
\includegraphics[width= \linewidth, trim = 0cm 0.2cm 0cm 0.5cm, clip=true]{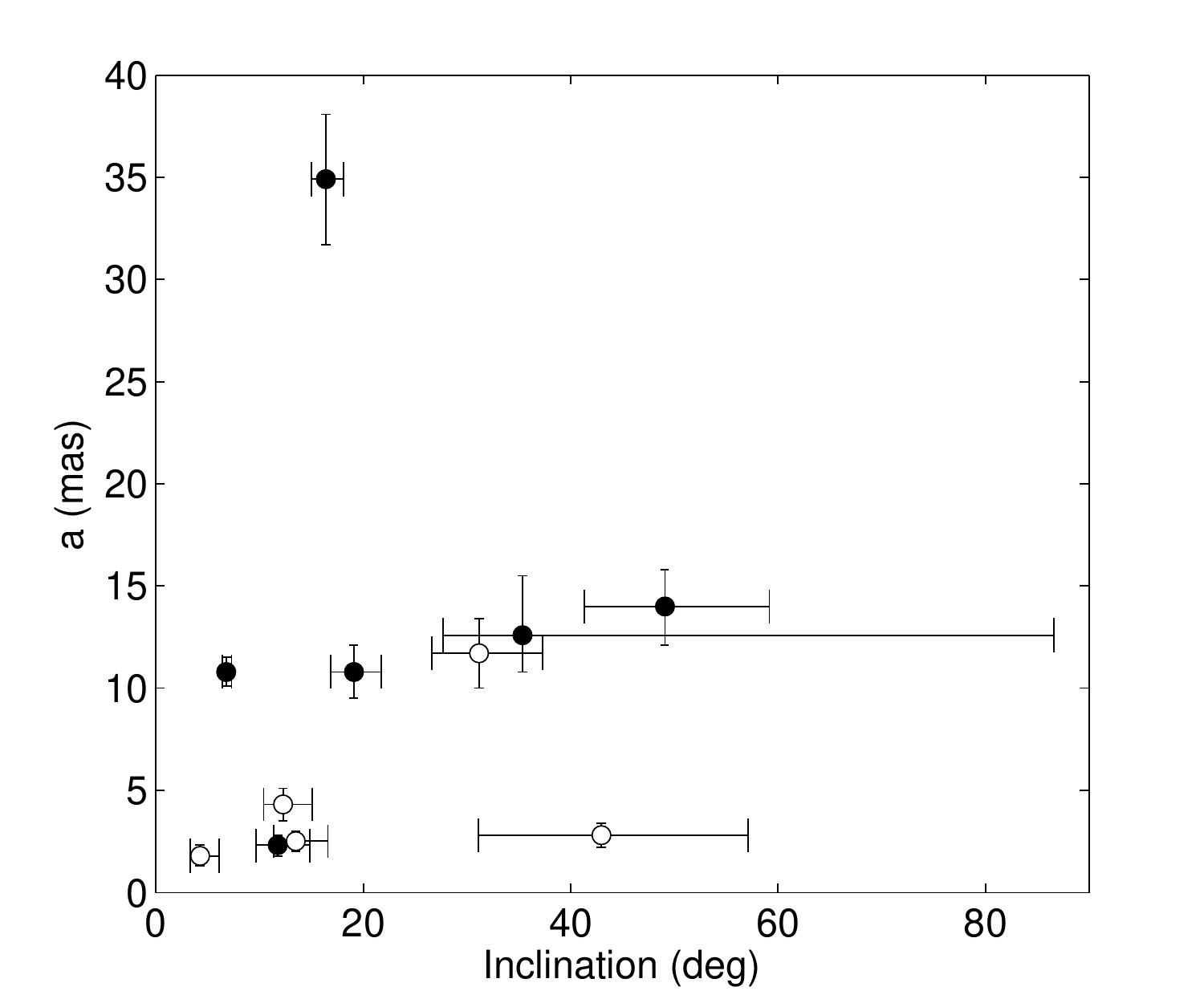}  
\caption{Inclinations and astrometric semimajor axes of the orbital solutions presented in Tables \ref{tab:sol2sigma} and \ref{tab:mass2sigma}. The inclination is displayed as angular deviation from a face-on orbit (corresponding to $i=0^{\circ}$ or $i=180^{\circ}$). Filled symbols indicate the orbits with high-significance above 3-$\sigma$, while open circles stand for orbits with moderate significance between 2-$\sigma$ and 3-$\sigma$.}
\label{fig:inclinations} 
  \end{center}\end{figure}

A problem of similar studies in the past was the appearance of biases towards small orbit inclinations and thus towards large companion masses \citep{Han:2001kx, Pourbaix:2001qe}. Using simulations, we showed that our method is not affected by such bias when using an appropriate detection limit in terms of orbit significance. Figure \ref{fig:inclinations} shows the distributions of orbital inclination and semimajor axis for all orbits with better than 2-$\sigma$ significance. No preference of very small inclinations can be observed, as it would be expected from biased solutions \citep{Pourbaix:2001qe}. However, we note that no orbit is detected at inclination above $50^{\circ}$. This is explicable, because in this study we have selected stars with minimum companion masses of $M_2 \sin i = 13-80\,M_J$. As shown by \cite{Grether:2006kx}, the frequency of brown-dwarf companions is very low compared to the stellar companions and therefore small inclinations are more likely to be detected. 

We showed that our analysis is ultimately limited by the astrometric precision of the satellite and therefore cannot constrain orbits with a size below typically $a=2$~mas, which in most cases will impede the detection of a true brown dwarf companion. In addition, the Hipparcos mission duration of about 1000 days limits the periods of detectable orbits to $\sim$2000 days.

In summary, we have developed a new method to combine a radial velocity solution with Hipparcos astrometric data that has several advantages compared to previous similar approaches: it uses the new Hipparcos reduction, which simplifies the mathematical formulation and the use of the permutation test compared to the implementation of e.g. \cite{Zucker:2001ve} with the original Hipparcos data. The new method defines rigorous detection limits in terms of orbit significance and yields companion mass limits for low-significance orbits, provided they are covered by astrometric measurements. It is stringent enough to avoid the output of biased results (an example is the case of HD~38529, see \citealt{Reffert:2006ly} and \citealt{Benedict:2010ph}). Because of the general format of the new reduction IAD, our method is universal and applicable to any astrometric data\footnote{This is true provided that the implicit approximations are valid, i.e. the astrometric precision is larger than about 1~mas, see e.g. \cite{Konacki:2002fk}.}.

\subsection{Brown-dwarf companions in the {\footnotesize CORALIE} survey}
The {\footnotesize CORALIE} planet-search sample contains about 1600 Sun-like stars within 50~pc. After a duration of 12 years, 21 potential brown companions have been observed in this survey. Eighteen are discussed in this work and three additional hosts of potential brown-dwarf companions are listed by \cite{Sozzetti:2010nx}\footnote{Three more stars from this list were not included because their minimum mass is $M_2 \sin i < 13\,M_J$.} and contained in the {\footnotesize CORALIE} sample. Assuming that most candidates contained in this survey have already been found, we can derive a typical frequency of 1.3~\% for potential brown-dwarf candidates in close orbit ($<10$~AU) around Sun-like stars. Seven of these candidates are identified as stellar companions through this study. Additionally, \cite{Zucker:2001ve} showed that the companions of \object{HD~18445}, \object{HD~112758}, and \object{HD~217580} are of stellar nature. Thus, at least 10 out of 21 brown-dwarf candidates are in fact low-mass stars, which corresponds to a rate of 48~\%. Based on the uniform stellar sample surveyed by the {\footnotesize CORALIE} planet search, we thus obtain an upper limit of 0.6~\% for the frequency of close-in brown-dwarf companions around Sun-like stars within 50~pc, which is close to the frequency of $<\!0.5$~\% that was initially quoted by \cite{Marcy:2000pb}. 

\begin{figure}\begin{center} 
\includegraphics[width= \linewidth, trim = 0cm 0cm 0cm 0.5cm, clip=true]{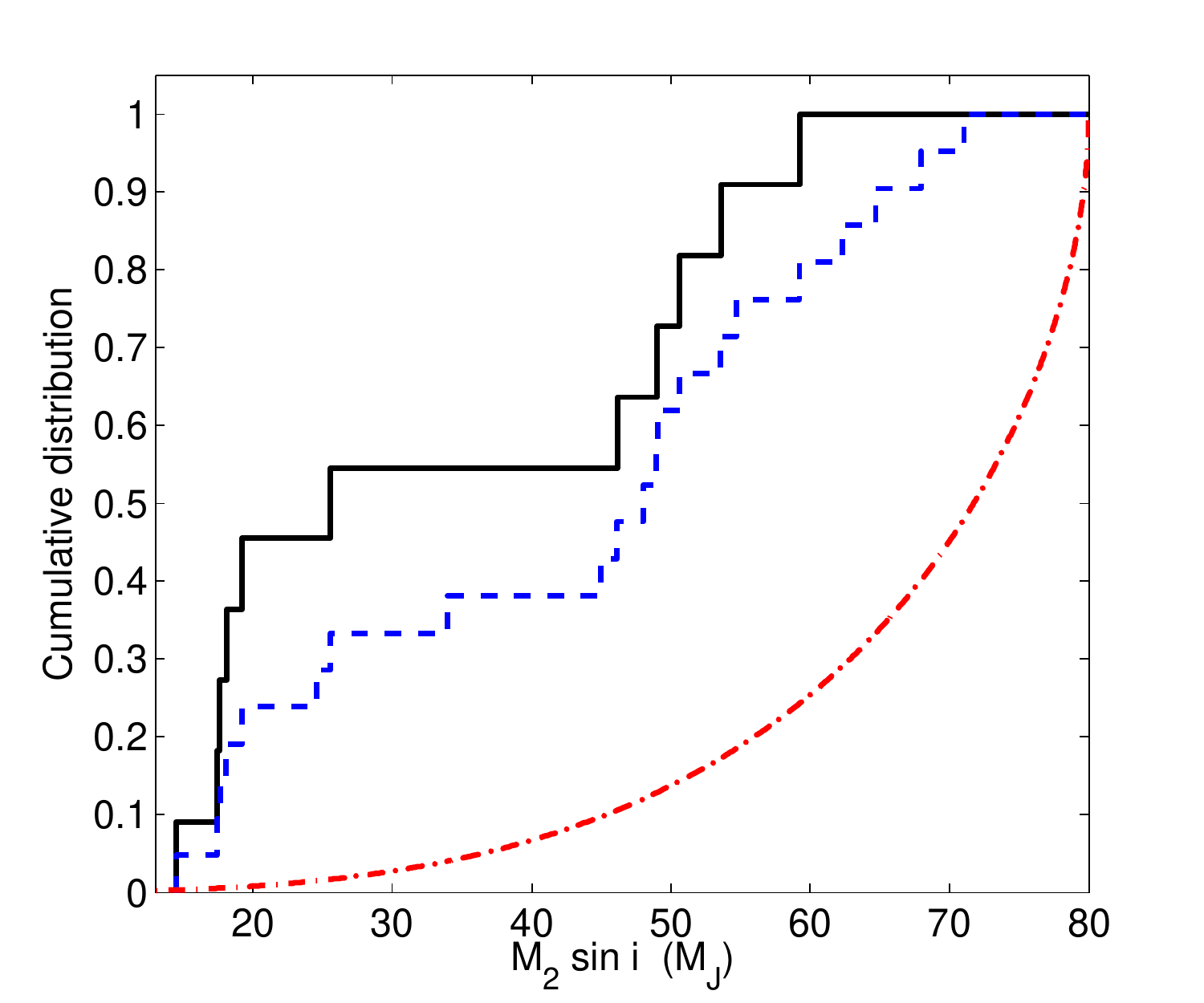}  
\caption{Cumulative mass distribution of potential brown-dwarf companions in the {\footnotesize CORALIE} survey. The blue dashed line shows the distribution of all 21 candidates. The black solid line shows the distribution for the 11 remaining candidates after removal of the 10 stellar companions. The companion of HD~38529 with $M_2=17.6\,M_J$ \citep{Benedict:2010ph} is included. For comparison, the red dash-dotted line shows the expected cumulative distribution if all companions had masses of $80\,M_J$ under the assumption that the orbits are randomly oriented in space.}
\label{fig:m2dist} 
  \end{center}\end{figure}

\begin{figure}\begin{center} 
\includegraphics[width= \linewidth, trim = 0cm 0cm 0cm 0.5cm, clip=true]{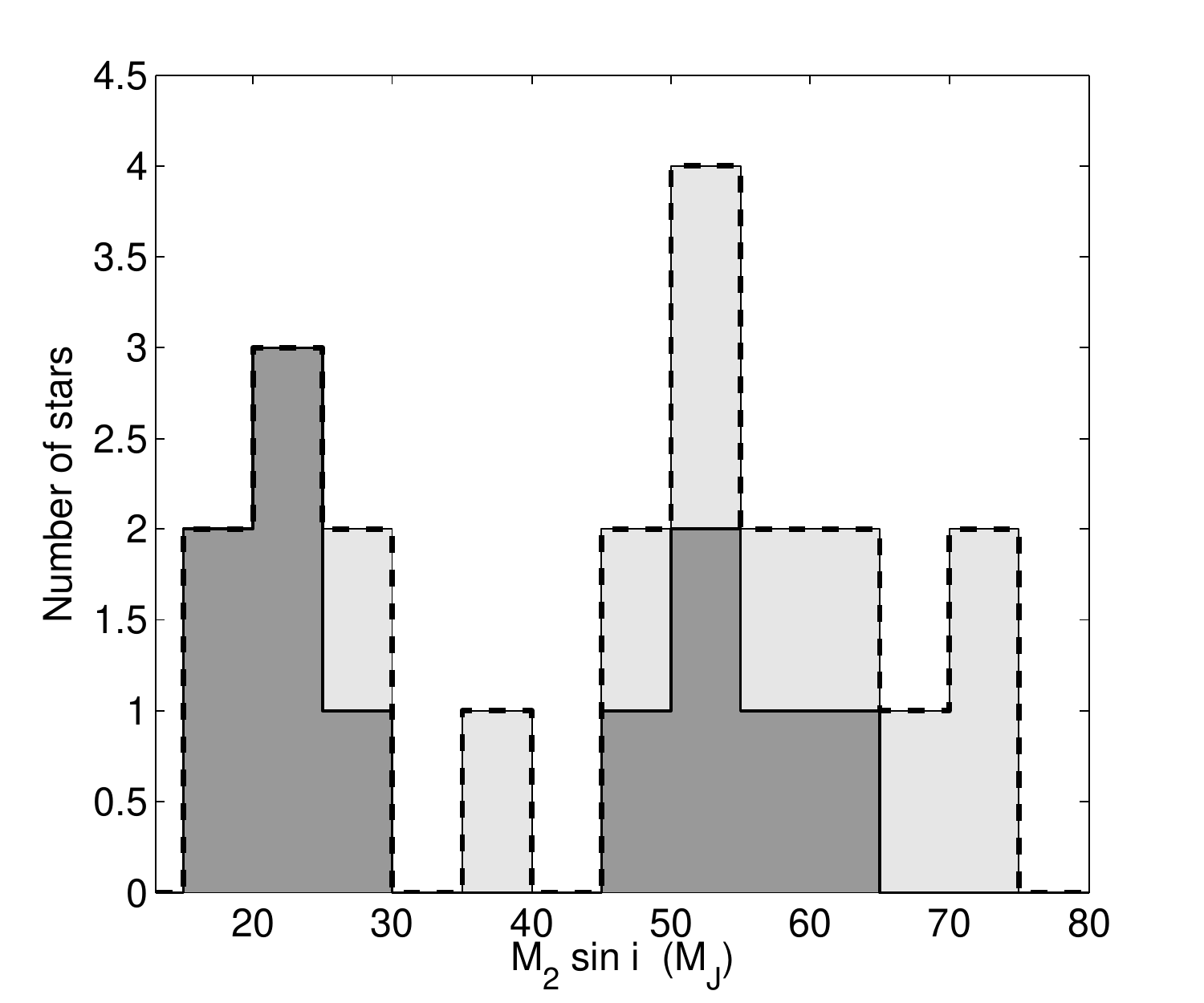}  
\caption{Histogram of minimum masses of potential brown-dwarf companions in the {\footnotesize CORALIE} survey. The light-grey histogram shows the distribution of all 21 candidates. The dark-grey histogram shows the distribution of the 11 remaining candidates after removal of the 10 stellar companions. }
\label{fig:m2histo} 
  \end{center}\end{figure}

Figure~\ref{fig:m2dist} shows the distribution of $M_2 \sin i$ in the brown-dwarf mass range for the {\footnotesize CORALIE} sample. While the cumulative distribution (blue dashed line) of all 21 candidates is reasonably compatible with a linear increase, i.e. a flat distribution function, the situation changes significantly after removal of the 10 stellar companions. As illustrated in Fig. \ref{fig:m2histo}, most stellar companions have $M_2 \sin i > 45 \,M_J$, and after their removal, a cumulative distribution with particular shape emerges (black solid line in Fig,~\ref{fig:m2dist}). It exhibits a steep ascent in the range of $13-25\,M_J$, which contains one half of the objects, followed by a slower increase up to $60\,M_J$. Surprisingly, there is no companion left with minimum mass larger than $60\,M_J$. 

In summary, the emerging cumulative distribution does not support a uniform distribution function of companion masses. The possibility, that all companions in the $13-25\,M_J$ range are in fact stellar companions seen at low inclination, is both statistically unlikely and also unplausible because their astrometric signature would be large and probably detectable. Instead, these companions supposedly have $M_2\approx M_2 \sin i$ and may represent the high-mass tail of the planetary distribution.  

We note that the distribution found here, which is based on radial velocities, is compatible with the first brown-dwarf companions found in transiting systems, such as \object{CoRoT-3b} ($M_2 = 21.7 \,M_J$, \citealt{Deleuil:2008it}) and \object{CoRoT-15b} ($M_2\! \sim\! 60\, M_J$, Bouchy et al., \emph{in preparation}).


\subsection{Metallicities of the brown-dwarf host stars}
We showed that the companion of HD~137510 is a brown dwarf. Additionally, the companion of HD~190228 can be considered a brown dwarf at 95~\% confidence. The host stars have metallicities of $\mathrm{[Fe/H]} = 0.373$~dex \citep{Butler:2006pi} and $\mathrm{[Fe/H]} = -0.24$~dex \citep{Perrier:2003dw}, respectively. Compared to the mean metallicity $<$[Fe/H]$> = -0.09$~dex with root-mean-square $\sigma_{<[Fe/H]>} = 0.24$~dex of the uniform sample of 451 nearby Sun-like stars from \cite{Sousa:2008nx}, the two brown-dwarf hosts are therefore metal-rich in one case and unremarkable in the other case. The question whether hosts of brown dwarf companions show a particular metallicity distribution, as the planet hosts do \citep{Santos:2001tg}, cannot be answered with this small sample and thus remains open. 

\section{Conclusions}\label{sec:conclusions}
We presented radial velocity solutions of 15 stars with potential brown dwarf companions, of which 9 are new discoveries. At this point, the {\footnotesize CORALIE} and {\footnotesize HARPS} surveys have increased the number of known potential brown-dwarf companions of Sun-like stars by 20~\%. 

Eight companions in our search sample were identified as being of stellar nature with masses above $80\,M_J$. These include one companion with a mass of 90~$M_J$ and thus very close to the brown-dwarf mass boundary.  The upper and lower mass limit for three companions was derived, but include the star-brown dwarf mass boundary and can therefore not be used to identify the object's nature. Yet, the companion of HD~190228 is a brown dwarf at 95~\% confidence. Upper mass limits were set for the companions of 9 stars and revealed a brown dwarf orbiting HD~137510. The masses of 12 companions could not be constrained. 
  
Our findings are in accordance with the presence of the brown-dwarf desert, because only 3 brown dwarfs are identified in our search sample of 33 companions: HD~190228b (at 95~\% confidence), HD~137510b, and HD~38529b \citep{Benedict:2010ph}. In contrast, 8 brown-dwarf candidates were identified as M dwarfs. Based on the {\footnotesize CORALIE} planet-search sample, we obtain an upper limit of 0.6~\% for the frequency of brown-dwarf companions around Sun-like stars and confirm their pronounced paucity. We find that the companion-mass distribution function is rising at the lower end of the brown-dwarf mass range, suggesting that in fact we are seeing the high-mass tail of the planetary distribution function.

We find good agreement with the previous similar studies of \cite{Halbwachs:2000rt} and \cite{Zucker:2001ve}. Our results are affected by the Hipparcos astrometric precision and mission duration, which limits the minimum detectable mass of the combined analysis. There are indications, that some of the remaining potential brown-dwarf companions are actual brown dwarfs. Those could be identified in the future, when astrometry at higher precision will be available. An example is the case of HD~38529c, whose mass we were not able to constrain with Hipparcos, but that was identified as brown dwarf by \cite{Benedict:2010ph} using {\footnotesize HST} astrometry. 

Higher-precision astrometry will be required to obtain an accurate census of low-mass companions in close orbit around Sun-like stars and to derive the relative frequencies of planets, brown dwarfs, and low-mass stars. The next major step in this direction is the {\footnotesize GAIA} astrometry satellite, which is the successor of Hipparcos and will make it possible to determine the masses of many radial-velocity companions.

\begin{acknowledgements}
We thank the numerous CORALIE observers and the Geneva observatory technical staff at the Swiss telescope for assuring high-quality radial-velocity measurements over the last decade. We thank the Swiss National Science Foundation and Geneva University for continuous support of our planet search programmes. NCS would like to thank the support by the European Research Council/European Community under the FP7 through a Starting Grant, as well from Funda\c{c}\~ao para a Ci\^encia e a Tecnologia (FCT), Portugal, through a Ci\^encia\,2007 contract funded by FCT/MCTES (Portugal) and POPH/FSE (EC), and in the form of grant reference PTDC/CTE-AST/098528/2008. SZ acknowledges financial support by the Israel Science Foundation (grant No. 119/07). This research has made use of the Smithsonian/NASA Astrophysics Data System (ADS) and of the Centre de Donn{\'e}es astronomiques de Strasbourg (CDS). All calculations were performed with the freeware \texttt{Yorick}.
\end{acknowledgements}
\bibliographystyle{aa} 
\bibliography{/Users/sahlmann/astro/papers} 
\end{document}